\documentclass[12pt]{article}
\usepackage{epsfig,amssymb,amsmath,psfrag,multirow,epstopdf,color}
\usepackage{breqn}
\usepackage{placeins}

\numberwithin{equation}{section}

\newcommand{\be}{\begin{equation}}
\newcommand{\ee}{\end{equation}}
\newcommand{\bea}{\begin{eqnarray}}
\newcommand{\eea}{\end{eqnarray}}
\newcommand{\eqn}[1]{eq.~\eqref{#1}}
\def\fig#1{fig.~{\ref{#1}}}
\def\Fig#1{Fig.~{\ref{#1}}}

\def\sect#1{section~{\ref{#1}}}
\def\eqn#1{eq.~(\ref{#1})}
\def\Eqn#1{Equation~(\ref{#1})}
\def\eqns#1#2{eqs.~(\ref{#1}) and~(\ref{#2})}

\def\tab#1{table~{\ref{#1}}}

\def\Eqn#1{Equation~(\ref{#1})}
\def\eqn#1{eq.~(\ref{#1})}
\def\eqns#1#2{eqs.~(\ref{#1}) and~(\ref{#2})}

\def\Li{{\rm Li}}

\def\ws{{w^\ast}}
\def\Su{{\mathcal{S}_u}}
\def\Sy{{\mathcal{S}_y}}
\def\GG{{\mathcal{G}}}

\def\to{\rightarrow}
\def\lr{\leftrightarrow}
\def\yn{s}

\def\PhiTilde{{\tilde{\Phi}_6}}
\def\Omegauvw{{\Omega^{(2)}(u,v,w)}}
\def\Omegavwu{{\Omega^{(2)}(v,w,u)}}
\def\Omegawuv{{\Omega^{(2)}(w,u,v)}}
\def\Huvw{{H_1(u,v,w)}}
\def\Juvw{{J_1(u,v,w)}}
\def\Kuvw{{K_1(u,v,w)}}
\def\Fuvw{{F_1(u,v,w)}}
\def\Fwuv{{F_1(w,u,v)}}
\def\Qep{Q_{\rm ep}}
\def\Rep{R_{\rm ep}}
\def\uF{{}^u\hspace{-0.8pt}F}
\def\vF{{}^v\hspace{-0.8pt}F}
\def\wF{{}^w\hspace{-0.8pt}F}

\font\cyr=wncyr8
\newcommand{\sha}{{\mbox{\cyr X}}}
\newfont{\scyr}{wncyr10 scaled 550}
\newcommand{\ssha}{\mbox{\bf \scyr X}}
\def\beq{\begin{equation}}
\def\eeq{\end{equation}}
 
\def\bsp#1\esp{\begin{split}#1\end{split}}

\newcommand{\rat}[2]{{#1\over#2}}
 
\begin{document}

%%%%%%%%%%%%%%%%%%%%%%%%%%%%%%%%%%%%%%%%%%%%%%%%%%%%%%%%%%%%%%%%%%%%%%%%
% Shamelessly stolen from Thorsten's thohacks.sty
%%%%%%%%%%%%%%%%%%%%%%%%%%%%%%%%%%%%%%%%%%%%%%%%%%%%%%%%%%%%%%%%%%%%%%%%
\catcode`\@=11
\font\manfnt=manfnt
\def\Watchout{\@ifnextchar [{\W@tchout}{\W@tchout[1]}}
\def\W@tchout[#1]{{\manfnt\@tempcnta#1\relax%
  \@whilenum\@tempcnta>\z@\do{%
    \char"7F\hskip 0.3em\advance\@tempcnta\m@ne}}}
\let\foo\W@tchout
\def\dubious{\@ifnextchar[{\@dubious}{\@dubious[1]}}
\let\enddubious\endlist
\def\@dubious[#1]{%
  \setbox\@tempboxa\hbox{\@W@tchout#1}
  \@tempdima\wd\@tempboxa
  \list{}{\leftmargin\@tempdima}\item[\hbox to 0pt{\hss\@W@tchout#1}]}
\def\@W@tchout#1{\W@tchout[#1]}
\catcode`\@=12
%%%%%%%%%%%%%%%%%%%%%%%%%%%%%%%%%%%%%%%%%%%%%%%%%%%%%%%%%%%%%%%%%%%%%%%%

%%%%%%%%%%%%%%%%%%%%%%%%%%%%%%%%%%%%%%%%%%%%%%%%%%%%

\thispagestyle{empty}

\begin{flushright}
SLAC--PUB--15545\hfill 
SU-ITP-13/19\hfill 
CERN-PH-TH/2013-142 \hfill
LAPTH-033/13
\end{flushright}

\begingroup\centering
{\Large\bfseries\mathversion{bold}
Hexagon functions and\\
the three-loop remainder function\par}%
\vspace{7mm}

\begingroup\scshape\large
Lance~J.~Dixon$^{(1)}$, James~M.~Drummond$^{(2,3)}$, 
Matt von Hippel$^{(1,4)}$,\\ and Jeffrey Pennington$^{(1)}$\\
\endgroup
\vspace{5mm}
\begingroup\small
$^{(1)}$\emph{SLAC National Accelerator Laboratory,
Stanford University, Stanford, CA 94309, USA} \\
$^{(2)}$\emph{CERN, Geneva 23, Switzerland} \\
$^{(3)}$\emph{LAPTH, CNRS et Universit\'e de Savoie,  F-74941 Annecy-le-Vieux
  Cedex, France}\\
$^{(4)}$\emph{Simons Center for Geometry and Physics,
Stony Brook University,
Stony Brook NY 11794 }\endgroup

\vspace{0.4cm}
\begingroup\small
E-mails:\\
{\tt lance@slac.stanford.edu}, {\tt drummond@cern.ch}, \\
{\tt matthew.vonhippel@stonybrook.edu}, {\tt jpennin@stanford.edu}\endgroup
\vspace{0.7cm}

\textbf{Abstract}\vspace{5mm}\par
\begin{minipage}{14.7cm}
We present the three-loop remainder function, which describes the
scattering of six gluons in the maximally-helicity-violating configuration
in planar ${\cal N}=4$ super-Yang-Mills theory, as a function of the three
dual conformal cross ratios.  The result can be expressed in terms of
multiple Goncharov polylogarithms.  We also employ a more restricted class
of {\it hexagon functions} which have the correct branch cuts and certain
other restrictions on their symbols.  We classify all the hexagon functions
through transcendental weight five, using the coproduct for their Hopf
algebra iteratively, which amounts to a set of first-order
differential equations.  The three-loop remainder function is a particular
weight-six hexagon function, whose symbol was determined previously.
The differential equations can be integrated
numerically for generic values of the cross ratios, or analytically in
certain kinematic limits, including the near-collinear and multi-Regge
limits.  These limits allow us to impose constraints from the operator
product expansion and multi-Regge factorization directly at the function
level, and thereby to fix uniquely a set of Riemann $\zeta$ valued constants
that could not be fixed at the level of the symbol.
The near-collinear limits agree precisely with recent predictions by
Basso, Sever and Vieira based on integrability.  The multi-Regge limits
agree with the factorization formula of Fadin and Lipatov, and determine
three constants entering the impact factor at this order.
We plot the three-loop remainder function for various slices of
the Euclidean region of positive cross ratios, and compare it to
the two-loop one.  For large ranges of the cross ratios, the ratio
of the three-loop to the two-loop remainder function is relatively constant,
and close to $-7$.  
\end{minipage}\par
\endgroup

\newpage

\tableofcontents

\newpage

%%%%%%%%%%%%%%%%%%%%%%%%%%%%%%%%%%%%%%%%%%%%%%%%%%%%%%%%%%%%%%%%%
\section{Introduction}

For roughly half a century we have known that many physical properties
of scattering amplitudes in quantum field theories are encoded in
different kinds of analytic behavior in various regions of the
kinematical phase space.  The idea that the amplitudes of a theory can
be reconstructed (or `bootstrapped') from basic physical principles
such as unitarity, by exploiting the link to the analytic behavior,
became known as the ``Analytic $S$-Matrix program''
(see {\it e.g.}~ref.~\cite{ELOP}).  In the narrow resonance approximation,
crossing symmetry duality led to the Veneziano formula~\cite{Veneziano1968yb}
for tree-level scattering amplitudes in string theory.

In conformal field theories, there exists a different kind of bootstrap
program, whereby correlation functions can be determined by imposing
consistency with the operator product expansion (OPE), crossing
symmetry, and unitarity~\cite{Ferrara1973yt,Polyakov1974gs}.
This program was most successful in two-dimensional conformal field
theories, for which conformal symmetry actually extends to an
infinite-dimensional Virasoro symmetry~\cite{Belavin1984vu}.
However, the basic idea can be applied in any dimension and recent progress
has been made in applying the program to conformal field theories in three
and four dimensions~\cite{Rattazzi2008pe,ElShowk2012ht,Beem2013qxa}.

In recent years, the scattering amplitudes of the planar $\mathcal{N}=4$
super-Yang-Mills theory have been seen to exhibit remarkable
properties. In particular, the amplitudes exhibit dual conformal
symmetry and a duality to light-like polygonal Wilson
loops~\cite{DualConformal}.  The dual description and its associated
conformal symmetry mean that CFT techniques can be applied to
calculating scattering amplitudes. In particular, the idea of imposing
consistency with the OPE applies.  However, since the dual observables
are non-local Wilson loop operators, a different OPE, involving the
near-collinear limit of two sides of the light-like polygon, has to be
employed~\cite{Alday2010ku,Gaiotto2010fk,Gaiotto2011dt,Sever2011da}.

Dual conformal symmetry implies that the amplitudes involving four or
five particles are fixed, because there are no invariant cross ratios
that can be formed from a five-sided light-like
polygon~\cite{Drummond2007cf}.  The four- and five-point amplitudes
are governed by the BDS ansatz~\cite{Bern2005iz}.
The amplitudes not determined by dual conformal symmetry begin at six points.
When the external gluons are in the maximally-helicity-violating (MHV)
configuration, such amplitudes can be expressed in terms of the BDS ansatz,
which contains all of the infrared divergences and transforms anomalously
under dual conformal invariance, and a so-called
``remainder function''~\cite{Bern2008ap}, which only depends
on dual-conformally-invariant cross ratios.
In the case of non-MHV amplitudes, one can define the ``ratio 
function''~\cite{Drummond2008vq}, which depends on the cross ratios
as well as dual superconformal invariants.
For six external gluons, the remainder and ratio functions are described
in terms of functions of three dual conformal cross ratios.

At low orders in perturbation theory, these latter functions can be
expressed in terms of multiple polylogarithms.  In general, multiple
polylogarithms are functions of many variables that can be defined as
iterated integrals over rational kernels.
A particularly useful feature of such functions is that they
can be classified according to their symbols~\cite{Chen,FBThesis,Gonch},
elements of the $n$-fold tensor product of the algebra of rational functions. 
The integer $n$ is referred to as the transcendental weight or degree.
The symbol can be defined iteratively in terms of the total derivative of
the function, or alternatively, in terms of the maximally iterated coproduct
by using the Hopf structure conjecturally satisfied by multiple  
polylogarithms~\cite{Gonch2,Brown2011ik}. Complicated functional identities among 
polylogarithms  become simple algebraic relations satisfied by their symbols, making
the symbol a very useful tool in the study of polylogarithmic functions.
The symbol can miss terms in the function that are proportional to transcendental 
constants (which in the present case are all multiple zeta values), so special care
must be given to account for these terms.
The symbol and coproduct have been particularly useful in recent field theory
applications~\cite{Goncharov2010jf,Gaiotto2011dt,Dixon2011pw,Dixon2011nj,%
Duhr2012fh}.  In the case of $\mathcal{N}=4$ super-Yang-Mills theory,
all amplitudes computed to date have exhibited a uniform maximal 
transcendentality, in which the finite terms (such as the remainder or ratio
functions) always have weight $n=2L$ at the $L$ loop order in
perturbation theory.

Based on the simplified form of the two-loop six-point remainder function
obtained in ref.~\cite{Goncharov2010jf} (which was first constructed
analytically in terms of multiple polylogarithms~\cite{DelDuca2009au}),
it was conjectured~\cite{Dixon2011pw,Dixon2011nj} that for multi-loop
six-point amplitudes, both the MHV remainder function and the next-to-MHV 
(NMHV) ratio function are described in terms of polylogarithmic functions
whose symbols are made from an alphabet of nine letters. 
The nine letters are related to the nine projectively-inequivalent
differences $z_{ij}$ of projective variables
$z_i$~\cite{Goncharov2010jf}, which can also be represented in terms
of momentum twistors~\cite{Hodges2009hk}.
Using this conjecture, the symbol for the three-loop six-point remainder
function was obtained up to two undetermined parameters~\cite{Dixon2011pw},
which were later fixed~\cite{CaronHuot2011kk} using a dual supersymmetry
``anomaly'' equation~\cite{CaronHuot2011kk,Bullimore2011kg}. 
The idea of ref.~\cite{Dixon2011pw} was to start with an ansatz for the
symbol, based on the above nine-letter conjecture, and then impose various
mathematical and physical consistency conditions. For example, imposing a simple 
integrability condition~\cite{FBThesis,Gonch} guarantees that the ansatz is actually
the symbol of some function, and demanding that the amplitude has physical branch cuts 
leads to a condition on the initial entries of the symbol. 

Because of the duality between scattering amplitudes and Wilson loops, one can also 
impose conditions on the amplitude that are more naturally expressed in terms of the 
Wilson loop, such as those based on the OPE satisfied by its near-collinear limit.  In
refs.~\cite{Alday2010ku,Gaiotto2010fk,Gaiotto2011dt,Sever2011da},
the leading-discontinuity terms in the OPE were computed.
In terms of the cross ratio variable that vanishes in the near-collinear
limit, the leading-discontinuity terms correspond to just the maximum powers
of logarithms of this variable ($L-1$ at $L$ loops), although they can
be arbitrarily power suppressed.  These terms require only the one-loop
anomalous dimensions of the operators corresponding to excitations
of the Wilson line, or flux tube.  That is, higher-loop corrections to the 
anomalous dimensions and to the OPE coefficients can only generate subleading
logarithmic terms.  While the leading-discontinuity information is sufficient to
determine all terms in the symbol at two loops, more information is necessary
starting at three loops~\cite{Dixon2011pw}.

Very recently, a new approach to polygonal Wilson loops has been set
forth~\cite{Basso2013vsa,Basso2013aha}, which is fully nonperturbative
and based on integrability.  The Wilson loop is partitioned into a
number of ``pentagon transitions'', which are labeled by flux tube
excitation states on either side of the transition.  (If one edge of the
pentagon coincides with an edge of the Wilson loop, then the corresponding
state is the flux tube vacuum.)  The pentagon transitions obey a set of
bootstrap consistency conditions.  Remarkably, they can be solved in terms
of factorizable $S$ matrices for two-dimensional scattering of the
flux tube excitations~\cite{Basso2013vsa,Basso2013aha}.

In principle, the pentagon transitions can be solved for arbitrary
excitations, but it is simplest to first work out the low-lying
excitations, which correspond to the leading power-suppressed terms in the 
near-collinear limit in the six-point case (and similar terms in 
multi-near-collinear limits for more than six particles).
Compared with the earlier leading-discontinuity data, now {\it all}
terms at a given power-suppressed order can be determined (to all
loop orders), not just the leading logarithms.  This information is
very powerful.
The first power-suppressed order in the six-point near-collinear limit
is enough to fix the two terms in the ansatz for the
symbol of the three-loop remainder function that could not be fixed 
using the leading discontinuity~\cite{Basso2013vsa}.
At four loops, the first power-suppressed order~\cite{Basso2013vsa}
and part of the second power-suppressed order~\cite{BSVPrivate}
are sufficient to fix all terms in the symbol~\cite{DDDPToAppear}.
At these orders, the symbol becomes heavily over-constrained,
providing strong cross checks on the assumptions about the letters
of the symbol, as well as on the solutions to the pentagon transition
bootstrap equations.

In short, the application of integrability to the pentagon-transition
decomposition of Wilson loops provides, through the OPE, all-loop-order
boundary-value information for the problem of determining Wilson
loops (or scattering amplitudes) at generic nonzero (interior) values of the
cross ratios. We will use this information in the six-point case to uniquely 
determine the three-loop remainder function, not just at symbol level, but
at function level as well. 

A second limit we study is the limit of multi-Regge kinematics (MRK), 
which has provided
another important guide to the perturbative structure of the six-point
remainder function~\cite{Bartels2008ce,Bartels2008sc,Schabinger2009bb,%
Lipatov2010qg,Lipatov2010ad,Bartels2010tx,Dixon2011pw,Fadin2011we},
as well as higher-point remainder 
functions~\cite{Prygarin2011gd,Bartels2011ge}
and NMHV amplitudes~\cite{Lipatov2012gk}.
The six-point remainder function and, more generally, 
the hexagon functions that we define shortly have simple behavior in
the multi-Regge limit. These functions depend on three dual-conformally-invariant
cross ratios, but in the multi-Regge limit they collapse~\cite{Dixon2012yy}
into single-valued harmonic polylogarithms~\cite{BrownSVHPLs}, which are
functions of two surviving real variables, or of a complex variable and
its conjugate.  The multi-Regge limit factorizes~\cite{Fadin2011we}
after taking the Fourier-Mellin transform of this complex variable.
This factorization imposes strong constraints on the remainder
function at high loop order~\cite{Fadin2011we,Dixon2012yy,Pennington2012zj}.

Conversely, determining the multi-loop remainder function, or just
its multi-Regge limit, allows the perturbative extraction of
the two functions that enter the factorized form of the amplitude,
the BFKL eigenvalue (in the adjoint representation) and a
corresponding impact factor.
This approach makes use of a map between the single-valued harmonic
polylogarithms and their Fourier-Mellin transforms, which can be constructed
from harmonic sums~\cite{Dixon2012yy}.  Using the three- and four-loop
remainder-function symbols, the BFKL eigenvalue has been determined
to next-to-next-to-leading-logarithmic accuracy (NNLLA), and the
impact factor at NNLLA and N$^3$LLA~\cite{Dixon2012yy}.  However, the
coefficients of certain transcendental constants in these three
quantities could not be fixed, due to the limitation of the symbol.
Here we will use the MRK limit at three loops to
fix the three undetermined constants in the NNLLA impact
factor.  Once the four-loop remainder function is determined, a
similar analysis will fix the undetermined constants in the NNLLA
BFKL eigenvalue and in the N$^3$LLA impact factor.

In general, polylogarithmic functions are not sufficient to describe
scattering amplitudes.  For example, an elliptic integral, in which
the kernel is not rational but contains a square root, enters the
two-loop equal-mass sunrise graph~\cite{Laporta2004rb},
and it has been shown that a very similar type of integral
enters a particular N$^3$MHV 10-point scattering amplitude
in planar ${\cal N}=4$ super-Yang-Mills theory~\cite{CaronHuot2012ab}. 
However, it has been argued~\cite{ArkaniHamed2012nw}, based on a 
novel form of the planar loop integrand, that MHV and NMHV amplitudes 
can all be described in terms of multiple polylogarithms alone.
Similar ``$d$log'' representations have appeared in a recent
twistor-space formulation~\cite{LipsteinMason}.
Because six-particle amplitudes are either MHV (or the parity conjugate 
${\overline {\rm MHV}}$) or NMHV, we expect that multiple polylogarithms
and their associated symbols should suffice in this case.
The nine letters that we assume for the symbol then follow naturally
from the fact that the kinematics can be described in terms of dual
conformally invariant combinations of six momentum
twistors~\cite{Hodges2009hk}.

Having the symbol of an amplitude is not the same thing as having the
function.  In order to reconstruct the function one first needs a 
representative, well-defined function in the class of multiple
polylogarithms which has the correct symbol.  Before enough physical
constraints are imposed, there will generally be multiple functions
matching the symbol, because of the symbol-level ambiguity associated
with transcendental constants multiplying well-defined functions of
lower weight.  Here we will develop techniques for building up the
relevant class of functions for hexagon kinematics, which we call
{\it hexagon functions}, whose symbols are as described above, but which 
are well-defined and have the proper branch cuts at the function level as well.
We will argue that the hexagon functions form the basis for a
perturbative solution to the MHV and NMHV six-point problem.

We will pursue two complementary routes toward the construction of
hexagon functions.  The first route is to express them explicitly in
terms of multiple polylogarithms.  This route has the advantage of being
completely explicit in terms of functions with well-known mathematical
properties, which can be evaluated numerically quite quickly, or
expanded analytically in various regions.  However, it also has the 
disadvantages that the representations are rather lengthy, and they are 
specific to particular regions of the full space of cross ratios.

The second route we pursue is to define each weight-$n$ hexagon function
iteratively in the weight, using the three first-order differential
equations they satisfy.  This information can also be codified by
the $\{n-1,1\}$ component of the coproduct of the function,
whose elements contain weight-$(n-1)$ hexagon functions (the source terms
for the differential equations).  The differential equations can
be integrated numerically along specific contours in the
space of cross ratios.  In some cases, they can be integrated analytically,
at least up to the determination of certain integration constants.

We can carry out numerical comparisons of the two approaches in regions
of overlapping validity.  We have also been able to determine the
near-collinear and multi-Regge limits of the functions analytically
using both routes.  As mentioned above, these limits are how we fix 
all undetermined constants in the function-level ansatz, and how we
extract additional predictions for both regimes.

We have performed a complete classification of hexagon functions through
weight five. Although the three-loop remainder function is a hexagon function
of weight six, its construction is possible given the weight-five basis.
There are other potential applications of our classification, beyond the
three-loop remainder function. One example is the three-loop six-point
NMHV ratio function, whose components are expected~\cite{Dixon2011nj}
to be hexagon functions of weight six.  Therefore, it should be possible
to construct the ratio function in an identical fashion to the remainder
function.

Once we have fixed all undetermined constants in the three-loop
remainder function, we can study its behavior in various
regions, and compare it with the two-loop function.
On several lines passing through the space of cross ratios,
the remainder function collapses to simple combinations of
harmonic polylogarithms of a single variable.
Remarkably, over vast swathes of the space of positive cross ratios,
the two- and three-loop remainder functions are strikingly similar, 
up to an overall constant rescaling. This similarity is in spite of the fact
that they have quite different analytic behavior along various edges of this
region.  We can also compare the perturbative remainder function with 
the result for strong coupling, computed using the AdS/CFT correspondence,
along the line where all three cross ratios are equal~\cite{Alday2009dv}.
We find that the two-loop, three-loop and strong-coupling results
all have a remarkably similar shape when the common cross ratio is
less than unity.  Although we have not attempted any kind of interpolation
formula from weak to strong coupling, it seems likely from the comparison
that the nature of the interpolation will depend very weakly on the
common cross ratio in this region.

The similarity of the weak and strong coupling limits of
remainder functions has been noticed before.
For the eight-point case in two-dimensional (AdS$_3$) kinematics,
there are two real kinematical parameters.
Refs.~\cite{Brandhuber2009da,DelDuca2010zp}
found impressive numerical agreement, to within 3\% or better,
between rescaled versions of the two-loop~\cite{Anastasiou2009kna}
and strong-coupling~\cite{Alday2009yn} octagon remainder function,
as a function of both parameters.  The octagon, decagon, and general
$2n$-point remainder functions were evaluated at strong coupling,
analytically in an expansion around the regular polygon limit,
and the rescaled functions were found to be very similar to the two-loop
result~\cite{Hatsuda}.  The six-point case we consider in this paper
has also been studied at strong coupling using a $Z_4$ symmetric
integrable model~\cite{Hatsuda2010vr} and using the homogeneous sine-Gordon
model and conformal perturbation theory~\cite{Hatsuda2012pb}.
In the latter work, the strong-coupling result was compared with
the two-loop one along a one-dimensional curve in the space of the
three cross ratios, corresponding to the trajectory of
an integrable renormalization group flow.  Again, good numerical
agreement was found between the two rescaled functions.

The remainder of this paper is organized as follows.  In \sect{sec:symbol}
we recall some properties of pure functions (iterated integrals)
and their symbols, as well as a representation of the two-loop remainder
function (and its symbol) in terms of an ``extra pure'' function and its cyclic
images.  We use this representation as motivation for an analogous
decomposition of the three-loop symbol.  In \sect{sec:hex_multi_poly}
we describe the first route to constructing hexagon functions, via
multiple polylogarithms.   In \sect{sec:integral_reps} we describe
the second route to constructing the same set of functions, via
the differential equations they satisfy.  In \sect{sec:collinear}
we discuss how to extract the near-collinear limits, and give results
for some of the basis functions and for the remainder function in 
this limit.  In \sect{sec:MRK} we carry out the analogous discussion
for the Minkowski multi-Regge limit.  In \sect{sec:Final} 
we give the final result for the three-loop remainder function,
in terms of a specific integral, as well as defining it through
the $\{5,1\}$ components of its coproduct.  We also present
the specialization of the remainder function onto various lines in the
three-dimensional space of cross ratios; along these lines its
form simplifies dramatically.  Finally, we plot the function on several lines
and two-dimensional slices.  We compare it numerically to the two-loop
function in some of these regions, and to the strong-coupling
result evaluated for equal cross ratios.  In \sect{sec:conclusions} we
present our conclusions and outline avenues for future research.
We include three appendices.  Appendix~\ref{sec:app_multi_poly}
provides some background material on multiple polylogarithms.
Appendix~\ref{sec:app_basis} gives the complete set of independent
hexagon functions through weight five in terms of the $\{n-1,1\}$ components
of their coproducts, and in appendix~\ref{sec:app_Rep} we
provide the same description of the extra pure weight six function $\Rep$
entering the remainder function.

In attached, computer-readable files we give the basis of hexagon
functions through weight five, as well as the three-loop
remainder function, expressed in terms of multiple polylogarithms
in two different kinematic regions. We also provide the near-collinear
and multi-Regge limits of these functions.

\vfill\eject

%%%%%%%%%%%%%%%%%%%%%%%%%%%%%%%%%%%%%%%%%%%%%%%%%%%%%%%%%%%%%%%%%%%%

\section{Extra-pure functions and the symbol of $R_6^{(3)}$}
\label{sec:symbol}

In this section, we describe the symbol of the three-loop remainder
function as obtained in ref.~\cite{Dixon2011pw}, which is the starting
point for our reconstruction of the full function.  Motivated by an
alternate representation~\cite{Dixon2011nj} of the two-loop remainder
function, we will rearrange the three-loop symbol.  In the new representation,
part of the answer will involve products of lower-weight (hence simpler)
functions, and the rest of the answer will be expressible as the sum of an
{\it extra-pure} function, called $\Rep$, plus its two images under 
cyclic permutations of the cross ratios.
An extra-pure function of $m$ variables, by definition,
has a symbol with only $m$ different final entries.  For the case of
hexagon kinematics, where there are three cross ratios, the symbol
of an extra-pure function has only three final entries, instead of the
potential nine.  Related to this, the three derivatives of the full function
can be written in a particularly simple form, which helps somewhat in its
construction.

All the functions we consider in this paper will be {\it pure functions}. 
The definition of a pure function $f^{(n)}$ of transcendental weight
(or degree) $n$ is that its first derivative obeys,
\be
d f^{(n)} = \sum_r f^{(n-1)}_r d \ln \phi_r \,,
\label{pure}
\ee
where $\phi_r$ are rational functions and the sum over $r$ is
finite. The only weight-zero functions are assumed to be rational
constants.  The $f_r^{(n-1)}$ and $\phi_r$ are not all independent of each
other because the integrability condition $d^2 f^{(n)} = 0$ imposes relations
among them,
\be
\sum_r d f^{(n-1)}_r \wedge d \ln \phi_r = 0\,.
\label{integrability}
\ee
Functions defined by the above conditions are iterated integrals of
polylogarithmic type.  Such functions have a {\it symbol}, defined
recursively as an element of the $n$-fold tensor product of the
algebra of rational functions, following \eqn{pure},
\be
\mathcal{S}\bigl( f^{(n)} \bigr) 
= \sum_r \mathcal{S} \bigl( f^{(n-1)}_r \bigr) \otimes \phi_r\,.
\ee
In the case of the six-particle amplitudes of planar 
$\mathcal{N}=4$ super Yang-Mills theory, we are interested in pure 
functions depending on the three dual conformally invariant cross ratios,
\be
u_1 = u = \frac{x_{13}^2 x_{46}^2}{x_{14}^2 x_{36}^2}\,, \qquad 
u_2 = v = \frac{x_{24}^2 x_{51}^2}{x_{25}^2 x_{41}^2}\,, \qquad 
u_3 = w = \frac{x_{35}^2 x_{62}^2}{x_{36}^2 x_{52}^2}\,.
\ee
The six particle momenta $k_i^\mu$ are differences of the
dual coordinates $x_i^\mu$:  $x_i^\mu - x_{i+1}^\mu = k_i^\mu$, 
with indices taken mod 6.

Having specified the class of functions we are interested in, we
impose further~\cite{Dixon2011pw,Dixon2011nj} that the entries of the
symbol are drawn from the following set of nine letters,
\be
\Su = \{u,v,w,1-u,1-v,1-w,y_u,y_v,y_w\}\,.
\label{nineletters}
\ee
The nine letters are related to the nine projectively-inequivalent 
differences of six $\mathbb{CP}^1$ variables $z_i$~\cite{Goncharov2010jf}
via
\be
u = \frac{(12)(45)}{(14)(25)}, \qquad 1-u = \frac{(24)(15)}{(14)(25)}\,, 
\qquad y_u = \frac{(26)(13)(45)}{(46)(12)(35)}\,,
\ee
and relations obtained by cyclically rotating the six points. 
The variables $y_u$, $y_v$ and $y_w$ can be expressed locally in terms of
the  cross ratios,
\be
y_u = \frac{u-z_+}{u-z_-}\,, \qquad y_v = \frac{v-z_+}{v-z_-}\,, 
\qquad y_w = \frac{w - z_+}{w - z_-}\,,
\label{yfromu}
\ee
where
\be
z_\pm = \frac{1}{2}\Bigl[-1+u+v+w \pm \sqrt{\Delta}\Bigr]\,, 
\qquad \Delta = (1-u-v-w)^2 - 4 uvw\,.
\ee

Note that under the cyclic permutation $z_i \to z_{i+1}$ we have 
$u \to v \to w \to u$, while the $y_i$ variables transform as
$y_u \to 1/y_v \to y_w \to 1/y_u$. 
A three-fold cyclic rotation amounts to a space-time parity 
transformation, under which the parity-even cross ratios are invariant,
while the parity-odd $y$ variables invert.
Consistent with the inversion of the $y$ variables under parity,
and with \eqn{yfromu}, the quantity $\sqrt{\Delta}$ must flip sign under parity,
so we have altogether,
\be
{\rm Parity:}\quad  u_i \to u_i \,, \quad y_i \to \frac{1}{y_i} \,, 
\quad \sqrt{\Delta} \to -\sqrt{\Delta} \,.
\label{parityuydelta}
\ee
The transformation of $\sqrt{\Delta}$ can also be seen from its 
representation in terms of the $z_{ij}$
variables,
\be
\sqrt{\Delta} = \frac{(12)(34)(56)-(23)(45)(61)}{(14)(25)(36)}\,,
\ee
upon letting $z_i\to z_{i+3}$.  It will prove very useful to classify
hexagon functions by their parity.  The remainder function is a parity-even
function, but some of its derivatives (or more precisely
coproduct components) are parity odd, so we need to understand
both the even and odd sectors.

Since the $y$ variables invert under parity, $y_u \to
1/y_u$, {\it etc.}, it is often better to think of the $y$ variables as
fundamental and the cross ratios as parity-even functions of them.
The cross ratios can be expressed in terms of the $y$ variables without
any square roots,
\bea
u &=& \frac{y_u (1 - y_v) (1 - y_w)}{(1 - y_u y_v) (1 - y_u y_w)}\,, 
\quad v = \frac{y_v (1 - y_w) (1 - y_u)}{(1 - y_v y_w) (1 - y_v y_u)}\,, 
\quad w = \frac{y_w (1 - y_u) (1 - y_v)}{(1 - y_w y_u) (1 - y_w y_v)}\,,
\nonumber\\
1-u &=& \frac{(1-y_u) (1-y_u y_v y_w)}{(1 - y_u y_v) (1 - y_u y_w)} \,,
\ \ {\rm etc.},
\quad
\sqrt{\Delta} = \frac{(1-y_u) (1-y_v) (1-y_w) (1-y_u y_v y_w)}
                       {(1-y_u y_v) (1-y_v y_w) (1-y_w y_u)} \,,
\label{u_from_y}
\eea
where we have picked a particular branch of $\sqrt{\Delta}$.

Following the strategy of ref.~\cite{Dixon2011pw}, we
construct all integrable symbols of the required weight, 
using the letters~(\ref{nineletters}),
subject to certain additional physical constraints. In the case of the
six-point MHV remainder function at $L$ loops, we require the symbol to
be that of a weight-$2L$ 
parity-even function with full $S_3$ permutation symmetry among the
cross ratios. The initial entries in the symbol can only be
the cross ratios themselves, in order to have physical branch
cuts~\cite{Gaiotto2011dt}:
\be
\hbox{first entry} \in \left\{ u,v,w \right\}\,.
\label{firstentry}
\ee
In addition we require that the final entries of the symbol are taken
from the following restricted set of six
letters~\cite{CaronHuot2011ky,Dixon2011pw}:
\be
\hbox{final entry} \in 
\left\{ \frac{u}{1-u}, \frac{v}{1-v}, \frac{w}{1-w}, y_u, y_v,y_w \right\}\,.
\label{finalentry}
\ee
Next one can apply constraints from the collinear OPE of Wilson
loops.  The leading-discontinuity
constraints~\cite{Alday2010ku,Gaiotto2010fk,Gaiotto2011dt} can be
expressed in terms of differential operators with a simple action on
the symbol~\cite{Dixon2011pw}.  At two loops, the leading (single) discontinuity
is the only discontinuity, and it is sufficient to determine
the full remainder function $R_6^{(2)}(u,v,w)$~\cite{Gaiotto2010fk}.
At three loops, the constraint on the leading (double) discontinuity leaves two free
parameters in the symbol, $\alpha_1$ and $\alpha_2$~\cite{Dixon2011pw}.
These parameters were determined in refs.~\cite{CaronHuot2011kk,Basso2013vsa},
but we will leave them arbitrary here to see what other information
can fix them.

The two-loop remainder function $R_6^{(2)}$ can be expressed simply
in terms of classical polylogarithms~\cite{Goncharov2010jf}.
However, here we wish to recall the form found in ref.~\cite{Dixon2011nj}
in terms of the infrared-finite double pentagon integral $\Omega^{(2)}$,
which was introduced in ref.~\cite{ArkaniHamed2010kv} and studied further in
refs.~\cite{Drummond2010cz,Dixon2011nj}:
\be
R_6^{(2)}(u,v,w) = \frac{1}{4} \Bigl[ \Omega^{(2)}(u,v,w) + \Omega^{(2)}(v,w,u) 
+ \Omega^{(2)}(w,u,v) \Bigr] + R_{6,\rm rat}^{(2)}(u,v,w) \,.
\label{R62decomp}
\ee
The function $R_{6,{\rm rat}}^{(2)}$ can be expressed in terms of 
single-variable classical polylogarithms,
\be\label{R62ratdef}
R_{6, {\rm rat}}^{(2)} = 
- \frac{1}{2} \biggl[ 
\frac{1}{4}\Bigl( {\rm Li}_2(1-1/u) + {\rm Li}_2(1-1/v)
                  + {\rm Li}_2(1-1/w)\Bigr)^2
+ r(u) + r(v) + r(w) - \zeta_{4} \biggr]\,,
\ee
with
\begin{align}\label{rudef}
r(u) = & - {\rm Li}_4(u) - {\rm Li}_4(1-u) + {\rm Li}_4(1-1/u)
- \ln u \, {\rm Li}_3(1-1/u) - \frac{1}{6} \, \ln^3 u \, \ln(1-u) \notag \\
& + \frac{1}{4} \Bigl({\rm Li}_2(1-1/u) \Bigr)^2 + \frac{1}{12} \ln^4 u
+ \zeta_{2} \Bigl({\rm Li}_2(1-u)+\ln^2 u \Bigr)+ \zeta_{3} \, \ln u \,.
\end{align}

We see that $R_{6,{\rm rat}}^{(2)}$ decomposes into a product of simpler,
lower-weight functions ${\rm Li}_2(1-1/u_i)$, plus the cyclic images
of the function $r(u)$, whose symbol can be written as,
\be
\mathcal{S}\bigl( r(u) \bigr) 
= -2\ u \otimes \frac{u}{1-u} \otimes \frac{u}{1-u} 
                \otimes \frac{u}{1-u}
\ +\ \frac{1}{2}\ u \otimes \frac{u}{1-u} \otimes u
                \otimes \frac{u}{1-u} \,.
\label{r_symbol}
\ee

The symbol of $\Omega^{(2)}$ can be deduced~\cite{Dixon2011nj} from
the differential equations it satisfies~\cite{Drummond2010cz,Dixon2011ng}. 
There are only three distinct
final entries of the symbol of $\Omega^{(2)}(u,v,w)$, namely
\be
\left\{ \frac{u}{1-u}, \frac{v}{1-v}, y_u y_v\right\}\,.
\label{Omfinals}
\ee
Note that three is the minimum possible number of distinct final
entries we could hope for, since $\Omega^{(2)}$ is genuinely dependent
on all three variables.  As mentioned above, we define extra-pure
functions, such as $\Omega^{(2)}$, to be those functions for which the number
of final entries in the symbol equals the number of variables on which they depend.
Another way to state the property (which also extends it from a property of 
symbols to a property of functions)
is that $p$-variable pure functions $f$ of weight $n$ are 
extra-pure if there exist $p$ independent commuting first-order differential
operators $\mathcal{O}_i$, such that $\mathcal{O}_i f$ are themselves all pure
of weight $(n-1)$.

More explicitly, the symbol of $\Omega^{(2)}$ can be written
as~\cite{Dixon2011nj},
\be
\mathcal{S}(\Omega^{(2)}(u,v,w))
= -\frac{1}{2} \Bigl[ \mathcal{S}(Q_{\phi}) \otimes \phi
                     + \mathcal{S}(Q_r) \otimes r
                     + \mathcal{S}(\tilde{\Phi}_6)\otimes y_u y_v \Bigr]\,,
\label{S(O2)}
\ee
where
\be
\phi = \frac{u v}{(1-u)(1-v)}, \qquad r = \frac{u(1-v)}{v(1-u)} \,.
\ee
The functions $Q_{\phi}$ and $Q_r$ will be defined below.
The function $\tilde{\Phi}_6$ is the weight-three, parity-odd
one-loop six-dimensional hexagon
function~\cite{Dixon2011ng,DelDuca2011ne}, whose symbol is given
by~\cite{Dixon2011ng},
\be
\mathcal{S}(\tilde{\Phi}_6)
= - \mathcal{S}\bigl( \Omega^{(1)}(u,v,w) \bigr) \otimes y_w
+ {\rm cyclic},
\label{tPhisymbol}
\ee
where $\Omega^{(1)}$ is a finite, four-dimensional one-loop hexagon
integral~\cite{ArkaniHamed2010kv,Drummond2010cz},
\be
\Omega^{(1)}(u,v,w) = \ln u \ln v
 + {\rm Li}_2(1-u) + {\rm Li}_2(1-v) + {\rm Li}_2(1-w) - 2 \zeta_{2}\,.
\label{Om1}
\ee

Although we have written \eqn{tPhisymbol} as an equation for the symbol
of $\tilde{\Phi}_6$, secretly it contains more information, because we have
written the symbol of a full function, $\Omega^{(1)}(u,v,w)$, in the first
two slots.  Later we will codify this extra information as corresponding
to the $\{2,1\}$ component of the coproduct of $\tilde{\Phi}_6$.
Another way of saying it is that all three derivatives of the function
$\tilde{\Phi}_6$, with respect to the logarithms of the $y$ variables,
are given by $-\Omega^{(1)}(u,v,w)$ or its permutations, including
the $\zeta_{2}$ term in \eqn{Om1}.
Any other derivative can be obtained by the chain rule.  For example,
to get the derivative with respect to $u$, we just need,
\be
\frac{\partial\ln y_u}{\partial u}
= \frac{1-u-v-w}{u\sqrt{\Delta}} \,, \quad
\frac{\partial\ln y_v}{\partial u}
= \frac{1-u-v+w}{(1-u)\sqrt{\Delta}} \,, \quad
\frac{\partial\ln y_w}{\partial u}
= \frac{1-u+v-w}{(1-u)\sqrt{\Delta}} \,, \quad
\label{yi_u_diff}
\ee
which leads to the differential equation found in ref.~\cite{Dixon2011ng},
\be
\partial_u \tilde{\Phi}_6 = 
- \frac{1-u-v-w}{u\sqrt{\Delta}} \Omega^{(1)}(v,w,u)
- \frac{1-u-v+w}{(1-u)\sqrt{\Delta}} \Omega^{(1)}(w,u,v)
- \frac{1-u+v-w}{(1-u)\sqrt{\Delta}} \Omega^{(1)}(u,v,w) \,.
\label{duPhi}
\ee
Hence $\tilde{\Phi}_6$ can be fully specified,
up to a possible integration constant, by promoting the first two slots
of its symbol to a function in an appropriate way. In fact, the ambiguity
of adding a constant of integration is actually fixed in this case, by
imposing the property that the function $\tilde{\Phi}_6$ is parity odd.

Note that for the solution to the differential equation (\ref{duPhi}) 
and its cyclic images to have physical branch cuts, the correct coefficients
of the $\zeta_2$ terms in~\eqn{Om1} are crucial. Changing the coefficients
of these terms in any of the cyclic images of $\Omega^{(1)}$ would correspond
to adding a logarithm of the $y$ variables to $\tilde{\Phi}_6$, which would
have branch cuts in unphysical regions.

The other weight-three symbols in \eqn{S(O2)}
can similarly be promoted to full functions.
To do this we employ the harmonic polylogarithms (HPLs) in one
variable~\cite{Remiddi1999ew}, $H_{\vec{w}}(u)$.
In our case, the weight vector $\vec{w}$ contains only 0's and 1's. 
If the weight vector is a string of $n$ 0's, $\vec{w}=0_n$, then we 
have $H_{0_n}(u) = \tfrac{1}{n!} \log^n u$. The remaining functions are 
defined recursively by
\be
H_{0,\vec{w}}(u) = \int_0^u \frac{dt}{t} H_{\vec{w}}(t), \quad
H_{1,\vec{w}}(u) = \int_0^u \frac{dt}{1-t} H_{\vec{w}}(t). 
\label{Hdef}
\ee
Such functions have symbols with only two letters, $\{u,1-u\}$.
We would like the point $u=1$ to be a regular point for the HPLs. 
This can be enforced by choosing the argument to be $1-u$, and restricting
to weight vectors whose last entry is 1.  The symbol and HPL definitions
have a reversed ordering, so to find an HPL with argument $1-u$ corresponding
to a symbol in $\{u,1-u\}$, one reverses the string, replaces $u\to1$ and 
$1-u\to0$, and multiplies by $(-1)$ for each 1 in the weight vector.
We also use a compressed notation where $(k-1)$ 0's followed by a 1 is
replaced by $k$ in the weight vector, and the argument $(1-u)$ 
is replaced by the superscript $u$.  For example, ignoring $\zeta$-value
ambiguities we have,
\bea
&& u \otimes (1-u)\ \to\ -H_{0,1}(1-u)\ \to\ -H_2^u \,, \nonumber\\
&& u \otimes u \otimes (1-u)\ \to\ H_{0,1,1}(1-u)\ \to\ H_{2,1}^u \,, \nonumber\\
&& v \otimes (1-v) \otimes v \otimes (1-v)\ \to\ H_{0,1,0,1}(1-v)\ 
\to\ H_{2,2}^v \,. \label{compressednotation}
\eea
The combination
\be
H_2^u + \tfrac{1}{2} \ln^2u = -{\rm Li}_2(1-1/u)
\ee
occurs frequently, because it is the lowest-weight extra-pure function, with
symbol $u \otimes u/(1-u)$.

In terms of HPLs, the functions corresponding to the weight-three, parity-even
symbols appearing in \eqn{S(O2)} are given by,
\be
\bsp
Q_\phi\ &=\ [-H^u_{3} - H^u_{2, 1} - H^v_{2} \ln u
 - \tfrac{1}{2} \ln^2 u \ln v + (H^u_{2} - \zeta_2) \ln w
 + (u \leftrightarrow v)] \\
&\quad + 2 H^w_{2, 1}  + H^w_{2}  \ln w +  \ln u \ln v \ln w\,, \\
Q_r\ &=\ [-H^u_{3} + H^u_{2, 1} + (H^u_{2} + H^w_{2} - 2\zeta_2) \ln u
 + \tfrac{1}{2} \ln^2 u \ln v - (u \leftrightarrow v)]   \,.
\esp
\label{qphiqr}
\ee
Here we have added some $\zeta_2$ terms with respect to
ref.~\cite{Dixon2011nj}, in order to match the $\{3,1\}$
component of the coproduct of $\Omega^{(2)}$ that we determine later.

Note that the simple form of the symbol of $R^{(2)}_{6, {\rm rat}}$
in \eqn{R62ratdef} means that it can be absorbed into the three cyclic images of
$\Omega^{(2)}(u,v,w)$ without ruining the extra-purity of the latter
functions.  Hence $R_6^{(2)}$ is the cyclic sum of an extra-pure function.

With the decomposition~(\ref{R62decomp}) in mind,
we searched for an analogous decomposition
of the symbol of the three-loop remainder function~\cite{Dixon2011pw}
into extra-pure components.  In other words, we looked for a representation
of $\mathcal{S}(R_6^{(3)})$ in terms  a function 
whose symbol has the same final entries~(\ref{Omfinals}) as
$\Omega^{(2)}(u,v,w)$, plus its cyclic rotations. After removing some 
products of lower-weight functions we find that this is indeed possible.
Specifically, we find that,
\be
\mathcal{S}(R_6^{(3)})
= \mathcal{S}( \Rep(u,v,w) + \Rep(v,w,u) + \Rep(w,u,v) )
+ \mathcal{S}(P_6(u,v,w)).
\label{R63decomp}
\ee
Here $P_6$ is the piece constructed from products of lower-weight functions,
\begin{align}
P_6(u,v,w) = &-\frac{1}{4} 
\Bigl[ \Omega^{(2)}(u,v,w) \, \Li_2(1-1/w) + {\rm cyclic} \Bigr]
- \frac{1}{16} (\tilde{\Phi}_6)^2\, \notag \\
&+ \frac{1}{4} \, \Li_2(1-1/u) \, \Li_2(1-1/v) \, \Li_2(1-1/w) \,.
\end{align}
The function $\Rep$ is very analogous to $\Omega^{(2)}$ in that it has
the same $(u\lr v)$ symmetry, and its symbol has the same final entries,
\be
\mathcal{S}\left( \Rep(u,v,w)\right) = 
\mathcal{S}\bigl(\Rep^u(u,v,w)\bigr)\otimes \frac{u}{1-u} 
+ \mathcal{S}\bigl(\Rep^u(v,u,w)\bigr)\otimes \frac{v}{1-v} 
+ \mathcal{S}\bigl(\Rep^{y_u}(u,v,w)\bigr) \otimes y_u y_v\,.
\label{Repcoprod1}
\ee

In the following we will describe a systematic construction of the function 
$\Rep$ and hence the three-loop remainder function. As in the case
just described for $\tilde{\Phi}_6$, and implicitly for $\Omega^{(2)}$,
the construction will involve promoting the quantities
$\mathcal{S}(\Rep^u)$ and $\mathcal{S}(\Rep^{y_u})$ to full functions,
with the aid of the coproduct formalism.  In fact, we will perform
a complete classification of all well-defined functions corresponding to 
symbols with nine letters and obeying the first entry
condition~(\ref{firstentry}) (but not the final entry 
condition~(\ref{finalentry})), iteratively in the weight through
weight five.  Knowing all such pure functions at weight five will then
enable us to promote the weight-five quantities $\mathcal{S}(\Rep^u)$ and
$\mathcal{S}(\Rep^{y_u})$ to well-defined functions, subject
to $\zeta$-valued ambiguities that we will fix using physical criteria.

\vfill\eject

%%%%%%%%%%%%%%%%%%%%%%%%%%%%%%%%%%%%%%%%%%%%%%%%%%%%%%%%%%%%%%%%%%%

\section{Hexagon functions as multiple polylogarithms}
\label{sec:hex_multi_poly}

The task of the next two sections is to build up an understanding
of the space of hexagon functions, using two complementary routes.
In this section, we follow the route of expressing the hexagon
functions explicitly in terms of multiple polylogarithms.  
In the next section, we will take a slightly more abstract route of
defining the functions solely through the differential equations
they satisfy, which leads to relatively compact integral
representations for them.

\subsection{Symbols}
\label{sec:hex_multi_poly_symb}

Our first task is to classify all integrable symbols at weight $n$ with
entries drawn from the set $\Su$ in~\eqn{nineletters} that also 
satisfy the first entry condition~(\ref{firstentry}).  We do not impose the
final entry condition~(\ref{finalentry})
because we need to construct quantities at intermediate weight,
from which the final results will be obtained by further integration;
their final entries correspond to intermediate entries of $\Rep$.

The integrability of a symbol may be imposed iteratively, first
as a condition on the first $n-1$ slots, and then as a separate condition on
the $\{n-1,n\}$ pair of slots, as in \eqn{integrability}.
Therefore, if $\mathcal{B}_{n-1}$ is the basis of integrable symbols at weight
$n-1$, then a minimal ansatz for the basis at weight $n$ takes the
form,
\be
\{b\otimes x\; |\; b\in \mathcal{B}_{n-1},\; x\in \Su\}\, ,
\ee
and $\mathcal{B}_n$ can be obtained simply by enforcing integrability
in the last two slots. This method for recycling lower-weight
information will also guide us toward an iterative construction of 
full functions, which we perform in the remainder of this section.

Integrability and the first entry condition together require the
second entry to be free of the $y_i$. Hence the maximum number of $y$
entries that can appear in a term in the symbol is $n-2$. In fact, the
maximum number of $y$'s that appear in any term in the symbol
defines a natural grading for the space of functions. In~\tab{tab:basis_count}, 
we use this grading to tabulate the
number of irreducible functions ({\it i.e.}~those functions that cannot be
written as products of lower-weight functions) through weight six.
The majority of the functions at low weight contain no $y$ entries.

The $y$ entries couple together $u,v,w$. In their absence, the symbols
with letters $\{u,v,w,1-u,1-v,1-w\}$ can be factorized, so that the
irreducible ones just have the letters $\{u,1-u\}$, plus cyclic
permutations of them.   The corresponding functions are the ordinary 
HPLs in one variable~\cite{Remiddi1999ew} introduced in the previous section, 
$H_{\vec{w}}^u$, with weight vectors $\vec{w}$ consisting only of 0's and 1's.
These functions are not all independent, 
owing to the existence of shuffle identities~\cite{Remiddi1999ew}. 
On the other hand, we may exploit Radford's theorem~\cite{Radford1979} 
to solve these identities in terms of a Lyndon basis, 
\be
\mathcal{H}_u = \left\{
H_{l_w}^{u}\;|\;l_w \in \textrm{Lyndon}(0,1)\backslash\{0\}\right\}\,,
\ee
where $H_{l_w}^{u}\equiv H_{l_w}(1 - u)$, and
$\textrm{Lyndon}(0,1)$ is the set of {\it Lyndon} words in the letters
$0$~and~$1$.  The Lyndon words are those words $w$ such that for 
every decomposition into two words $w=\{u,v\}$, the left word $u$
is smaller\footnote{%
We take the ordering of words to be lexicographic. The ordering of 
the letters is specified by the order in which they appear in the 
argument of ``$\textrm{Lyndon}(0,1)$'', i.e. $0<1$. Later we will 
encounter words with more letters for which this specification is 
less trivial.} 
than the right word $v$, {\it i.e.}~$u<v$.  Notice that we exclude the case
$l_w=0$ because it corresponds to $\ln(1-u)$, which has an unphysical 
branch cut.  Further cuts of this type occur whenever $l_w$ has a 
trailing zero, but such words are excluded from the Lyndon basis by 
construction.

The Lyndon basis of HPLs with proper branch cuts through weight 
six can be written explicitly as,
\bea
\mathcal{H}_u|_{n\leq6} &=& 
\{ \ln u,\ H_2^u,\ H_3^u, H_{2,1}^u,\ H_4^u, H_{3,1}^u, H_{2,1,1}^u,\
H_5^u, H_{4,1}^u, H_{3,2}^u, H_{3,1,1}^u, H_{2,2,1}^u, H_{2,1,1,1}^u,
\nonumber\\ &&\null \hskip5mm
H_6^u, H_{5,1}^u, H_{4,2}^u, H_{4,1,1}^u, H_{3,2,1}^u, H_{3,1,2}^u,
H_{3,1,1,1}^u, H_{2,2,1,1}^u, H_{2,1,1,1,1}^u \} \,.
\label{HPLbasis6}
\eea
\Eqn{HPLbasis6} and its two cyclic permutations, 
$\mathcal{H}_v$ and $\mathcal{H}_w$, account entirely
for the $y^0$ column of~\tab{tab:basis_count}.
Although the $y$-containing functions are not very numerous through
weight five or so, describing them is considerably more involved.

\renewcommand{\arraystretch}{1.05}
\begin{table}[t]
\centering
\begin{tabular}[t]{c||c|c|c|c|c}
\hline\hline
\textrm{Weight} & $y^0$ & $y^1$ & $y^2$ & $y^3$ & $y^4$ \\
\hline
1 & 3 & - & - & - & -\\
2 & 3 & - & - & - & -\\
3 & 6 & 1 & - & - & -\\
4 & 9 & 3 & 3 & - & -\\
5 & 18 & 4 & 13 & 6 & -\\
6 & 27 & 4 & 27 & 29 & 18\\
\hline\hline
\end{tabular}
\caption{The dimension of the irreducible basis of hexagon functions,
graded by the maximum number of $y$ entries in their symbols.}
\label{tab:basis_count}
\end{table}

In order to parametrize the full space of functions whose symbols can be
written in terms of the elements in the set $\Su$, it is useful to
reexpress those elements in terms of three independent variables. 
The cross ratios themselves are not a convenient choice of variables because
rewriting the $y_i$ in terms of the $u_i$ produces explicit square
roots. A better choice is to consider the $y_i$ as
independent variables, in terms of which the $u_i$ are given by \eqn{u_from_y}.
In this representation, the symbol has letters drawn from the ten-element set,
\be
\label{eq:symb_ten}
\Sy = \{y_u, y_v, y_w, 1-y_u, 1-y_v, 1-y_w, 1-y_u y_v, 1-y_u y_w, 1-y_v y_w, 
1 - y_u y_v y_w\} \,.
\ee

We appear to have taken a step backward since there is an extra letter
in $\Sy$ relative to $\Su$. Indeed, writing the symbol of a typical
function in this way greatly increases the length of its expression. 
Also, the first entry condition becomes more complicated in the $y$
variables. On the other hand, $\Sy$
contains purely rational functions of the $y_i$, and as such it is
easy to construct the space of functions that give rise to symbol
entries of this type. We will discuss these functions in the next
subsection.

%%%%%%%%%%%%%%%%%%%%%%%%%%%%%%

\subsection{Multiple polylogarithms}
\label{sec:hex_multi_poly_multi_poly}

Multiple polylogarithms are a general class of multi-variable iterated
integrals, of which logarithms, polylogarithms, harmonic
polylogarithms, and various other iterated integrals are special
cases. They are defined recursively by $G(z)=1$, and,
\be
\label{eq:main_G_def}
G(a_1,\ldots,a_n; z) = 
\int_0^z\; \frac{dt}{t-a_1}\,G(a_2,\ldots,a_n;t)\,, 
\quad\quad G(\underbrace{0,\ldots,0}_{p}; z) = \frac{\ln^p z}{p!} \,.
\ee
Many of their properties are reviewed in appendix~\ref{sec:app_multi_poly}, 
including an expression for their symbol, which is
also defined recursively~\cite{Gonch3},
\begin{align}
\label{eq:G_symb}
\mathcal{S}\big(G(a_{n-1},\ldots,a_1;a_n)\big) 
= \sum_{i=1}^{n-1} \biggl[&\mathcal{S}\big(G(a_{n-1},\ldots,\hat{a}_i,\ldots,a_1;a_n)\big)
\otimes\left(a_{i}-a_{i+1}\right) \notag\\
-& \mathcal{S}\big(G(a_{n-1},\ldots,\hat{a}_i,\ldots,a_1;a_n)\big)
\otimes (a_{i}-a_{i-1}) \biggr]
\,,
\end{align}
where $a_0=0$ and the hat on $a_i$ on the right-hand side indicates that this index
should be omitted.

Using \eqn{eq:G_symb}, it is straightforward to write down a set
of multiple polylogarithms whose symbol entries span $\Sy$,
\be
\GG = \biggl\{ G(\vec{w};y_u) | w_i \in \{0,1\} \biggr\} 
\cup \biggl\{ G(\vec{w};y_v) \Big| w_i \in \Bigl\{0,1,\frac{1}{y_u} \Bigr\} \biggr\}
\cup \biggl\{ G(\vec{w};y_w) \Big| w_i \in 
\Bigl\{0,1,\frac{1}{y_u},\frac{1}{y_v},\frac{1}{y_u y_v} \Bigr\} 
\biggr\}\,,
\ee
The set $\GG$ also emerges naturally from a simple procedure by
which symbols are directly promoted to polylogarithmic functions. For
each letter $\phi_i(y_u,y_v,y_w) \in \Sy$ we write $\omega_i = d \log
\phi_i(t_u,t_v,t_w)$. Then following refs.~\cite{FBThesis,Bogner2012dn},
which are in turn based on ref.~\cite{Chen}, we use the integration map,
\be
\phi_1 \otimes \ldots \otimes \phi_n \mapsto 
\int_\gamma \omega_n \circ \ldots \circ \omega_1\,.
\label{intmap}
\ee
The integration is performed iteratively along the contour $\gamma$
which we choose to take from the origin $t_i=0$ to the point
$t_i=y_i$. The precise choice of path is irrelevant, provided the
symbol we start from is integrable~\cite{Chen,FBThesis}. So we may
choose to take a path which goes sequentially along the $t_u,t_v,t_w$
directions. Near the axes we may find some divergent integrations of
the form $\int_0^y dt/t \circ \ldots \circ dt/t$. We regularize these
divergences in the same way as in the one-dimensional HPL case (see
the text before \eqn{Hdef}) by replacing them with $\tfrac{1}{n!}
\log^n y$. In this way we immediately obtain an expression in terms of
the functions in $\GG$, with the three subsets corresponding to the
three segments of the contour.

The set $\GG$ is larger than what is required to construct the basis
of hexagon functions.  One reason for this is that $\GG$ generates
unwanted symbol entries outside of the set $\Su$, such as the
differences $y_i-y_j$, as is easy to see from \eqn{eq:G_symb}; the
cancellation of such terms is an additional constraint that any valid
hexagon function must satisfy. Another reason is that multiple
polylogarithms satisfy many identities, such as the shuffle and
stuffle identities (see Appendix~\ref{sec:app_multi_poly} or
refs.~\cite{Gonch3,Duhr2011zq} for a review).  While there are no
relevant stuffle relations among the functions in $\GG$, there are
many relations resulting from shuffle identities. Just as for the
single-variable case of HPLs, these shuffle relations may be resolved
by constructing a Lyndon basis, $\GG_I^L \subset \GG$,
\be
\bsp
\GG_I^L &= \biggl\{ G(\vec{w};y_u) | w_i \in \textrm{Lyndon}(0,1) \biggr\} 
\cup \biggl\{ G(\vec{w};y_v) \Big| 
w_i \in\textrm{Lyndon}\Bigl(0,1,\frac{1}{y_u} \Bigr)
\biggr\} \\
&\quad\quad
\cup \biggl\{ G(\vec{w};y_w) \Big| w_i 
\in \textrm{Lyndon}\Bigl(0,1,\frac{1}{y_u},\frac{1}{y_v},\frac{1}{y_u y_v} \Bigr) 
\biggr\}\,.
\label{GL_I}
\esp
\ee

\begin{figure}
\begin{center}
\includegraphics[width=6.5in]{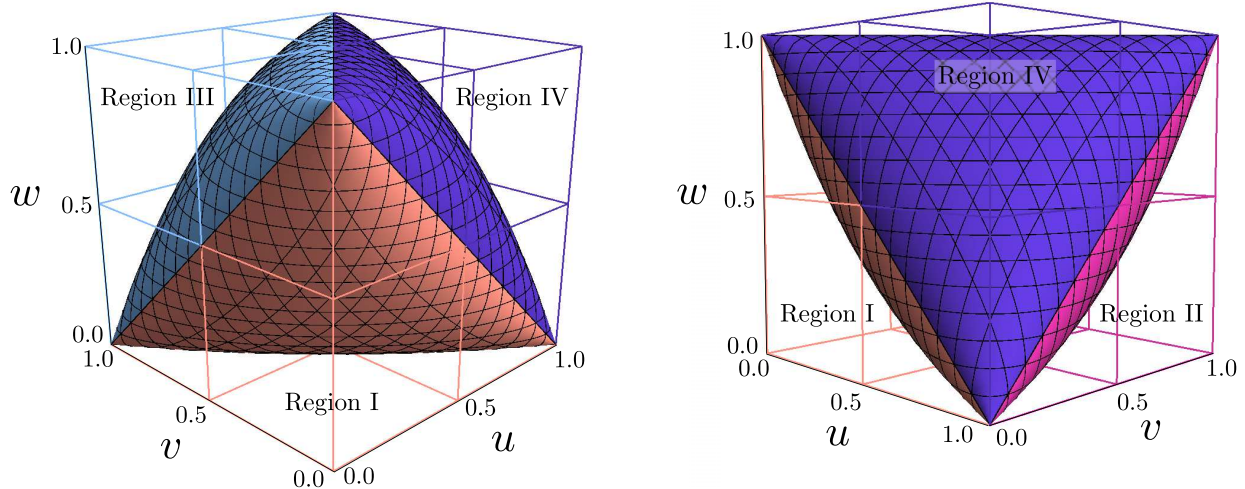}
\end{center}
\caption{Illustration of Regions I, II, III and IV.  Each region lies between
the colored surface and the respective corner of the unit cube.}
\label{fig:regionsmesh}
\end{figure}

A multiple polylogarithm $G(w_1,\ldots,w_n;z)$ admits a convergent series
expansion if 
$|z| \le |w_i|$ for all nonzero $w_i$, and it is manifestly real-valued if 
the nonzero $w_i$ and $z$ are real and positive. Therefore, the set $\GG_I^L$
is ideally suited for describing configurations for which $0 < y_i < 1$. In terms
of the original cross ratios, this region is characterized by,
\be
\textrm{Region I}:\quad
\left\{
\begin{array}{l}
\Delta > 0\,,\quad 0<u_i<1\,, \quad~\textrm{and}~\quad u+v+w<1,\\
0< y_i < 1 \, .
\end{array}
\right.
\label{RegionIDef}
\ee
We will construct the space of hexagon functions in Region I as a subspace
of $\GG_I^L$ with good branch-cut properties.

What about other regions?
As we will discuss in the next subsection, multiple polylogarithms in
the $y$ variables are poorly suited to regions where $\Delta<0$;
in these regions the $y_i$ are complex. For such cases, we turn to certain
integral representations that we will describe
in~\sect{sec:integral_reps}.  In this section,
we restrict ourselves to the subspace of the unit
cube for which $\Delta>0$.  As shown in~\fig{fig:regionsmesh}, there are 
four disconnected regions with $\Delta>0$, which we refer to as
Regions I, II, III, and IV. They are the regions that extend respectively from 
the four points $(0,0,0)$, $(1,1,0)$, $(0,1,1)$, and $(1,0,1)$
to the intersection with the $\Delta=0$ surface.
Three of the regions (II, III and IV) are related
to one another by permutations of the $u_i$, so it suffices to
consider only one of them,
\be
\textrm{Region II}:\quad
\left\{
\begin{array}{l}
\Delta > 0\,,\quad 0<u_i<1\,, \quad~\textrm{and}~\quad u+v-w>1,\\
0< y_w < \frac{1}{y_u y_v} < \frac{1}{y_u} , \frac{1}{y_v} < 1\, .
\end{array}
\right.
\label{RegionIIDef}
\ee

In Region II, the set $\GG_I^L$ includes functions
$G(w_1,\ldots,w_n;z)$ for which $|w_i| < |z|$ for some $i$. As
mentioned above, such functions require an analytic continuation and
are not manifestly real-valued. On the other hand, it is
straightforward to design an alternative basis set that does not
suffer from this issue,
\be
\bsp
\GG_{II}^L &= \biggl\{ G\Bigl(\vec{w};\frac{1}{y_u}\Bigr) \Big| 
w_i \in \textrm{Lyndon}(0,1) \biggr\} 
\cup \biggl\{ G\Bigl(\vec{w};\frac{1}{y_v}\Bigr) \Big| 
w_i \in\textrm{Lyndon}(0,1,y_u)\biggr\} \\
&\quad\quad
\cup \biggl\{ G(\vec{w};y_w) \Big| w_i 
\in \textrm{Lyndon}\Bigl(0,1,\frac{1}{y_u},\frac{1}{y_v},\frac{1}{y_u y_v} \Bigr) 
\biggr\}\,.
\label{GL_II}
\esp
\ee
Like $\GG_I^L$, $\GG^L_{II}$ also generates symbols with the desired
entries.  It is therefore a good starting point for constructing
a basis of hexagon functions in Region II.

%%%%%%%%%%%%%%%%%%%%%%%%%%%%%%%%%%%%%%%%%%%%%%%%%%%%%%%
\subsection{The coproduct bootstrap}
\label{sec:coproduct}

The space of multiple polylogarithms enjoys various nice properties,
many of which are reviewed in appendix~\ref{sec:app_multi_poly}. 
For example, it can be
endowed with the additional structure necessary to promote it to a
Hopf algebra. For the current discussion, we make use of one element
of this structure, namely the coproduct. The coproduct on multiple
polylogarithms has been used in a variety of
contexts~\cite{Duhr2011zq,Duhr2012fh,Chavez2012kn,Drummond2013nda,%
vonManteuffel2013uoa,Schlotterer2012ny}. 
It serves as a powerful tool to help lift symbols to full functions and to
construct functions or identities iteratively in the weight.

Let $\mathcal{A}$ denote the Hopf algebra of multiple polylogarithms
and $\mathcal{A}_n$ the weight-$n$ subspace, so that,
\be
\mathcal{A} = \bigoplus_{n=0}^{\infty} \mathcal{A}_n\,.
\ee
Then, for $G_n \in \mathcal{A}_n$, the coproduct decomposes as,
\be
\Delta(G_n) = \sum_{p+q=n} \Delta_{p,q}(G_n)\, ,
\ee
where $\Delta_{p,q}\in \mathcal{A}_p\otimes\mathcal{A}_q$. It is
therefore sensible to discuss an individual $\{p,q\}$ component of the
coproduct, $\Delta_{p,q}$. In fact, we will only need two cases, $\{p,q\}=\{n-1,1\}$
and $\{p,q\}=\{1,n-1\}$, though the other components carry additional information
that may be useful in other contexts.

A simple (albeit roundabout) procedure to extract the coproduct of a generic 
multiple polylogarithm, $G$,
is reviewed in appendix~\ref{sec:app_multi_poly}. One first rewrites $G$ in
the notation of a slightly more general function, usually denoted by $I$ in the 
mathematical literature. Then one applies the main coproduct formula,~\eqn{eq:coprI},
and finally converts back into the $G$ notation. 

Let us discuss how the coproduct can be used to construct identities
between multiple polylogarithms iteratively. Suppose we know all relevant
identities up to weight $n-1$, and we would like to establish the validity of some
potential weight-$n$ identity, which can always be written in the
form,
\be
A_n = 0\,,
\label{An_vanish}
\ee
for some combination of weight-$n$ functions, $A_n$. If this identity
holds, then we may further conclude that each component of the
coproduct of $A_n$ should vanish. In particular,
\be
\Delta_{n-1,1}(A_n) = 0\,.
\label{Delta_nm1_1_vanish}
\ee

Since this is an equation involving functions of weight less than or
equal to $n-1$, we may check it explicitly. \Eqn{Delta_nm1_1_vanish} 
does not imply \eqn{An_vanish}, because $\Delta_{n-1,1}$ has a nontrivial
kernel. For our purposes, the only relevant elements of the kernel are
multiple zeta values, zeta values, $i \pi$, and their
products. Through weight six, the elements of the kernel are the
transcendental constants,
\be
\mathcal{K} = \{ i\pi,\ \zeta_2,\ \zeta_3, i \pi^3,\ \zeta_4, i\pi \zeta_3,
\ \zeta_2 \zeta_3, \zeta_5, i \pi^5,
\ \zeta_6, \zeta_3^2, i \pi \zeta_5, i\pi^3 \zeta_3,\ \ldots \}\,.
\ee
At weight two, for example, we may use this information to write,
\be
\Delta_{1,1}(A_2) = 0 \qquad \Rightarrow \qquad A_2=c\,\zeta_2\,,
\ee
for some undetermined rational number $c$, which we can fix
numerically or by looking at special limits. Consider the following
example for some real positive $x\leq1$,
\be
\bsp
A_2 &= - \, G(0,x;1) - G\left(0,\frac{1}{x};1\right)
         + \frac{1}{2}G(0;x)^2 - i \pi G(0;x) \\
& = \textrm{Li}_2\left(\frac{1}{x}\right) + \textrm{Li}_2(x)
         + \frac{1}{2}\ln^2 x - i \pi \ln x \,.
\label{wt2example}
\esp
\ee
Using \eqn{wt2example} and simple identities among logarithms, it is
easy to check that
\be
\Delta_{1,1}(A_2) = 0\,,
\ee
so we conclude that $A_2 = c\, \zeta_2$. Specializing to $x=1$, 
we find $c=2$ and therefore $A_2 =2\,\zeta_2$. Indeed, this confirms 
the standard inversion relation for dilogarithms.

The above procedure may be applied systematically to generate all
identities within a given ring of multiple polylogarithms and
multiple zeta values. Denote this ring by $\mathcal{C}$ and its
weight-$n$ subspace by $\mathcal{C}_n$. Assume that we have found 
all identities through weight $n-1$. To find the identities at weight $n$,
we simply look for all solutions to the equation,
\be
\Delta_{n-1,1}\bigg(\sum_{i} c_i \,G_i\bigg) 
= \sum_{i}c_i\, \Big(\Delta_{n-1,1}(G_i)\Big)= 0\,,
\ee
where $G_i\in \mathcal{C}_n$ and the $c_i$ are rational numbers.
Because we know all identities through
weight $n-1$, we can write each $\Delta_{n-1,1}(G_i)$ as a combination
of linearly-independent functions of weight $n-1$. The problem is then
reduced to one of linear algebra. The nullspace encodes the set of new
identities, modulo elements of the kernel $\mathcal{K}$.  The latter 
transcendental constants can be fixed numerically, or perhaps analytically
with the aid of an integer-relation algorithm like PSLQ~\cite{PSLQRef}.

For the appropriate definition of $\mathcal{C}$, the above procedure
can generate a variety of interesting relations. For example, we can
choose $\mathcal{C}=\GG^L_I$ or $\mathcal{C}=\GG^L_{II}$ and confirm that there
are no remaining identities within these sets. 

We may also use this method to express all harmonic polylogarithms
with argument $u_i$ or $1-u_i$ in terms of multiple polylogarithms in the set
${\GG}^L_{I}$ or the set ${\GG}^L_{II}$. The only
trick is to rewrite the HPLs as multiple polylogarithms. For example,
using the uncompressed notation for the HPLs,
\be
H_{a_1,\ldots,a_n}(u) = (-1)^{w_1}\, G(a_1,\ldots,a_n; u)
= (-1)^{w_1}\, G\left(a_1,\ldots,a_n; 
\frac{y_u (1 - y_v) (1 - y_w)}{(1 - y_u y_v) (1 - y_u y_w)}\right)\,,
\ee
where $w_1$ is the number of $a_i$ equal to one. With this understanding,
we can simply take,
\be
\mathcal{C} = \{H_{\vec{w}}(u_i)\} \cup {\GG}^L_{I}
\qquad~\textrm{or}~\qquad
\mathcal{C} = \{H_{\vec{w}}(u_i)\} \cup {\GG}^L_{II}\,,
\ee
and then proceed as above to generate all identities within
this expanded ring.

In all cases, the starting point for the iterative procedure for generating
identities is the set of identities at weight one, {\it i.e.}~the set of
identities among logarithms.  All identities among logarithms are of course
known, but in some cases they become rather cumbersome, and one
must take care to properly track various terms that depend on the ordering
of the $y_i$.  For example, consider the following
identity, which is valid for all complex $y_i$,
\be
\bsp
\ln u &= \ln \left(\frac{y_u(1 - y_v)(1 - y_w)}{(1 - y_u y_v)(1 - y_u y_w)}
\right)\\
&=\ln y_u+\ln(1-y_v)+\ln(1-y_w)-\ln(1-y_u y_v)-\ln(1-y_u y_w) \\
&\quad + i \bigg[
\operatorname{Arg}\left(\frac{y_u(1 - y_v)(1 - y_w)}{(1 - y_u y_v)(1 - y_u y_w)}
\right)
- \operatorname{Arg}(y_u) 
- \operatorname{Arg}(1-y_v) - \operatorname{Arg}(1-y_w) \\
&\quad\quad\quad + \operatorname{Arg}(1-y_u y_v)
                 + \operatorname{Arg}(1-y_u y_w) \bigg] \, ,
\esp
\ee
where $\operatorname{Arg}$ denotes the principal value of the complex
argument. In principle, this identity can be used to seed the iterative
procedure for constructing higher-weight identities, which would also be
valid for all complex $y_i$. Unfortunately, the bookkeeping quickly becomes
unwieldy and it is not feasible to track the proliferation of
$\operatorname{Arg}$'s for high weight.

To avoid this issue, we will choose to focus on Regions I and II,
defined by \eqns{RegionIDef}{RegionIIDef}.  In both regions, $\Delta > 0$, so 
the $y$ variables are real, and the $\operatorname{Arg}$'s take on specific
values.  In Region I, for example, we may write,
\be
\bsp
\ln u &\stackrel{\textrm{\scriptsize{ Region I }}}{=}
\ln y_u+\ln(1-y_v)+\ln(1-y_w)-\ln(1-y_u y_v)-\ln(1-y_u y_w) \\
&\stackrel{\textrm{\scriptsize{\phantom{ Region I }}}}{=}
G(0; y_u) + G\left(1;y_v\right)+ G\left(1;y_w\right)
- G\left(\frac{1}{y_u };y_v\right)- G\left(\frac{1}{y_u};y_w\right)\,.
\esp
\label{lnuI}
\ee
In the last line, we have rewritten the logarithms in terms of multiple 
polylogarithms in the set ${\GG}_I^L$, which, as we argued in the 
previous subsection, is the appropriate basis for this region. In Region II, 
the expression for $\ln u$ looks a bit different,
\be
\bsp
\ln u &\stackrel{\textrm{\scriptsize{ Region II }}}{=}
\ln\left(1-\frac{1}{y_v}\right) + \ln(1-y_w) 
- \ln\left(1-\frac{1}{y_u y_v}\right) - \ln(1-y_u y_w)\\
&\stackrel{\textrm{\scriptsize{\phantom{ Region II }}}}{=}
G\left(1;\frac{1}{y_v}\right) + G\left(1;y_w\right) 
- G\left(y_u;\frac{1}{y_v}\right) - G\left(\frac{1}{y_u};y_w\right)\, .
\esp
\ee
In this case, we have rewritten the logarithms as multiple
polylogarithms belonging to the set ${\GG}^L_{II}$.

We now show how to use these relations and the coproduct to deduce
relations at weight two. In particular, we will derive an expression
for $H_2^u = H_2(1-u)$ in terms of multiple polylogarithms in the basis
${\GG}^L_I$ in Region I.  A similar result holds in Region II.
First, we need one more weight-one identity,
\be
\ln(1-u) \stackrel{\textrm{\scriptsize{ Region I }}}{=} 
G\left(1;y_u\right) -   G\left(\frac{1}{y_u};y_v\right) 
-   G\left(\frac{1}{y_u};y_w\right) 
+ G\left(\frac{1}{y_u y_v};y_w\right)\, .
\label{lnomuI}
\ee
Next, we take the $\{1,1\}$ component of the coproduct,
\be
\Delta_{1,1}(H_2^u)  = -\ln u \otimes \ln (1-u)\,,
\ee
and substitute \eqns{lnuI}{lnomuI},
\be
\bsp
\Delta_{1,1}(H_2^u) & =  
- \left[  G(0; y_u) + G\left(1;y_v\right)+ G\left(1;y_w\right)
- G\left(\frac{1}{y_u };y_v\right)- G\left(\frac{1}{y_u};y_w\right) \right]\\
&\qquad \otimes\left[G\left(1;y_u\right) - G\left(\frac{1}{y_u};y_v\right) 
-  G\left(\frac{1}{y_u};y_w\right) 
+ G\left(\frac{1}{y_u y_v};y_w\right)
\right] \,.
\esp
\label{H2u11}
\ee
Finally, we ask which combination of weight-two functions in
${\GG}^L_I$ has the $\{1,1\}$ component of its coproduct
given by \eqn{H2u11}. There is a unique answer, modulo elements in
$\mathcal{K}$,
\be
\bsp
H_2^u &\stackrel{\textrm{\scriptsize{ Region I }}}{=}   
-   G\left(\frac{1}{y_u},\frac{1}{y_u y_v};y_w\right) 
+   G\left(1,\frac{1}{y_u y_v};y_w\right) -   G\left(1,\frac{1}{y_u};y_w\right)
-   G\left(1,\frac{1}{y_u};y_v\right) \\
&\qquad  +   G\left(0,1;y_u\right) 
+   G\left(\frac{1}{y_u y_v};y_w\right)G\left(\frac{1}{y_u};y_w\right)
+   G\left(\frac{1}{y_u};y_v\right)G\left(\frac{1}{y_u y_v};y_w\right)  \\
&\qquad -   G\left(1;y_w\right)G\left(\frac{1}{y_u y_v};y_w\right) 
-   G\left(1;y_v\right)G\left(\frac{1}{y_u y_v};y_w\right) 
-   G\left(0;y_u\right)G\left(\frac{1}{y_u y_v};y_w\right) \\
&\qquad - \frac{1}{2}  G\left(\frac{1}{y_u};y_w\right)^2 
-   G\left(\frac{1}{y_u};y_v\right)G\left(\frac{1}{y_u};y_w\right)
+   G\left(1;y_w\right)G\left(\frac{1}{y_u};y_w\right)  \\
&\qquad +   G\left(1;y_v\right)G\left(\frac{1}{y_u};y_w\right) 
+   G\left(0;y_u\right)G\left(\frac{1}{y_u};y_w\right)
- \frac{1}{2}  G\left(\frac{1}{y_u};y_v\right)^2  \\
&\qquad+   G\left(1;y_v\right)G\left(\frac{1}{y_u};y_v\right) 
+   G\left(0;y_u\right)G\left(\frac{1}{y_u};y_v\right)
-   G\left(0;y_u\right)G\left(1;y_u\right) + \zeta_2 \,.
\esp
\label{H2umodK}
\ee
We have written a specific value for the coefficient of $\zeta_2$, though 
at this stage it is completely arbitrary since $\Delta_{1,1}(\zeta_2)=0$. 
To verify that we have chosen the correct value, we specialize to the surface 
$y_u=1$, on which $u=1$ and $H_2^u=0$. It is straightforward to check that the 
right-hand side of~\eqn{H2umodK} does indeed vanish in this limit.
%%%

An alternative way to translate expressions made from HPLs of
arguments $u_i$ into expressions in terms of the $y$ variables is as
follows.  Any expression in terms of HPLs of arguments $u,v,w$ may be
thought of as the result of applying the integration map to words made
from the letters $u_i$ and $1-u_i$ only. For example,
\begin{align}
H_2^u = H_2(1-u) &= H_{10}(u)+\zeta_2 \\
&= - \int_\gamma d \log (1-s_1) \circ d \log  s_1 + \zeta_2 \,,
\label{exampleintmap}
\end{align}
where, to verify the final equality straightforwardly, we may choose
the contour $\gamma$ to run from $s_i=0$ to $(s_1=u,s_2=v,s_3=w)$
sequentially along the $s_1,s_2,s_3$ axes. In the above simple example
the second and third parts of the contour are irrelevant since the
form to be integrated only depends on $s_1$ anyway. Then we can change
variables from $u,v,w$ to $y_u,y_v,y_w$ by defining
\be
s_1 = \frac{t_1 (1-t_2)(1-t_3)}{(1-t_1 t_2)(1-t_1 t_3)} \,,
\ee
and similarly for $s_2,s_3$. Since the result obtained depends only
on the end points of the contour, and not the precise path taken, we
may instead choose the contour as the one which goes from the origin
$t_i=0$ to the point $t_1=y_u, t_2=y_v, t_3=y_w$ sequentially along
the $t_1,t_2,t_3$ axes, as in the discussion around \eqn{intmap}.
Then expression~(\ref{exampleintmap}) yields an
expression equivalent to~\eqn{H2umodK}.

Continuing this procedure on to higher weights is straightforward, although
the expressions become increasingly complicated. For example, the
expression for $H_{4,2}^w$ has 9439 terms. 
It is clear that
${\GG}^L_I$ is not an efficient basis, at least for
representing harmonic polylogarithms with argument $u_i$. Despite this
inefficiency, ${\GG}^L_I$ and ${\GG}^L_{II}$ have
the virtue of spanning the space of hexagon functions, although
they still contain many more functions than desired.
In the next subsection, we describe how we can
iteratively impose constraints in order to construct a basis for just the
hexagon functions.

%%%%%%%%%%%%%%%%%%%%%%%%%%%%%%%%%%%%%%%%%%%%%%%
\subsection{Constructing the hexagon functions}
\label{sec:construct}

Unitarity requires the branch cuts of physical quantities
to appear in physical channels. For dual conformally-invariant functions
corresponding to the scattering of massless particles, the only permissible
branching points are when a cross ratio vanishes or approaches infinity.
The location of branch points in an iterated integral is controlled by
the first entry of the symbol; hence the first entry should be one of
the cross ratios, as discussed previously.
However, it is not necessary to restrict our attention to the symbol:
it was argued in ref.~\cite{Duhr2012fh} that the condition of only
having physical branch points can be promoted to the coproduct.
Then the monodromy operator $\mathcal{M}_{z_k=z_0}$ (which gives the phase in
analytically continuing the variable $z_k$ around the point $z_0$)
acts on the first component of the coproduct $\Delta$
(see appendix~\ref{sec:app_multi_poly2}),
\be
\Delta\circ \mathcal{M}_{z_k=z_0}
= \left(\mathcal{M}_{z_k=z_0} \otimes \textrm{id}\right) \circ\, \Delta\,.
\label{eq:monodromy}
\ee
We conclude that if $F_n$ is a weight-$n$ function with the 
proper branch-cut locations, and 
\be
\Delta_{n-1,1}(F_n) = \sum_r F_{n-1}^r \otimes d \ln \phi_r \,,
\label{Fncoprod}
\ee
then $F_{n-1}^r$ must also be a weight-$(n-1)$ function with the proper branch-cut
locations, for every $r$ (which labels the possible letters in the symbol).
Working in the other direction, suppose we know the basis of
hexagon functions through weight $n-1$. We may then use \eqn{Fncoprod}
and the coproduct bootstrap of \sect{sec:coproduct} to build the basis at
weight $n$.

There are a few subtleties that must be taken into account before
applying this method directly. To begin with, the condition that all
the $F_{n-1}^r$ belong to the basis of hexagon functions guarantees that they
have symbol entries drawn from $\Su$.  However, it does not guarantee
that $F_n$ has this property since the $\phi_r$ are drawn from the set $\Sy$,
which is larger than $\Su$. This issue is easily remedied by simply
disregarding those functions whose symbols have final entries outside of
the set $\Su$.

In pushing to higher weights, it becomes necessary to pursue a more
efficient construction. For this purpose, it is useful to decompose
the space of hexagon functions, which we denote by $\mathcal{H}$, into its
parity-even and parity-odd components,
\be
\mathcal{H}\ =\ \mathcal{H}^+ \oplus \mathcal{H}^-\, .
\ee
The coproduct can be taken separately on each component,
\be
\bsp
\Delta_{n-1,1}(\mathcal{H}_n^+)\ &\subseteq\ 
\big(\mathcal{H}_{n-1}^+ \otimes \mathcal{L}_1^+\big) \, \oplus 
\, \big(\mathcal{H}_{n-1}^- \otimes \mathcal{L}_1^-\big)\,,\\
\Delta_{n-1,1}(\mathcal{H}_n^-)\ &\subseteq\ 
\big(\mathcal{H}_{n-1}^+ \otimes \mathcal{L}_1^-\big) \, \oplus
\, \big(\mathcal{H}_{n-1}^- \otimes \mathcal{L}_1^+\big)\,,
\esp
\label{ParitySplitting}
\ee
where $\mathcal{L}_1^+$ and $ \mathcal{L}_1^-$ are the parity-even and
parity-odd functions of weight one, 
\be
\bsp 
\mathcal{L}_1^+\ &=\
\left\{\ln u,\,\ln(1-u),\,\ln v,\,\ln(1-v),\,\ln w,\,\ln(1-w)\right\}\,,\\ 
\mathcal{L}_1^-\ &=\ \left\{\ln y_u,\, \ln y_v,\, \ln y_w\right\}\,.
\esp 
\ee

To construct $\mathcal{H}_n^{\pm}$, we simply write down the most
general ansatz for both the left-hand side and the right-hand side of
\eqn{ParitySplitting} and solve the linear system. The ansatz for
$\mathcal{H}_n^{\pm}$ will be constructed from the either
${\GG}^L_{I}$ or ${\GG}^L_{II}$, supplemented by 
multiple zeta values, while a
parametrization of the right-hand side is known by assumption. For
high weights, the linear system becomes prohibitively large, which is
one reason why it is useful to construct the even and odd sectors
separately, since it effectively halves the computational burden. We
note that not every element on the right hand side of \eqn{ParitySplitting}
is actually in the image of $\Delta_{n-1,1}$. For such cases, we will
simply find no solution to the linear equations. Finally, this
parametrization of the $\{n-1,1\}$ component of the coproduct
guarantees that the symbol of any function in $\mathcal{H}_n$ will
have symbol entries drawn from $\Su$.

Unfortunately, the procedure we just have outlined does not actually
guarantee proper branch cuts in all cases. The obstruction is related
to the presence of weight-$(n-1)$ multiple zeta values in
the space $\mathcal{H}_{n-1}^+$. Such terms may become problematic
when used as in \eqn{ParitySplitting} to build the weight-$n$ space,
because they get multiplied by logarithms, which may contribute improper
branch cuts. For example,
\be
\zeta_{n-1} \otimes \ln(1-u) \in \mathcal{H}_{n-1}^+ \otimes \mathcal{L}_1^+ \,,
\ee
but the function $\zeta_{n-1} \, \ln(1-u)$ has a spurious branch point
at $u=1$.  Naively, one might think such terms must be excluded from 
our ansatz, but this turns out to be incorrect. In some cases, they are
needed to cancel off the bad behavior of other, more complicated
functions.

We can exhibit this bad behavior in a simple one-variable function,
\be
f_2(u) = \textrm{Li}_{2}(u) + \ln u\, \ln(1-u)\; \in\; \mathcal{H}_2^+\,.
\label{f2example}
\ee
It is easy to write down a weight-three function $f_3(u)$ that satisfies,
\be
\Delta_{2,1}(f_3(u)) = f_2(u)\otimes\ln(1-u)\,.
\ee
Indeed, one may easily check that
\be
f_3(u) = H_{2,1}(u) + \textrm{Li}_{2}(u) \ln(1-u) + \frac{1}{2}\ln^2(1-u)\ln u
\ee
does the job. The problem is that $f_3(u) \not\in \mathcal{H}_3^+$
because it has a logarithmic branch cut starting at $u=1$. In fact,
the presence of this cut is indicated by a simple pole at $u=1$ in its
first derivative,
\be
f_3'(u)\big|_{u\to 1}\ \to\ - \frac{\zeta_2}{1-u}\,.
\ee
The residue of the pole is just $f_2(1)$ and can be read directly from 
\eqn{f2example} without ever writing down $f_3(u)$.  This suggests that
the problem can be remedied by subtracting $\zeta_2$ from $f_2(u)$.
Indeed, for
\be
\tilde{f}_2(u) = f_2(u) - \zeta_2 = -\textrm{Li}_2(1-u)\,,
\ee
there does exist a function,
\be
\tilde{f}_3(u)= - \textrm{Li}_3(1-u) \in \mathcal{H}_3^+\, , 
\ee
for which,
\be
\Delta_{2,1}(\tilde{f}_3(u)) = \tilde{f}_2(u)\otimes\ln(1-u)\,.
\ee

More generally, any function whose first derivative yields a simple
pole has a logarithmic branch cut starting at the location of that
pole. Therefore, the only allowed poles in the $u_i$-derivative are at
$u_i=0$. In particular, the absence of poles at $u_i=1$ provides
additional constraints on the space $\mathcal{H}_n^{\pm}$.

These constraints were particularly simple to impose in the above
single-variable example, because the residue of the pole at $u=1$ could
be directly read off from a single term in the coproduct, namely the
one with $\ln(1-u)$ in the last slot. In the full multiple-variable
case, the situation is slightly more complicated. The coproduct 
of any hexagon function will generically have nine terms,
\be
\Delta_{n-1,1}(F) \equiv \sum_{i=1}^3 \Bigl[ F^{u_i} \otimes \ln u_i 
+ F^{1-u_i} \otimes \ln (1-u_i) + F^{y_i} \otimes \ln y_i \Bigr] \,,
\label{FullFcoprod}
\ee
where $F$ is a function of weight $n$ and the nine functions 
$\{F^{u_i},F^{1-u_i}, F^{y_i}\}$ are of weight $(n-1)$ and completely
specify the $\{n-1,1\}$ component of the coproduct. The derivative
with respect to $u$ can be evaluated using \eqns{yi_u_diff}{FullFcoprod}
and the chain rule,
\be
\label{eq:diff_basis}
\frac{\partial F}{\partial u}\bigg|_{v,w} 
\ =\ \frac{F^u}{u} - \frac{F^{1-u}}{1-u} 
+ \frac{1-u-v-w}{u\sqrt{\Delta}} F^{y_u}
+ \frac{1-u-v+w}{(1-u)\sqrt{\Delta}}F^{y_v}
+ \frac{1-u+v-w}{(1-u)\sqrt{\Delta}} F^{y_w}\,.
\ee
Clearly, a pole at $u=1$ can arise from $F^{1-u}$, $F^{y_v}$ or $F^{y_w}$,
or it can cancel between these terms.

The condition that \eqn{eq:diff_basis} has no pole at $u=1$ is a strong one,
because it must hold for any values of $v$ and $w$. In fact, this condition 
mainly provides consistency checks, because a much weaker set of constraints 
turns out to be sufficient to fix all undetermined constants in our ansatz.

It is useful to consider the constraints in the even and odd subspaces 
separately. Referring to~\eqn{parityuydelta}, parity sends 
$\sqrt{\Delta}\to -\sqrt{\Delta}$, and, therefore, any parity-odd 
function must vanish when $\Delta=0$. Furthermore, recalling \eqn{u_from_y},
\be
\sqrt{\Delta} = \frac{(1-y_u) (1-y_v) (1-y_w) (1-y_u y_v y_w)}
                       {(1-y_u y_v) (1-y_v y_w) (1-y_w y_u)}
\,,
\ee
we see that any odd function must vanish when $y_i\to 1$ or when 
$y_u y_v y_w \to 1$. It turns out that these conditions are sufficient
to fix all undetermined constants in the odd sector. One may then verify
that there are no spurious poles in the $u_i$-derivatives.

There are no such vanishing conditions in the even sector, and to fix 
all undetermined constants we need to derive specific constraints from
\eqn{eq:diff_basis}.  We found it convenient
to enforce the constraint for the particular values of $v$ and $w$ such 
that the $u\to 1$ limit coincides with the limit of 
Euclidean multi-Regge kinematics (EMRK).  In this limit, $v$ and $w$
vanish at the same rate that $u$ approaches 1,
\be
\hbox{EMRK:}\quad u\to1,\ v\to0,\ w\to0;\quad
\frac{v}{1-u} \equiv x,\ \frac{w}{1-u} \equiv y,
\label{EMRK}
\ee
where $x$ and $y$ are fixed.  In the $y$ variables, the EMRK limit
takes $y_u\to1$, while $y_v$ and $y_w$ are held fixed, and can be related to
$x$ and $y$ by,
\be
x = \frac{y_v(1-y_w)^2}{(1-y_v y_w)^2} \,,
\qquad
y = \frac{y_w(1-y_v)^2}{(1-y_v y_w)^2} \,.
\label{xyEMRK}
\ee
This limit can also be called the (Euclidean) soft limit, in which
one particle gets soft.  The final point, $(u,v,w)=(1,0,0)$, also
lies at the intersection of two lines representing different
collinear limits: $(u,v,w)=(x,1-x,0)$ and $(u,v,w)=(x,0,1-x)$,
where $x\in[0,1]$.

In the case at hand, $F$ is an even function and so the coproduct components 
$F^{y_i}$ are odd functions of weight $n-1$, and as such have already been
constrained to vanish when $y_i\to 1$. 
(Although the coefficients of $F^{y_v}$ and $F^{y_w}$ in \eqn{eq:diff_basis}
contain factors of $1/\sqrt{\Delta}$, which diverge in the limit $y_u\to1$,
the numerator factors $1-u\mp (v-w)$ can be seen from \eqn{EMRK}
to vanish in this limit, cancelling the $1/\sqrt{\Delta}$ divergence.)
Therefore, the constraint that \eqn{eq:diff_basis} have no pole at $u=1$
simplifies considerably:
\be
F^{1-u}(y_u=1,y_v,y_w) = 0\,.
\label{Fomu_vanish}
\ee
Of course, two additional constraints can be obtained by taking
cyclic images. These narrower constraints turn out to be sufficient to
completely fix all free coefficients in our ansatz in the even sector.

Finally, we are in a position to construct the functions of the
hexagon basis. At weight one, the basis simply consists of the three
logarithms, $\ln u_i$. Before proceeding to weight two, we must
rewrite these functions in terms of multiple polylogarithms. This
necessitates a choice between Regions I and II, or between the bases
${\GG}^L_{I}$ and ${\GG}^L_{II}$. We construct
the basis for both cases, but for definiteness let us work
in Region I.

Our ansatz for $\Delta_{1,1}(\mathcal{H}_{2}^+$) consists of the 18
tensor products,
\be
\{\ln u_i \otimes x\;\big|\; x\in \mathcal{L}_1^+\}\, ,
\ee
which we rewrite in terms of multiple polylogarithms in
${\GG}^L_{I}$. Explicit linear algebra shows that only a
nine-dimensional subspace of these tensor products can be written as
$\Delta_{1,1}(G_2)$ for $G_2\in {\GG}^L_{I}$. Six of these
weight-two functions can be written as products of logarithms. The
other three may be identified with $H_2(1-u_i)$ by using the methods of
\sect{sec:coproduct}. (See {\it e.g.}~\eqn{H2umodK}.)

Our ansatz for $\Delta_{1,1}(\mathcal{H}_{2}^-$) consists of the nine
tensor products,
\be
\{\ln u_i \otimes x\;\big|\; x\in \mathcal{L}_1^-\}\, ,
\ee
which we again rewrite in terms of multiple polylogarithms in
${\GG}^L_{I}$. In this case, it turns out that there is no
linear combination of these tensor products that can be written as
$\Delta_{1,1}(G_2)$ for $G_2\in {\GG}^L_{I}$. This confirms
the analysis at symbol level as summarized in \tab{tab:basis_count},
which shows three parity-even irreducible functions of weight two
(which are identified as HPLs), and no parity-odd functions.

A similar situation unfolds in the parity-even sector at weight three,
namely that the space is spanned by HPLs of a single variable.
However, the parity-odd sector reveals a new function.  To find it,
we write an ansatz for $\Delta_{2,1}(\mathcal{H}_{3}^-$) consisting
of the 39 objects,
\be
\{f_2 \otimes x \;\big|\; f_2 \in \mathcal{H}_2^+,\, 
x\in \mathcal{L}_1^-\}
\ee
(where $\mathcal{H}_2^+ = \{ \zeta_2, \ln u_i \ln u_j, H_2^{u_i} \}$), 
and then look for a linear
combination that can be written as $\Delta_{2,1}(G_3)$ for
$G_3\in {\GG}^L_{I}$. After imposing the constraints that the
function vanish when $y_i\to 1$ and when $y_u y_v y_w \to 1$,
there is a unique solution,
\be
\bsp
\PhiTilde &\stackrel{\textrm{\scriptsize{ Region I }}}{=}
-  G\left(0;y_u\right)G\left(0;y_v\right)G\left(0;y_w\right)
+  G\left(0,1;y_u\right)G\left(0;y_u\right) -  G\left(0,1;y_u\right)G\left(0;y_v\right) \\
 &\phantom{\stackrel{\textrm{\scriptsize{ Region I }}}{=}}
-  G\left(0,1;y_u\right)G\left(0;y_w\right) 
-  G\left(0,1;y_v\right)G\left(0;y_u\right) +  G\left(0,1;y_v\right)G\left(0;y_v\right) \\
 &\phantom{\stackrel{\textrm{\scriptsize{ Region I }}}{=}}
-  G\left(0,1;y_v\right)G\left(0;y_w\right) 
-  G\left(0,\frac{1}{y_u};y_v\right)G\left(0;y_u\right) 
-  G\left(0,\frac{1}{y_u};y_v\right)G\left(0;y_v\right) \\
 &\phantom{\stackrel{\textrm{\scriptsize{ Region I }}}{=}}
+  G\left(0,\frac{1}{y_u};y_v\right)G\left(0;y_w\right) 
-  G\left(0,1;y_w\right)G\left(0;y_u\right)
-  G\left(0,1;y_w\right)G\left(0;y_v\right) \\
 &\phantom{\stackrel{\textrm{\scriptsize{ Region I }}}{=}}
+  G\left(0,1;y_w\right)G\left(0;y_w\right) 
+ 2 G\left(0,1;y_w\right)G\left(\frac{1}{y_u};y_v\right) 
-  G\left(0,\frac{1}{y_u};y_w\right)G\left(0;y_u\right) \\
 &\phantom{\stackrel{\textrm{\scriptsize{ Region I }}}{=}}
+  G\left(0,\frac{1}{y_u};y_w\right)G\left(0;y_v\right) 
-  G\left(0,\frac{1}{y_u};y_w\right)G\left(0;y_w\right) 
- 2 G\left(0,\frac{1}{y_u};y_w\right)G\left(1;y_v\right) \\
 &\phantom{\stackrel{\textrm{\scriptsize{ Region I }}}{=}}
+  G\left(0,\frac{1}{y_v};y_w\right)G\left(0;y_u\right) 
-  G\left(0,\frac{1}{y_v};y_w\right)G\left(0;y_v\right) 
-  G\left(0,\frac{1}{y_v};y_w\right)G\left(0;y_w\right) \\
 &\phantom{\stackrel{\textrm{\scriptsize{ Region I }}}{=}}
- 2 G\left(0,\frac{1}{y_v};y_w\right)G\left(1;y_u\right) 
+  G\left(0,\frac{1}{y_u y_v};y_w\right)G\left(0;y_u\right) 
+  G\left(0,\frac{1}{y_u y_v};y_w\right)G\left(0;y_v\right) \\
 &\phantom{\stackrel{\textrm{\scriptsize{ Region I }}}{=}}
+  G\left(0,\frac{1}{y_u y_v};y_w\right)G\left(0;y_w\right) 
+ 2 G\left(0,\frac{1}{y_u y_v};y_w\right)G\left(1;y_u\right) 
+ 2 G\left(0,\frac{1}{y_u y_v};y_w\right)G\left(1;y_v\right) \\
 &\phantom{\stackrel{\textrm{\scriptsize{ Region I }}}{=}}
- 2 G\left(0,\frac{1}{y_u y_v};y_w\right)G\left(\frac{1}{y_u};y_v\right) 
- 2 G\left(0,0,1;y_u\right) - 2 G\left(0,0,1;y_v\right) \\
 &\phantom{\stackrel{\textrm{\scriptsize{ Region I }}}{=}}
+ 2 G\left(0,0,\frac{1}{y_u};y_v\right) - 2 G\left(0,0,1;y_w\right) 
+ 2 G\left(0,0,\frac{1}{y_u};y_w\right) + 2 G\left(0,0,\frac{1}{y_v};y_w\right) \\
 &\phantom{\stackrel{\textrm{\scriptsize{ Region I }}}{=}}
- 2 G\left(0,0,\frac{1}{y_u y_v};y_w\right) - 2 G\left(0,1,1;y_u\right) 
- 2 G\left(0,1,1;y_v\right) + 2 G\left(0,\frac{1}{y_u},\frac{1}{y_u};y_v\right) \\
 &\phantom{\stackrel{\textrm{\scriptsize{ Region I }}}{=}}
- 2 G\left(0,1,1;y_w\right) + 2 G\left(0,\frac{1}{y_u},\frac{1}{y_u};y_w\right) 
+ 2 G\left(0,\frac{1}{y_v},\frac{1}{y_v};y_w\right) \\
 &\phantom{\stackrel{\textrm{\scriptsize{ Region I }}}{=}}
+ 2 G\left(0,\frac{1}{y_u y_v},1;y_w\right) 
- 2 G\left(0,\frac{1}{y_u y_v},\frac{1}{y_u};y_w\right) 
- 2 G\left(0,\frac{1}{y_u y_v},\frac{1}{y_v};y_w\right)\\
 &\phantom{\stackrel{\textrm{\scriptsize{ Region I }}}{=}} 
- \zeta_2\, G\left(0;y_u\right)-\zeta_2\, G\left(0;y_v\right) 
- \zeta_2\, G\left(0;y_w\right) \,.
\esp
\label{Phimultipoly}
\ee
The normalization can be fixed by comparing to the differential equation for
$\PhiTilde$, \eqn{duPhi}.
This solution is totally symmetric under the $S_3$ permutation group
of the three cross ratios $\{u,v,w\}$, or equivalently of the three
variables $\{y_u,y_v,y_w\}$.  However, owing to our choice of basis $\GG_I^L$, 
this symmetry is broken in the representation~(\ref{Phimultipoly}).

In principle, this procedure may be continued and used to construct a
basis for the space $\mathcal{H}_n$ for any value of $n$. In practice, it
becomes computationally challenging to proceed beyond moderate weight,
say $n=5$. The three-loop remainder function is a weight-six function,
but, as we will see shortly, to find its full functional form we do
not need to know anything about the other weight-six functions. On the
other hand, we do need a complete basis for all functions of weight
five or less. We have constructed all such functions using the methods
just described. Referring to \tab{tab:basis_count}, there are 69
functions with weight less than or equal to five. However, any
function with no $y$'s in its symbol can be written in terms of
ordinary HPLs, so there are only 30 genuinely new functions. The expressions for
these functions in terms of multiple polylogarithms are quite lengthy, so we
present them in computer-readable format in the attached files.

\renewcommand{\arraystretch}{1.25}
\begin{table}[!t]
\centering
\begin{tabular}[t]{c||c|c|c|c|c}
\hline\hline
\textrm{Weight} & $y^0$ & $y^1$ & $y^2$ & $y^3$ & $y^4$ \\
\hline
1 & 3 HPLs& - & - & - & -\\
\hline
2 & 3 HPLs & - & - & - & -\\
\hline
3 & 6 HPLs & $\PhiTilde$ & - & - & -\\
\hline
4  & 9 HPLs & $3\times F_1$ & $3 \times \Omega^{(2)}$ & -& -\\
\hline
5 & 18 HPLs & $G$, $3\times K_1$ & $5\times M_1$, $N$, $O$, $6 \times \Qep$
  & $3\times H_1$, $3\times J_1$  & - \\
\hline
6 & 27 HPLs & 4 & 27 & 29 & $3\times \Rep$ + 15\\
\hline\hline
\end{tabular}
\caption{Irreducible basis of hexagon functions, graded by the maximum 
number of $y$ entries in the symbol. The indicated multiplicities 
specify the number of independent functions obtained by applying the 
$S_3$ permutations of the cross ratios.}
\label{tab:basis_labeled}
\end{table}

The 30 new functions can be obtained from the permutations of 11 basic functions
which we call $\tilde{\Phi}_6$, $F_1$, $\Omega^{(2)}$, $G$, $H_1$,
$J_1$, $K_1$, $M_1$, $N$, $O$, and $Q_{\textrm{ep}}$. 
Two of these functions, $\tilde{\Phi}_6$ and $\Omega^{(2)}$,
have appeared in other contexts, as mentioned in \sect{sec:symbol}.
Also, a linear combination of $F_1$ and its cyclic image can be
identified with the odd part of the two-loop ratio function, denoted
by $\tilde{V}$~\cite{Dixon2011nj}.  (The precise relation is given
in \eqn{FromF1}.)  We believe that the remaining functions are new.
In \tab{tab:basis_labeled}, we organize these functions
by their weight and $y$-grading.  We also indicate how many independent
functions are generated by permuting the cross ratios.  For example,
$\tilde{\Phi}_6$ is totally symmetric, so it generates a unique entry,
while $F_1$ and $\Omega^{(2)}$ are symmetric under exchange of two variables,
so they sweep out a triplet of independent functions under cyclic
permutations.  The function $\Qep$ has no symmetries, so under $S_3$ 
permutations it sweeps out six independent functions.  The same would
be true of $M_1$, except that a totally antisymmetric
linear combination of its $S_3$ images and those of $\Qep$ are
related, up to products of lower-weight functions and ordinary HPLs 
(see \eqn{eq:M1_combo}).  Therefore we count only five independent
functions arising from the $S_3$ permutations of $M_1$.

We present the $\{n-1,1\}$ components of the coproduct of these 11
basis functions in appendix~\ref{sec:app_basis}. This information,
together with the value of the function at the point $(1,1,1)$ (which
we take to be zero in all but one case), is sufficient to uniquely
define the basis of hexagon functions.  We will elaborate on these
ideas in the next section.

\vfill\eject

%%%%%%%%%%%%%%%%%%%%%%%%%%%%%%%%%%%%%%%%%%%%%%%%%%%%%%%%%%%%%%%%%%%%%%%

\section{Integral Representations}
\label{sec:integral_reps}

In the previous section, we described an iterative procedure to
construct the basis of hexagon functions in terms of multiple
polylogarithms in the $y$ variables. The result is a fully analytic,
numerically efficient representation of any given basis
function. While convenient for many purposes, this representation is
not without some drawbacks. Because $\Sy$ has one more element than
$\Su$, and because the first entry condition is fairly opaque in the
$y$ variables, the multiple polylogarithm representation is often
quite lengthy, which in turn sometimes obscures interesting
properties. Furthermore, the iterative construction and the numerical
evaluation of multiple polylogarithms are best performed when the
$y_i$ are real-valued, limiting the kinematic regions in which these
methods are practically useful.

For these reasons, it is useful to develop a parallel representation
of the hexagon functions, based directly on the system of first-order
differential equations they satisfy. These differential equations can
be solved in terms of (iterated) integrals over lower-weight
functions.  Since most of the low weight functions are HPLs, which are
easy to evaluate, one can obtain numerical representations for the
hexagon functions, even in the kinematic regions where the $y_i$ are
complex.  The differential equations can also be solved in terms of
simpler functions in various limits, which will be the subject of
subsequent sections.

\subsection{General setup}
\label{sec:general_setup}
One benefit of the construction of the basis of hexagon functions in
terms of multiple polylogarithms is that we can explicitly calculate
the coproduct of the basis functions. We tabulate the $\{n-1,1\}$
component of the coproduct for each of these functions in
appendix~\ref{sec:app_basis}. This data exposes how the various
functions are related to one another, and, moreover, this web of
relations can be used to define a system of differential equations
that the functions obey. These differential equations, together with
the appropriate boundary conditions, provide an alternative definition
of the hexagon functions.  In fact, as we will soon argue, it is
actually possible to derive these differential equations iteratively,
without starting from an explicit expression in terms of multiple
polylogarithms. It is also possible to express the differential equations
compactly in terms of a Knizhnik-Zamolodchikov equation along the lines studied
in ref.~\cite{FBThesis}. Nevertheless, the coproduct on multiple
polylogarithms, in particular the $\{n-1,1\}$ component as given in
\eqn{FullFcoprod}, is useful to frame the discussion of the
differential equations and helps make contact with
\sect{sec:hex_multi_poly}. 

It will be convenient to consider not just derivatives with respect to
a cross ratio, as in \eqn{eq:diff_basis}, but also derivatives with
respect to the $y$ variables. For that purpose, we need the following
derivatives, which we perform holding $y_v$ and $y_w$ constant,
\be
\bsp
\frac{\partial\ln u}{\partial y_u} &= \frac{(1-u)(1-v-w)}{y_u\sqrt{\Delta}} \,,
\qquad
\frac{\partial\ln v}{\partial y_u}\ =\ -\frac{u(1-v)}{y_u\sqrt{\Delta}} \,,
\\
\frac{\partial\ln(1-u)}{\partial y_u} &= -\frac{u(1-v-w)}{y_u\sqrt{\Delta}} \,,
\qquad
\frac{\partial\ln(1-v)}{\partial y_u}\ =\ \frac{uv}{y_u\sqrt{\Delta}} \,.
\esp
\label{partials_yu}
\ee
We also consider the following linear combination,
\be
\frac{\partial}{\partial\ln(y_u/y_w)}\ \equiv\ 
  y_u \frac{\partial}{\partial y_u}\bigg|_{y_v,y_w}
-\;\, y_w \frac{\partial}{\partial y_w}\bigg|_{y_v,y_u} \,.
\label{lnyuywdef}
\ee
Using \eqns{yi_u_diff}{partials_yu}, as well as the
definition~(\ref{lnyuywdef}), we obtain three differential
equations (plus their cyclic images) relating a function $F$ to its 
various coproduct components,
\bea
\frac{\partial F}{\partial u}\bigg|_{v,w} \!\! &=& \!\!
\frac{F^u}{u} -\frac{F^{1-u}}{1-u}
+ \frac{1-u-v-w}{u\,\sqrt{\Delta}} F^{y_u}
+ \frac{1-u-v+w}{(1-u)\sqrt{\Delta}}F^{y_v}
+\frac{1-u+v-w}{(1-u)\sqrt{\Delta}} F^{y_w} \,, 
\nonumber\\
&&~\label{u_diffeq}\\
\sqrt{\Delta}\, y_u \frac{\partial F}{\partial y_u}\bigg|_{y_v,y_w} \!\! &=& \!\!
(1-u)(1-v-w) F^u - u(1-v) F^v - u(1-w)F^w - u(1-v-w)F^{1-u}
\nonumber\\
&&\quad+uv\,F^{1-v}+uw\,F^{1-w} + \sqrt{\Delta}\, F^{y_u} \,, 
\label{yu_diffeq}\\
\sqrt{\Delta}\frac{\partial F}{\partial \ln(y_u/y_w)} \!\! &=& \!\!
(1-u)(1-v) F^u - (u-w)(1-v) F^v - (1-v)(1-w)F^w - u(1-v)F^{1-u} 
\nonumber\\
&&\quad + (u-w)v\,F^{1-v} + w(1-v)\,F^{1-w}
+ \sqrt{\Delta}\, F^{y_u}-\sqrt{\Delta}\, F^{y_w} \,.
\label{yuyw_diffeq}
\eea

Let us assume that we somehow know the coproduct components of $F$,
either from the explicit representations given in
appendix~\ref{sec:app_basis}, or from the iterative approach that we
will discuss in the next subsection. We then know the right-hand sides
of eqs.~(\ref{u_diffeq})-(\ref{yuyw_diffeq}), and we can integrate any
of these equations along the appropriate contour to obtain an integral
representation for the function $F$. While \eqn{u_diffeq} integrates
along a very simple contour, namely a line that is constant in $v$ and
$w$, this also means that the boundary condition, or initial data, must
be specified over a two-dimensional plane, as a function of $v$ and
$w$ for some value of $u$. In contrast, we will see that the other two
differential equations have the convenient property that the initial
data can be specified on a single point.

Let us begin with the differential equation~(\ref{yu_diffeq}) and its
cyclic images. For definiteness, we consider the differential equation
in $y_v$. To integrate it, we must find the contour in $(u,v,w)$ that
corresponds to varying $y_v$, while holding $y_u$ and $y_w$ constant.
Following ref.~\cite{Dixon2011nj}, we define the three ratios,
\be
\bsp
r&=\frac{w(1-u)}{u(1-w)}=\frac{y_w(1-y_u)^2}{y_u(1-y_w)^2} \,, \\
s&=\frac{w(1-w)u(1-u)}{(1-v)^2}=\frac{y_w(1-y_w)^2y_u(1-y_u)^2}{(1-y_wy_u)^4}
 \,, \\
t&=\frac{1-v}{uw}=\frac{(1-y_wy_u)^2(1-y_uy_vy_w)}{y_w(1-y_w)y_u(1-y_u)(1-y_v)}
 \,.
\esp
\ee
Two of these ratios, $r$ and $s$, are actually independent of $y_v$, 
while the third, $t$, varies. Therefore, we can let $t$ parameterize the 
contour, and denote by $(u_t,v_t,w_t)$ the values of the cross ratios 
along this contour at generic values of $t$. Since $r$ and $s$ are 
constants, we have two constraints,
\be
\bsp
\frac{w_t(1-u_t)}{u_t(1-w_t)}&=\frac{w(1-u)}{u(1-w)} \,, \\
\frac{w_t(1-w_t)u_t(1-u_t)}{(1-v_t)^2}&=\frac{w(1-w)u(1-u)}{(1-v)^2} \,.
\esp
\ee
We can solve these equations for $v_t$ and $w_t$, giving,
\be
\label{eq:y_v_contour}
v_t = 1-\frac{(1-v)u_t(1-u_t)}{u(1-w)+(w-u)u_t} \,, \qquad\qquad
w_t = \frac{(1-u)wu_t}{u(1-w)+(w-u)u_t} \,.
\ee
Finally, we can change variables so that $u_t$ becomes the integration 
variable. Calculating the Jacobian, we find,
\be
\frac{d\ln y_v}{du_t} = \frac{d\ln y_v}{d\ln t} \, \frac{d\ln t}{du_t}
 = \frac{(1-y_v)(1-y_uy_vy_w)}{y_v(1-y_w y_u)}\frac{1}{u_t(u_t-1)}
 = \frac{\sqrt{\Delta_t}}{v_t\,u_t(u_t-1)}\, ,
\ee
where $\Delta_t\equiv\Delta(u_t,v_t,w_t)$. There are two natural
basepoints for the integration: $u_t=0$, for which $y_v=1$ and
$(u,v,w)=(0,1,0)$; and $u_t=1$, for which $y_v=1/(y_u y_w)$ and
$(u,v,w)=(1,1,1)$. Both choices have the convenient property that they
correspond to a surface in terms of the variables $(y_u,y_v,y_w)$, but
only to a single point in terms of the variables $(u,v,w)$. This
latter fact allows for the simple specification of boundary data.

For most purposes, we choose to integrate off of the point $u_t=1$, in which
case we find the following solution to the differential equation,
\be
\bsp
F(u,v,w) &= F(1,1,1) + \int_{\frac{1}{y_u y_w }}^{y_v}d \ln \hat{y}_v
\,\frac{\partial F}{\partial \ln y_v}\big(y_u,\hat{y}_v,y_w\big)\\
&= F(1,1,1) + \int^u_1\frac{du_t}{u_t(u_t-1)}
\frac{\sqrt{\Delta_t}}{v_t}
\frac{\partial F}{\partial\ln y_v}(u_t,v_t,w_t)\\
&= F(1,1,1) - \sqrt{\Delta}\int^u_1\frac{du_t}{v_t[u(1-w)+(w-u)u_t]}
\frac{\partial F}{\partial\ln y_v}(u_t,v_t,w_t) \,.
\esp
\label{yv_int_rep}
\ee
The last step follows from the observation that $\sqrt{\Delta}/(1-v)$ is 
independent of $y_v$, which implies
\be
\frac{\sqrt{\Delta_t}}{1-v_t}=\frac{\sqrt{\Delta}}{1-v}\,.
\ee

The integral representation~(\ref{yv_int_rep}) for $F$ may be
ill-defined if the integrand diverges at the lower endpoint of
integration, $u_t=1$ or $(u,v,w)=(1,1,1)$. On the other hand, for $F$ to be a
valid hexagon function, it must be regular near this point, and therefore no
such divergence can occur. In fact, this condition is closely related
to the constraint of good branch-cut behavior near $u=1$ discussed in
\sect{sec:construct}. As we build up integral representations for
hexagon functions, we will use this condition to help fix various
undetermined constants.

Furthermore, if $F$ is a parity-odd function, we may immediately conclude 
that $F(1,1,1)=0$, since this point corresponds to the surface $y_u y_v y_w=1$.
If $F$ is parity even, we are free to define the function by the condition 
that $F(1,1,1)=0$. We use this definition for all basis functions, except for 
$\Omegauvw$, whose value at $(1,1,1)$ is specified by its correspondence 
to a particular Feynman integral.

While~\eqn{yv_int_rep} gives a representation that can be evaluated
numerically for most points in the unit cube of cross ratios $0\le u_i
\le 1$, it is poorly suited for Region I. The problem is that the
integration contour leaves the unit cube, requiring a cumbersome
analytic continuation of the integrand. One may avoid this issue by
integrating along the same contour, but instead starting at the point
$u_t=0$ or $(u,v,w)=(0,1,0)$. The resulting representation is,
\be
F(u,v,w) = F(0,1,0) - \sqrt{\Delta}\int^u_0\frac{du_t}{v_t[u(1-w)+(w-u)u_t]}
\frac{\partial F}{\partial\ln y_v}(u_t,v_t,w_t) \,.
\label{yv_int_rep_2}
\ee
If $F$ is a parity-odd function, then the boundary value $F(0,1,0)$ must
vanish, since this point corresponds to the EMRK limit $y_v\to 1$. In
the parity-even case, there is no such condition, and in many cases
this limit is in fact divergent. Therefore, in contrast to
\eqn{yv_int_rep}, this expression may require some regularization near
the $u_t=0$ endpoint in the parity-even case.

It is also possible to integrate the differential
equation~(\ref{yuyw_diffeq}). In this case, we look for a contour
where $y_v$ and $y_u y_w$ are held constant, while the ratio $y_u/y_w$
is allowed to vary. The result is a contour $(u_t,v_t,w_t)$ defined
by,
\be
\label{eq:y_u_y_w_contour}
v_t=\frac{vu_t(1-u_t)}{uw+(1-u-w)u_t} \,, \qquad
w_t=\frac{uw(1-u_t)}{uw+(1-u-w)u_t} \,.
\ee
Again, there are two choices for specifying the boundary data:
either we set $y_u/y_w=y_u y_w$ for which we may take $u_t=0$ and
$(u,v,w)=(0,0,1)$; or we set $y_u/y_w=1/(y_u y_w)$, for which we may take
$u_t=1$ and $(u,v,w)=(1,0,0)$. We therefore obtain two different
integral representations,
\be
\bsp
F(u,v,w) &= F(0,0,1) +  \int_0^u \frac{du_t \, \sqrt{\Delta_t}}
   {u_t(1-u_t)(1-v_t)} \frac{\partial F}{\partial\ln(y_u/y_w)}(u_t,v_t,w_t)\\
&=  F(0,0,1) + \sqrt{\Delta}\int^u_0\frac{du_t}{(1-v_t)[uw+(1-u-w)u_t]}
  \frac{\partial F}{\partial\ln(y_u/y_w)}(u_t,v_t,w_t) \,,
\label{yuyw_int_rep}
\esp
\ee
and,
\be
F(u,v,w) = F(1,0,0) + \sqrt{\Delta}\int^u_1\frac{du_t}{(1-v_t)[uw+(1-u-w)u_t]}
  \frac{\partial F}{\partial\ln(y_u/y_w)}(u_t,v_t,w_t) \,.
\label{yuyw_int_rep_2}
\ee
Here we used the relation,
\be
\frac{\sqrt{\Delta_t}}{v_t} = \frac{\sqrt{\Delta}}{v}\, ,
\ee
which follows from the observation that $\sqrt{\Delta}/v$ is constant
along either integration contour. Finally, we remark that the boundary
values $F(1,0,0)$ and $F(0,0,1)$ must vanish for parity-odd functions,
since the points $(1,0,0)$ and $(0,0,1)$ lie on the $\Delta=0$
surface.  In the parity-even case, there may be issues of
regularization near the endpoints, just as discussed for \eqn{yv_int_rep_2}.

\begin{figure}
\begin{center}
\includegraphics[width=5.0in]{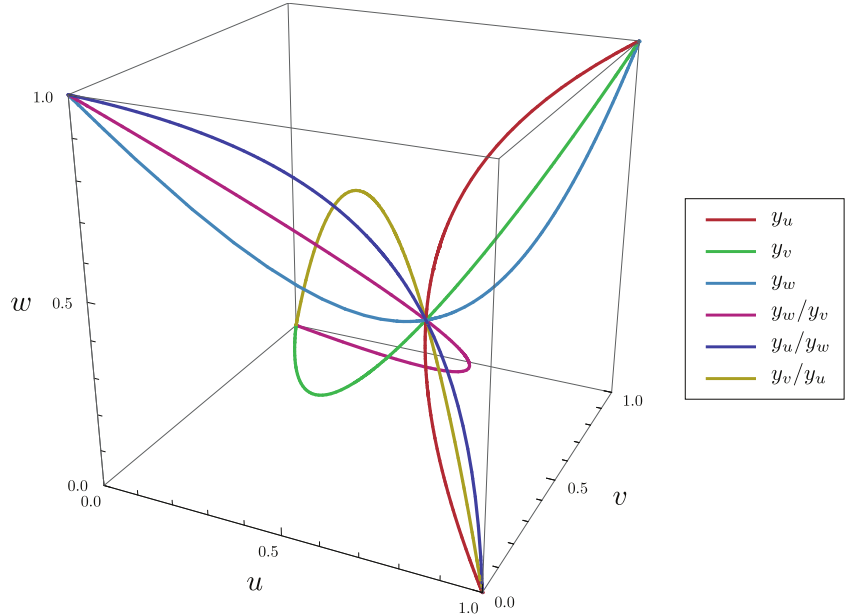}
\end{center}
\caption{The six different integration contours for the point 
$(u,v,w) = (\frac{3}{4},\frac{1}{4},\frac{1}{2})$, labeled by the $y$-variables 
(or their ratios) that vary along the contour.}
\label{fig:contours}
\end{figure}

Altogether, there are six different contours, corresponding to the
three cyclic images of the two types of contours just described. They
may be labeled by the $y$-variables or their ratios that are allowed
to vary along the contour: $\{y_u, y_v, y_w, y_u/y_w, y_v/y_u,
y_w/y_v\}$. The base points for these contours together encompass
$(1,1,1)$, $(0,1,0)$, $(1,0,0)$ and $(0,0,1)$, the four corners of a
tetrahedron whose edges lie on the intersection of the surface $\Delta=0$
with the unit cube.  See~\fig{fig:contours} for an illustration of the contours
passing through the point $(u,v,w)=
(\frac{3}{4},\frac{1}{4},\frac{1}{2})$.

%%%%%%%%%%%%%%%%%%%%%%%%%%%%%%%%%%%%%%%%%%%%%%
\subsection{Constructing the hexagon functions}
\label{sec:construct_int}

In this subsection, we describe how to construct differential
equations and integral representations for the basis of hexagon
functions. We suppose that we do not have any of the function-level
data that we obtained from the analysis of \sect{sec:hex_multi_poly};
instead, we will develop a completely independent alternative method
starting from the symbol. The two approaches are complementary and
provide important cross-checks of one another.

In~\sect{sec:hex_multi_poly_symb}, we presented the construction of
the basis of hexagon functions at symbol level. Here we will promote
these symbols to functions in a three-step iterative process:
\begin{enumerate}
\item Use the symbol of a given weight-$n$ function to write down an ansatz for 
the $\{n-1,1\}$ component of its coproduct in terms of a function-level basis 
at weight $n-1$ that we assume to be known.
\item Fix the undetermined parameters in this ansatz by imposing various 
function-level consistency conditions. These conditions are:
\begin{enumerate}
\item Symmetry. The symmetries exhibited by the symbol should carry over 
to the function.
\item Integrability. The ansatz should be in the image of $\Delta_{n-1,1}$. 
This condition is equivalent to the consistency of mixed partial derivatives.
\item Branch cuts. The only allowed branch cuts start when a cross ratio 
vanishes or approaches infinity. 
\end{enumerate}
\item Integrate the resulting coproduct using the methods of the previous 
subsection, specifying the boundary value and thereby obtaining a well-defined 
function-level member of the hexagon basis.
\end{enumerate}

Let us demonstrate this procedure with some examples. Recalling the
discussion in~\sect{sec:hex_multi_poly_symb}, any function whose
symbol contains no $y$ variables can be written as products of
single-variable HPLs. Therefore, the first nontrivial example occurs
at weight three. As previously mentioned, this function corresponds to
the one-loop six-dimensional hexagon integral, $\PhiTilde$. Its symbol
is given by,
\be
\mathcal{S}(\PhiTilde) = 
\Bigl[-u\otimes v-v\otimes u+u\otimes(1-u) +v\otimes(1-v) +w\otimes(1-w)\Bigr]
\otimes y_w \; + \; \textrm{cyclic}\,.
\ee
It is straightforward to identify the object in brackets as the symbol of a 
linear combination of weight-two hexagon functions (which are just HPLs),
allowing us to write an ansatz for the $\{2,1\}$ component of the coproduct,
\be
\Delta_{2,1}\left(\PhiTilde\right) = 
- \Bigl[ \ln u \ln v + {\rm Li}_2(1-u)+ {\rm Li}_2(1-v) 
+ {\rm Li}_2(1-w) + a \zeta_{2}\Bigr]\otimes \ln y_w \;+\; \text{cyclic}\, ,
\label{eq:PhiTilde21}
\ee
for some undetermined rational number $a$.

The single constant, $a$, can be fixed by requiring that $\PhiTilde$
have the same symmetries as its symbol. In particular, we demand that
$\PhiTilde$ be odd under parity.  As discussed in the previous
section, this implies that it must vanish in the limit that one of the
$y_i$ goes to unity.  In this EMRK limit~(\ref{EMRK}), the
corresponding $u_i$ goes to unity while the other two cross ratios go
to zero. The right-hand side of~\eqn{eq:PhiTilde21} vanishes in this
limit only for the choice $a=-2$. So we can write,
\be
\Delta_{2,1}\left(\PhiTilde\right) = 
-\Omega^{(1)}(u,v,w)\otimes\ln y_w \;+\; \text{cyclic}\, ,
\ee
where,
\be
\bsp
\Omega^{(1)}(u,v,w) &= \ln u \ln v
 + {\rm Li}_2(1-u) + {\rm Li}_2(1-v) + {\rm Li}_2(1-w) - 2 \zeta_{2} \\
&= H_2^u + H_2^v + H_2^w + \ln u \ln v - 2 \zeta_{2} \,,
\esp
\label{Om1final}
\ee
confirming the expression given in~\eqn{tPhisymbol}. It is also
straightforward to verify that~\eqn{eq:PhiTilde21} is integrable and
that it does not encode improper branch cuts. We will not say more
about these conditions here, but we will elaborate on them shortly, in
the context of our next example.

Now that we have the coproduct, we can
use~\eqns{yu_diffeq}{yuyw_diffeq} to immediately write down the
differential equations,
\begin{eqnarray}
\label{eq:diff_phi6_1}
\frac{\partial\PhiTilde}{\partial\ln y_v} &=& - \Omega^{(1)}(w,u,v)\, ,\\
\label{eq:diff_phi6_2}
\frac{\partial\PhiTilde}{\partial\ln(y_u/y_w)} 
&=& - \Omega^{(1)}(v,w,u) + \Omega^{(1)}(u,v,w) = \ln(u/w) \ln v \,.
\end{eqnarray}
These derivatives lead, via \eqns{yv_int_rep}{yuyw_int_rep},
to the following integral representations:
\be
\PhiTilde = \sqrt{\Delta(u,v,w)} 
\int^u_1\frac{du_t\ \Omega^{(1)}(w_t,u_t,v_t)}{v_t[u(1-w)+(w-u)u_t]} \,,
\ee
with $(u_t,v_t,w_t)$ as in \eqn{eq:y_v_contour}, or
\be
\PhiTilde=\sqrt{\Delta(u,v,w)}\int^u_0
\frac{du_t \, \ln(u_t/w_t)\ln v_t}{(1-v_t)[uw+(1-u-w)u_t]} \,,
\ee
with $(u_t,v_t,w_t)$ as in \eqn{eq:y_u_y_w_contour}. We have set the
integration constants to zero because $\PhiTilde$ is a parity-odd
function.

We have now completed the construction of the hexagon basis through
weight three. Moving on to weight four, the symbol-level
classification reveals one new parity-even function, $\Omegauvw$, and
one new parity-odd function, $F_1(u,v,w)$, as well as their cyclic
images. We will discuss the parity-even function $\Omegauvw$ since it
exhibits a variety of features that the parity-odd functions lack.

As discussed in~\sect{sec:symbol}, $\Omegauvw$ is an extra-pure
function, and as such its symbol has only three distinct final
entries, which were given in~\eqn{Omfinals},
\be
\hbox{final entry} \in \left\{ \frac{u}{1-u}, \frac{v}{1-v}, y_u y_v\right\}\,.
\ee
Furthermore, the symbol is symmetric under the exchange of $u$ with
$v$. Taken together, these symmetry properties dictate the form of the
$\{3,1\}$ component of the coproduct,
\be
\Delta_{3,1}(\Omegauvw) = 
\Omega^{(2),u}\otimes\ln\Bigl(\frac{u}{1-u}\Bigr)
+\Omega^{(2),u}\Big|_{u\leftrightarrow v}\otimes\ln\Bigl(\frac{v}{1-v}\Bigr)
+\Omega^{(2),y_u} \otimes\ln y_uy_v \,.
\label{Om2_31}
\ee
There are two independent functions in~\eqn{Om2_31},
$\Omega^{(2),u}$ and $\Omega^{(2),y_u}$.  The symbols of these
functions can be read off from the symbol of $\Omegauvw$. Both
functions must be valid hexagon functions of weight three. The symbol
indicates that $\Omega^{(2),u}$ is parity-even and $\Omega^{(2),y_u}$
is parity-odd.

The most general linear combination of parity-even hexagon functions
of weight three whose symbol is consistent with that of
$\Omega^{(2),u}$ is
\be
\bsp
\Omega^{(2),u} &= H_3^{u} + H_{2,1}^v - H_{2,1}^w 
- \frac{1}{2} \ln(u w/v) \big(H_2^u + H_2^w\big) 
+\frac{1}{2}\,\ln(uv/w)\,H_2^v \\
&\quad + \frac{1}{2} \ln{u}\,\ln{v}\,\ln(v/w)
+ a_1\,\zeta_2\ln u + a_2 \,\zeta_2\ln v + a_3\,\zeta_2 \ln w + a_4\,\zeta_3\, ,
\esp
\label{Om2_u}
\ee
for four arbitrary rational numbers $a_i$. There is only a single parity-odd
hexagon function of weight three, so $\Omega^{(2),y_u}$ is uniquely determined from
its symbol,
\be
\Omega^{(2),y_u} = -\frac{1}{2}\, \PhiTilde\,.
\ee

It is not necessarily the case that the right hand side
of~\eqn{Om2_31} is actually the $\{3,1\}$ component of the coproduct
of a well-defined function for arbitrary values of the parameters
$a_i$. This integrability condition can be formalized by the
requirement that the operator
\be
(\textrm{id}\otimes d\wedge d)(\Delta_{2,1}\otimes\textrm{id})
\label{eq:int_copr}
\ee
annihilate the right hand side of~\eqn{Om2_31}. To see this, note
that $(\Delta_{2,1}\otimes \textrm{id}) \circ \Delta_{3,1} =
\Delta_{2,1,1}$, and therefore $d\wedge d$ acts on the last two
slots, which are just weight-one functions (logarithms). This can be
recognized as the familiar symbol-level integrability
condition,~\eqn{integrability}, promoted to function-level.

Another way of thinking about the integrability condition is that it
guarantees the consistency of mixed partial derivatives. Since there
are three variables, there are three pairs of derivatives to check. To
illustrate the procedure, we will examine one pair of derivatives by
verifying the equation,
\be
\sqrt{\Delta}\frac{\partial}{\partial\ln y_w}
\left[\frac{\partial\Omegauvw}{\partial\ln(y_v/y_u)}\right]
=  \sqrt{\Delta}\frac{\partial}{\partial\ln(y_v/y_u)}
\left[\frac{\partial\Omegauvw}{\partial\ln y_w}\right]\, .
\label{eq:mixed1}
\ee
We have multiplied by an overall factor of $\sqrt{\Delta}$ for
convenience. To simplify the notation, let us define,
\be
U \equiv\Omega^{(2),u} \quad~\textrm{and}~\quad V 
\equiv \Omega^{(2),u}|_{u\leftrightarrow v}\,.
\ee
Then, using \eqns{yu_diffeq}{yuyw_diffeq}, we can immediately write down
an expression for the left-hand side of~\eqn{eq:mixed1},
\be
\bsp
\sqrt{\Delta}\frac{\partial}{\partial\ln y_w}
\left[\frac{\partial\Omegauvw}{\partial\ln(y_v/y_u)}\right]
&=\sqrt{\Delta}\frac{\partial}{\partial\ln y_w}
\left[-\frac{1-w}{\sqrt{\Delta}}(U - V)\right]\\
&=\ (1-w)^2(1-u-v)(V^{w}-U^{w})\\
&+w(1-w)(U^{u}+U^{v}+U^{1-w}-V^{u}-V^{v}-V^{1-w})\\
&-uw(1-w)(U^{u}+U^{1-u}+U^{1-w}-V^{u}-V^{1-u}-V^{1-w})\\
&-vw(1-w)(U^{v}+U^{1-v}+U^{1-w}-V^{v}-V^{1-v}-V^{1-w}) \,. \\
\esp
\ee
The algebra leading to the second line may be simplified by using 
the fact that $(1-w)/\sqrt{\Delta}$ is independent of $y_w$. Similarly, 
it is straightforward to write down an expression for the right-hand side
of~\eqn{eq:mixed1},
\be
\bsp
\sqrt{\Delta}\frac{\partial}{\partial\ln(y_v/y_u)}
\left[\frac{\partial\Omegauvw}{\partial\ln y_w}\right]
&=\sqrt{\Delta}\frac{\partial}{\partial\ln(y_v/y_u)}
\left[-\frac{w}{\sqrt{\Delta}}(U + V)\right]\\
&=\ -w(1-w)(U^{v}-U^{u}+V^{v}-V^{u})\\
&-uw(1-w)(U^{w}+U^{u}+U^{1-u}+V^{w}+V^{u}+V^{1-u})\\
&+vw(1-w)(U^{w}+U^{v}+U^{1-v}+V^{w}+V^{v}+V^{1-v})\\
&+w^2(u-v)(U^{1-w}+V^{1-w}) \,,
\label{eq:mixedpartials}
\esp
\ee
where we have used the fact that $w/\sqrt{\Delta}$ is annihilated by
$\partial/\partial\ln(y_v/y_u)$.

As usual, the superscripts indicate the various coproduct
components. A special feature of this example is that the functions
$U$ and $V$ are built entirely from single-variable HPLs, so it is
straightforward to extract these coproduct components using the
definitions in appendix~\ref{sec:app_multi_poly}. More generally, the
functions may contain non-HPL elements of the hexagon basis. For these
cases, the coproduct components are already known from previous steps
in the iterative construction of the basis.

The nonzero coproduct components of $U$ are,
\be
\begin{aligned}
U^u &= -\frac{1}{2}\,\Big(H_2^u - H_2^v + H_2^w
- \ln v\,\ln(v/w)\Big) + a_1\,\zeta_2\,,\\
U^v &= \frac{1}{2}\,\Big(H_2^u + H_2^v + H_2^w +2 \ln u\,\ln v
- \ln u\,\ln w\Big) + a_2\,\zeta_2\,,\\
U^w &=-\frac{1}{2}\,\Big(H_2^u + H_2^v + H_2^w + \ln u\,\ln v\Big)
+ a_3\,\zeta_2\, ,
\end{aligned}
\qquad
\begin{aligned}
U^{1-u} &= H_2^u + \frac{1}{2}\, \ln u\,\ln(uw/v)\,, \\
U^{1-v} &= -\frac{1}{2}\,\ln v\ln(u/w)\, ,\\
U^{1-w} &= \frac{1}{2}\ln w\ln(u/v)\, ,
\end{aligned}
\label{eq:coprU}
\ee
while those of $V$ are related by symmetry,
\be
\begin{aligned}
&V^u = U^v\, , \\
&V^{1-u} = U^{1-v}\,,
\end{aligned}
\quad
\begin{aligned}
&V^{v} = U^{u}\, ,\\
&V^{1-v} = U^{1-u}\, ,\\
\end{aligned}
\quad
\begin{aligned}
&V^{w} = U^w|_{u\leftrightarrow v}\, ,\\
&V^{1-w} = U^{1-w}|_{u\leftrightarrow v}\,.
\end{aligned}
\label{eq:coprV}
\ee
Using eqs.~(\ref{eq:mixedpartials}), (\ref{eq:coprU}), and (\ref{eq:coprV}),
it is straightforward to check that the equality of mixed-partial
derivatives, \eqn{eq:mixed1}, is satisfied if and only if $a_2=-a_3$.

Continuing in this way, we can derive similar constraints from the 
remaining two mixed partial derivative consistency conditions. The result 
is that
\be
a_2 = -1\,, \quad\textrm{and}\quad a_3 = 1\,.
\ee
Finally, we must impose good branch-cut behavior. As discussed
in~\sect{sec:construct}, this constraint can be implemented by
imposing~\eqn{Fomu_vanish}, or, in this case,
\be
U(1,0,0) = 0\,,
\ee
which implies that $a_4=0$.

The one remaining parameter, $a_1$, corresponds to an ambiguity that
cannot be fixed by considering mathematical consistency
conditions. Indeed, it arises from a well-defined weight-four function
with all the appropriate symmetries and mathematical properties. In
particular, it is the product of $\zeta_2$ with an extra-pure
weight-two hexagon function that is symmetric under $u \leftrightarrow
v$,
\be
-\zeta_2 \Bigl[\textrm{Li}_2(1-1/u) + \textrm{Li}_2(1-1/v)\Bigr]\,.
\ee
In general, we would resolve such an ambiguity by making an arbitrary
(though perhaps convenient) choice in order to define the new hexagon
function. But because $\Omegauvw$ corresponds to a particular Feynman
integral, the value of $a_1$ is not arbitrary, and the only way to fix
it is to bring in specific data about that integral. We are not
interested in determining the value of $a_1$ directly from the Feynman
integral since this integral has been evaluated
previously~\cite{Dixon2011nj}. Instead, we will be satisfied simply to
verify that a consistent value of $a_1$ exists.

From~\eqn{eq:mixedpartials} we have,
\bea
\sqrt{\Delta} \, \frac{\partial\Omegauvw}{\partial\ln y_w}
&=& - w \left(U + V \right)\,, 
\label{Om2_yw}\\
\sqrt{\Delta} \, \frac{\partial\Omegauvw}{\partial\ln(y_v/y_u)}
&=& - (1-w) \left(U - V \right)\,.
\label{Om2_yvyu}
\eea
\Eqn{Om2_yw} is consistent with the differential equations of section 4 of
ref.~\cite{Dixon2011nj} only if the function $Q_\phi$ from that reference
(and \eqn{qphiqr}) is related to $U$ and $V$ by,
\be
Q_\phi = - ( U + V ) \,.
\label{QasminusUplusV}
\ee
This equation is satisfied, provided that $a_1=1$. Having fixed all $a_i$, we have
uniquely determined the $\{3,1\}$ component of the coproduct of $\Omegauvw$.
Indeed,~\eqn{Om2_31} is consistent with the expressions in~\eqns{S(O2)}{qphiqr},
as of course it must be.

We remark that the antisymmetric combination
appearing in \eqn{Om2_yvyu} is related to another
function defined in ref.~\cite{Dixon2011nj},
\be
\tilde{Z}(v,w,u) = - 2 (U-V) \,,
\label{Om2_uv_neg}
\ee
where $\tilde{Z}$ appears in a derivative of the odd part of
the NMHV ratio function (see \eqn{Ztilde}).

Following the discussion in~\sect{sec:general_setup}, the differential
equation~(\ref{Om2_yw}) gives rise to the integral representation,
\be
\Omegauvw = - 6\zeta_4 
+ \int^u_1 du_t\, \frac{Q_{\phi}(u_t,v_t,w_t)}{u_t(u_t-1)} \,,
\label{Om2firstint}
\ee
where,
\be
\label{eq:y_w_contour}
v_t = \frac{(1-u)vu_t}{u(1-v)+(v-u)u_t} \,, \qquad\qquad
w_t = 1-\frac{(1-w)u_t(1-u_t)}{u(1-v)+(v-u)u_t} \,.
\ee
While our conventions for generic hexagon functions require the
functions to vanish at the boundary value $(1,1,1)$, in this specific
case we must specify a nonzero value $\Omega^{(2)}(1,1,1)=-6\zeta_4$
in order to match a prior definition of the function.

The differential equation~(\ref{Om2_yvyu}) gives rise to another integral
representation for $\Omega^{(2)}$,
\be
\Omegauvw = \frac{1}{2} \int^v_0
\frac{dv_t \, \tilde{Z}(v_t,w_t,u_t)}{v_t(1-v_t)}\,,
\label{Om2secondint}
\ee
where,
\be
u_t = \frac{u v(1-v_t)}{u v+(1-u-v)v_t} \,, \qquad\qquad
w_t = \frac{w v_t(1-v_t)}{u v+(1-u-v)v_t} \,.
\ee
There is no constant of integration in \eqn{Om2secondint} because in this case
$\Omega^{(2)}$ vanishes at the lower endpoint,
$\Omega^{(2)}(1,0,0)=0$~\cite{Drummond2010cz,Dixon2011nj}.

Continuing onward, we construct the remaining functions of the hexagon basis in
an iterative fashion, using the above methods. We collect the results through
weight five in appendix~\ref{sec:app_basis}. We present the data by the $\{n-1,1\}$
component of the coproduct, plus the constraint that the functions vanish
at $(u,v,w)=(1,1,1)$ (except for the special case of $\Omega^{(2)}$). With this 
information, we can build an ansatz for the three-loop remainder function, as we
discuss in the next subsection.

%%%%%%%%%%%%%%%%%%%%%%%%%%%%%%%%%%%%%%%%%%%%%%%%%%%%%%%%%%%%%%
\subsection{Constructing the three-loop remainder function}
\label{sec:constructthreeloop}

In this subsection, we complete the construction of an ansatz
for the three-loop remainder function.  We use the 
decomposition~(\ref{R63decomp}) of the symbol of $R_6^{(3)}$
as a template, and extend it to a definition of the function
using the same steps as in~\sect{sec:construct_int}:
\begin{enumerate}
\item From the symbol of the extra-pure function $\Rep(u,v,w)$,
which depends on $\alpha_1$ and $\alpha_2$, we expand the $\{5,1\}$
components of its coproduct in terms of our weight-five basis functions.
These functions can be given as multiple polylogarithms, as in~\sect{sec:construct},
or as integral representations, as in~\sect{sec:construct_int}. 
We also allow for the addition of zeta values multiplying lower-weight basis
functions.
\item  We fix as many undetermined parameters in this ansatz as possible by
enforcing various mathematical consistency conditions. In particular,
\begin{enumerate}
\item We impose extra-purity and symmetry in the exchange of $u$ and $v$ as
function-level conditions on the coproduct entries, since these conditions are
satisfied at symbol level:
\be
\bsp
&\Rep^{v} = -\Rep^{1-v} =  -\Rep^{1-u}(u\leftrightarrow v) 
=  \Rep^{u}(u\leftrightarrow v)\,,\\
&\qquad \Rep^{y_v} 
= \Rep^{y_u}\,,\quad \Rep^{w}=\Rep^{1-w}=\Rep^{y_w} = 0\,.
\label{Repep}
\esp
\ee
In principle, beyond-the-symbol terms do not need to obey the extra-purity relations.
At the end of \sect{sec:collinear}, we will relax this assumption and
use the near-collinear limits to show that the potential additional terms
vanish.
\item We demand that the ansatz be integrable. For the multiple polylogarithm
approach,
this amounts to verifying that there is a weight-six function with our ansatz as the
$\{5,1\}$ component of its coproduct. For the approach based on integral
representations, we check that there are consistent mixed partial derivatives.
\item We require that the resulting function have the proper branch-cut structure.
We impose this constraint by verifying that there are no spurious poles in the
first derivatives, just as we did in the construction of the hexagon basis.
\end{enumerate}
After imposing these constraints, there are still nine undetermined beyond-the-symbol
parameters. They correspond to well-defined extra-pure hexagon functions of weight
six, and cannot be fixed by mathematical consistency conditions.
\item We integrate the resulting coproduct. This result is a weight-six function,
$\Rep^{(\alpha_1,\alpha_2)}$, which depends on the symbol-level constants, $\alpha_1$ and
$\alpha_2$,  and nine lower-weight functions $r_1,\ldots,r_9$, which come multiplied
by zeta values. The $r_i$ may be expressed in terms of previously-determined hexagon
functions, while $\Rep^{(\alpha_1,\alpha_2)}$ may be given as an integral representation
or explicitly in terms of multiple polylogarithms.
\end{enumerate}
This procedure leaves us with the following ansatz for $R_6^{(3)}$:
\be
R_6^{(3)}(u,v,w)
= \Bigl[ \Bigl( \Rep^{(\alpha_1,\alpha_2)}(u,v,w) + \sum_{i=1}^9 c_i \,r_i(u,v) \Bigr)
  + {\rm cyclic} \Bigr]
+ P_6(u,v,w) + c_{10} \,\zeta_6 + c_{11} \,(\zeta_3)^2 \,,
\label{R63fndecomp}
\ee
where the $c_i$ are undetermined rational numbers, and
\be
\bsp
r_1 &= \zeta_4 \, \Bigl[ H_2^u + \frac{1}{2} \ln^2 u 
  + (u\leftrightarrow v)\Bigr] \,, \\
r_2 &= \zeta_3 \, \Bigl[ H_{2,1}^u - \frac{1}{6} \ln^3 u 
  + (u\leftrightarrow v)\Bigr] \,, \\
r_3 &= \zeta_3 \, \Bigl[ H_{3}^u - 2 H_{2,1}^u + H_1^u H_2^u 
  + (u\leftrightarrow v)\Bigr] \,, \\
r_4 &= \zeta_2 \, \Bigl[ H_{4}^u + H_1^u H_3^u - \frac{1}{2} (H_2^u)^2
  + (u\leftrightarrow v)\Bigr] \,, \\
r_5 &= \zeta_2 \, \Bigl[ H_{4}^u - 3 H_{2,1,1}^u +  H_1^u H_3^u 
 + \frac{1}{2} (H_1^u)^2 H_2^u
  + (u\leftrightarrow v)\Bigr] \,, \\
r_6 &= \zeta_2 \, \Bigl[ H_{3,1}^u - 3 H_{2,1,1}^u +  H_1^u H_{2,1}^u
  + (u\leftrightarrow v)\Bigr] \,, \\
r_7 &= \zeta_2 \, \Bigl[ H_{2,1,1}^u + \frac{1}{24} (H_1^u)^4
  + (u\leftrightarrow v)\Bigr] \,, \\
r_8 &= \zeta_2 \, \Bigl( H_2^u + \frac{1}{2} \ln^2 u \Bigr)
                 \Bigl( H_2^v + \frac{1}{2} \ln^2 v \Bigr)\,, \\ 
r_9 &= \zeta_2 \, \Omegauvw \,.
\esp
\label{r_ambig}
\ee
In the following section we will use the collinear limits of
this expression to fix $\alpha_1$, $\alpha_2$ and the $c_i$.
After fixing these parameters, we can absorb all but the constant
terms into a redefinition of $\Rep$.  The $\{5,1\}$ component of its
coproduct is given in appendix~\ref{sec:app_Rep}. The final integral
representation for $R_6^{(3)}$, having fixed also $c_{10}$ and $c_{11}$,
is given in \sect{sec:Final}, \eqn{R63Final}. The final expression in
terms of multiple polylogarithms is quite lengthy, but it is provided
in a computer-readable format in the attached files.

\vfill\eject

%%%%%%%%%%%%%%%%%%%%%%%%%%%%%%%%%%%%%%%%%%%%%%%%%%%%%%%%%%%%%%%%%%%%%%

\section{Collinear limits}
\label{sec:collinear}

In the previous section, we constructed a 13-parameter ansatz for the three-loop 
remainder function. It has the correct symbol, proper branch structure, and 
total $S_3$ symmetry in the cross ratios. In other words, the ansatz obeys
all relevant mathematical consistency conditions. So in order to fix the 
undetermined constants, we need to bring in some specific physical data.

Some of the most useful data available comes from the study of the
collinear limit. In the strict collinear limit in which two gluons are
exactly collinear, the remainder function must vanish to all loop
orders. This condition fixes many, but not all, of the parameters in
our ansatz. To constrain the remaining constants, we expand in the
near-collinear limit, keeping track of the power-suppressed
terms. These terms are predicted by the OPE for flux tube excitations.
In fact, the information about the leading discontinuity terms in the
OPE~\cite{Alday2010ku,Gaiotto2010fk,Gaiotto2011dt} was already
incorporated at symbol level and used to constrain the symbol for the
three-loop remainder function up to two undetermined
parameters~\cite{Dixon2011pw}.

Here we take the same limit
at function level, and compare to the recent work of Basso, Sever and
Vieira (BSV)~\cite{Basso2013vsa}, which allows us to uniquely constrain all of
the beyond-the-symbol ambiguities, as well as the two symbol-level parameters.
The two symbol-level parameters were previously fixed by using dual
supersymmetry~\cite{CaronHuot2011kk}, and also by studying the near-collinear
limit at symbol level~\cite{Basso2013vsa}, and we agree with both of these
determinations.  

%%%%%%%%%%%%%%%%%%%%%%%%%%%%%%%%%%%%%%%%%%%%%%%%%%%%%%%%
\subsection{Expanding in the near-collinear limit}

In the (Euclidean) limit that two gluons become collinear, 
one of the cross ratios goes to zero and the sum of the other two
cross ratios goes to one.  For example, if we let $k_2$ and $k_3$ become
parallel, then $x_{24}^2 \equiv (k_2+k_3)^2 \to 0$, corresponding to $v\to0$, and 
$u+w\to1$.  BSV~\cite{Basso2013vsa} provide a convenient set of variables
$(\tau,\sigma,\phi)$ with which one can approach this collinear limit.
They are related to the $(u_i,y_i)$ variables by~\cite{BSVPrivate}:
\be
\bsp
u &= \frac{FS^2}{(1+T^2)(F+FS^2+ST+F^2ST+FT^2)} \,,\\
v &= \frac{T^2}{1+T^2} \,,\\
w &= \frac{F}{F+FS^2+ST+F^2ST+FT^2} \,,\\
y_u &= \frac{FS+T}{F(S+FT)} \,,\\
y_v &= \frac{(S+FT)(1+FST+T^2)}{(FS+T)(F+ST+FT^2)} \,,\\
y_w &= \frac{F+ST+FT^2}{F(1+FST+T^2)} \,,
\esp
\label{BSVparam}
\ee
where $T = e^{-\tau}$, $S = e^{\sigma}$, and $F = e^{i\phi}$.

As $T\to0$ ($\tau\to\infty$) we approach the collinear limit.
The parameter $S$ controls the partitioning of the momentum between
the two collinear gluons, according to $k_2/k_3 \sim S^2$, or 
$k_2/(k_2+k_3) \sim S^2/(1+S^2)$.  The parameter $F$ controls the
azimuthal dependence as the two gluons are rotated around their
common axis with respect to the rest of the scattering process.
This dependence is related to the angular momentum of flux-tube
excitations in the OPE interpretation.

By expanding an expression in $T$ we can probe its behavior in the
near-collinear limit, order by order in $T$.  Each order in $T$ also contains
a polynomial in $\ln T$.  In general, the expansions of parity-even and
parity-odd hexagon functions $f^{\rm even}$ and $f^{\rm odd}$ have the form,
\bea
f^{\rm even}(T,F,S) &=& \sum_{m=0}^\infty \sum_{n=0}^N \sum_{p=0}^m 
T^m \, (-\ln T)^n \, \cos^p\phi \ f^{\rm even}_{m,n,p}(S) \,,
\label{EvenTexpansion}\\
f^{\rm odd}(T,F,S) &=& 2i\sin\phi \sum_{m=1}^\infty \sum_{n=0}^N \sum_{p=0}^{m-1} 
T^m \, (-\ln T)^n \, \cos^p\phi \ f^{\rm odd}_{m,n,p}(S) \,.
\label{OddTexpansion}
\eea
Odd parity necessitates an extra overall factor of $\sin\phi$.
The maximum degree of the polynomial in $e^{\pm i\phi}$ is $m$,
the number of powers in the $T$ expansion, which is related to the
twist of a flux tube excitation in the final answer.
The maximum degree $N$ of the polynomial in $\tau \equiv -\ln T$
satisfies $N=w-2$ for the non-HPL hexagon functions with weight $w$ in
appendix~\ref{sec:app_Rep}, although in principle it could be as large
as $N=w$ for the $m=0$ term (but only from the function $\ln^w v$),
and as large as $N=w-1$ when $m>0$.  For the final remainder function
at $L$ loops, with weight $w=2L$, the leading discontinuity terms in the OPE
imply a relatively small value of $N$ compared to the maximum possible,
namely $N = L-1 = 2L - (L+1)$ for $R_6^{(L)}$, or $N=2$ for $R_6^{(3)}$.

BSV predict the full order $T^1$ behavior of the remainder
function~\cite{Basso2013vsa}.  The part of the $T^2$ behavior that
is simplest for them to predict (because it is purely gluonic)
contains azimuthal variation proportional to $\cos^2\phi$, {\it i.e.} the
$T^2F^2$ or $T^2F^{-2}$ terms; however, they can also extract the
$T^2F^0$ behavior, which depends upon the scalar and fermionic
excitations as well~\cite{BSVPrivate}.  To compare with this data, 
we must expand our expression for $R_6^{(3)}$ to this order.

The expansion of an expression is relatively straightforward when its
full analytic form is known, for example when the expression is given
in terms of multiple polylogarithms.  In this case, one merely needs
to know how to take a derivative with respect to $T$ and how to
evaluate the functions at $T=0$. The derivative of a generic multiple
polylogarithm can be read off from its coproduct, which is given in
appendix~\ref{sec:app_multi_poly}. Evaluating the functions at $T=0$
is more involved because it requires taking $y_u\to 1$ and $y_w\to 1$
simultaneously. However, the limit of all relevant multiple
polylogarithms can be built up iteratively using the coproduct
bootstrap of~\sect{sec:coproduct}.

If the expression is instead represented in integral form, or is defined
through differential equations, then it becomes necessary to integrate up
the differential equations, iteratively in the transcendental weight,
and order by order in the $T$ expansion.  Recall that for any function
in our basis we have a complete set of differential equations whose
inhomogeneous terms are lower weight hexagon functions. 
The change of variables~(\ref{BSVparam}) and its Jacobian
allow us to go from differential equations
in the $u_i$ or $y$ variables to differential equations in $(F,S,T)$.

The structure of the $T\to0$ expansion makes most
terms very straightforward to integrate.  In 
\eqns{EvenTexpansion}{OddTexpansion}, $T$ only appears as powers of $T$,
whose coefficients are polynomials of fixed order in $\ln T$.  The 
variable $F$ only appears as a polynomial in $\cos\phi$ and $\sin\phi$, 
{\it i.e.}~as powers of $F$ and $F^{-1}$.  Hence any $T$ or $F$ derivative 
can be integrated easily, up to a constant of integration, which can depend
on $S$.  The $S$ derivatives require a bit of extra work.  However,
the differential equation in $S$ is only required for the $T$- and
$F$-independent term arising in the parity-even case, $f^{\rm even}_{0,0,0}(S)$.
This coefficient is always a pure function of the same transcendental
weight as $f$ itself, and it can be constructed from a complete set of HPLs in 
the argument $-S^2$.  Thus we can integrate the one required differential
equation in $S$ by using a simple ansatz built out of HPLs.

There is still one overall constant of integration to determine
for each parity-even function, a term that is completely independent of
$T$, $F$ and $S$.  It is a linear combination of zeta values.  
(The parity-odd functions all vanish as $T\to0$, so they do not
have this problem.)  The constant of integration can be determined
at the endpoint $S=0$ or $S=\infty$, with the aid of a second limiting line,
$(u,v,w)=(u,u,1)$.  On this line, all the hexagon functions are very simple,
collapsing to HPLs with argument $(1-u)$.
In the limit $u\to0$ this line approaches the point $(0,0,1)$, which can
be identified with the $S\to0$ ``soft-collinear''
corner of the $T\to0$ collinear limit
in the parametrization~(\ref{BSVparam}).  Similarly, the $S\to\infty$
corner of the $T\to0$ limit intersects the line $(1,v,v)$ at $v=0$.
Both lines $(u,u,1)$ and $(1,v,v)$ pass through the point $(1,1,1)$.  At
this point, (most of) the hexagon functions are defined to vanish, which
fixes the integration constants on the $(u,u,1)$ and $(1,v,v)$ lines.
HPL identities then give the desired values of the functions in the
soft-collinear corner, which is enough to fix the integration constant 
for the near-collinear limit.  We will illustrate this method with an
example below.

The coefficients of the power-suppressed
terms that also depend on $T$ and $F$, namely $f_{m,n,p}(S)$ 
in \eqns{EvenTexpansion}{OddTexpansion} for $m>0$,
are functions of $S$ that involve HPLs with the same argument $-S^2$,
but they also can include prefactors to the HPLs that are rational
functions of $S$. The $f_{m,n,p}(S)$ for $m>0$ generally have a mixed
transcendental weight.  Mixed transcendentality
is common when series expanding generic HPLs around particular points.
For example, expanding $\textrm{Li}_2(1-x)$ around $x=0$ gives
\be
\textrm{Li}_2(1-x)\ \sim\
\frac{\pi ^2}{6} + x (\ln x-1)
+ x^2 \Big(\frac{\ln x}{2}-\frac{1}{4}\Big)
+ x^3 \Big(\frac{\ln x}{3}-\frac{1}{9}\Big) + {\cal O}(x^4)\,.
\ee

Using an HPL ansatz for the pure $S$-dependent terms, we use the 
differential equations to fix any unfixed parameters and cross-check
the ansatz.  Repeating this process order by order we build up the
near-collinear limiting behavior of each element of the basis of 
hexagon functions as a series expansion.

\subsection{Examples}

In order to illustrate the collinear expansion, it is worthwhile to
present a few low-weight examples. We begin with the simplest nontrivial
example, the weight-three parity-odd function $\PhiTilde$.
Since $\PhiTilde$ is fully symmetric in the $u_i$ and vanishes in the
collinear limit (like any parity-odd function), its expansion is 
particularly simple.  To conserve space in later formulas, we adopt 
the notation,
\be
\label{eq:coll_vars}
\yn = S^2 \,, \qquad L = \ln S^2 \,, \qquad H_{\vec{w}} = H_{\vec{w}}(-S^2) \,.
\ee
The expansion of $\PhiTilde$ is then
\be
\bsp
\PhiTilde &= \frac{2 i T \sin \phi }{S} \biggl[
  2 \ln T  \Bigl( (1+\yn) H_1 + \yn  L \Bigr)
    - (1+\yn) \Bigl( H_1^2 + (L+2) H_1  \Bigr) - 2 \yn L \biggr]\\
&\quad + \frac{2 i T^2 \cos \phi  \sin \phi }{S^2} \biggl[
 - 2 \ln T  \Bigl( (1+\yn^2) H_1 + \yn  (\yn L + 1) \Bigr)
  + (1+\yn^2) ( H_1^2 + L H_1 )\\
&\quad\hskip3.5cm + (1+\yn)^2 H_1 + \yn \Bigl( (1+\yn) L + 1 \Bigr) \biggr]
+ {\cal O}(T^3) \,.
\esp
\label{PhiTildeCollimit}
\ee
The sign of \eqn{PhiTildeCollimit}, and of the collinear expansions
of all of the parity-odd functions, 
depend on the values of the $y$ variables used.  This sign is
appropriate to approaching the collinear limit from Region I, with $0 < y_i < 1$.

Because $\Omega^{(2)}$ lacks the symmetries of $\PhiTilde$, its expansion
must be evaluated in multiple channels, and it is substantially lengthier.
Through order $T^2$, we find,
\be
\bsp
\Omegauvw &=
\ln^2 T \Bigl( 2 ( H_2 + \zeta_2 ) + L^2 \Bigr)
+ 2 \ln T \Bigl( H_3 - 2 ( H_{2,1} - \zeta_3 ) - L H_2 \Bigr)
+ H_4 - 4 H_{3,1}\\
&\quad + 4 H_{2,1,1}
+ \frac{1}{2} \Bigl( H_2^2 + L^2 ( H_2 + \zeta_2 ) \Bigr)
- L \Bigl( H_3 - 2 (H_{2,1} + \zeta_3) \Bigr)
+ 2 \zeta_2 H_2 + \frac{5}{2} \zeta_4\\
&\quad 
+ \frac{T\cos\phi}{S} \biggl[ 
  - 4 \ln^2 T\ \yn (H_1+L)
  + 4 \ln T\ \Bigl( (1+\yn) H_1 + \yn ( H_1^2 + L ( H_1 + 1 ) ) \Bigr)\\
&\quad\hskip1.5cm
  + \yn \Bigl( 4 (H_{2,1} - \zeta_3) - \frac{4}{3} H_1^3 - H_1 (2 H_2+L^2)
      - 2 L (H_1^2+2) - 2 \zeta_2 (2 H_1 + L) \Bigr)\\
&\quad\hskip1.5cm
  - 2 (1+\yn) ( H_1^2 + H_1 (L+2) ) \biggr]\\
&
+ \frac{T^2}{S^2} \Biggl\{ \cos^2\phi \biggl[
  4 \ln^2 T\ \yn^2 \biggl( H_1 + L + \frac{1}{1+\yn} \biggr)\\
&\quad\hskip1.5cm
 - 2 \ln T \biggl( 2 \yn^2 H_1 (H_1+L) + H_1 + \yn
               + \frac{\yn^2}{1+\yn} \Bigl( (5+\yn)   H_1 + (3+\yn)   L \Bigr) 
 \biggr)\\
&\quad\hskip1.5cm
 + \yn^2 \Bigl( - 4 (H_{2,1} - \zeta_3) + H_1 (2 H_2 + L^2 + 4 \zeta_2 )
         + \frac{4}{3} H_1^3 + 2 L  ( H_1^2 + \zeta_2 ) \Bigr)\\
&\quad\hskip1.5cm
 + H_1 (H_1+L) + (1 + 3 \yn) \Bigl( (1+\yn) H_1 + \yn L \Bigr) + \yn\\
&\quad\hskip1.5cm
 + \frac{\yn^2}{1+\yn} \Bigl( (5+\yn) H_1 (H_1+L) - \yn (2 H_2 + L^2) \Bigr)
 + 2 \zeta_2 \frac{\yn^2(1-\yn)}{1+\yn} \biggr]\\
&\quad\hskip0.8cm
 - 2 \ln^2 T\  \yn ( (2+\yn) (H_1+L) + 1 )\\
&\quad\hskip0.8cm
 + \ln T \Bigl( \yn^2   ( 2 H_1 (H_1+L) + 3 (H_1+L) )
         + \yn ( 4 H_1 (H_1+L+1) + 2 L  + 3 ) + H_1 \Bigr)\\
&\quad\hskip0.8cm
 + \yn (2+\yn) \Bigl( 2 H_{2,1} - H_1 H_2 - L H_1^2 - \frac{2}{3} H_1^3
             - \frac{1}{2} L^2 H_1 - \zeta_2 (2 H_1+L) - 2 \zeta_3 \Bigr)\\
&\quad\hskip0.8cm
 - \yn \Bigl( H_2 + \frac{1}{2} L^2 + 2 \zeta_2 + \frac{3}{2} \Bigr)
 - \frac{1}{2} \Bigl( (1+\yn)(1+3\yn) H_1 (H_1+L)
                    + (1+5\yn) H_1 \\
&\quad\hskip1.5cm
                    + \yn (3+7 \yn) (H_1+L) \Bigr) \Biggr\} 
+ {\cal O}(T^3) \,.
\label{Om2uvw}
\esp
\ee

The integral $\Omegauvw$ is symmetric under the exchange of $u$ and $v$.
This implies that the limiting behavior of $\Omegavwu$ can be determined
from that of $\Omegauvw$ by exchanging the roles of $u$ and $w$ in the
collinear limit.  At leading order in $T$, this symmetry corresponds
to letting $S \leftrightarrow 1/S$.  This symmetry is broken by the
parametrization~(\ref{BSVparam}) at order $T^2$; nevertheless, the correction
at order $T^2$ is relatively simple,
\be
\bsp
\Omegavwu&=\Omegauvw\Big|_{S\to1/S}
+ 4 T^2\biggl[
 \ln^2 T \, H_1 - \ln T \, H_1 (H_1+L) - H_3 + H_{2,1}\\
&\quad\hskip3.5cm
 - \frac{1}{2} \Bigl( H_1 (H_2-\zeta_2) - L (H_2+H_1^2) \Bigr) 
 + \frac{1}{3} H_1^3 \biggr]
+ {\cal O}(T^3) \,.
\label{Om2vwu}
\esp
\ee

The last independent permutation is $\Omegawuv$. It is symmetric under
$u\leftrightarrow w$ and vanishes at order $T^0$, which together imply
that its near-collinear expansion is symmetric under $S\leftrightarrow
1/S$ through order $T^2$, although that symmetry is not manifest in
the HPL representation,
\be
\bsp
\Omegawuv &= \frac{T \cos \phi}{S}
 (1+\yn) \Bigl( 2 L H_2 - H_1 ( L^2 + 2 \zeta_2 ) \Bigr)\\
&\quad\hskip0cm
+ \frac{T^2}{S^2} \biggl\{ \cos^2\phi \biggl[ 
    (1+\yn^2) \Bigl( - 2 L H_2 + H_1 (L^2 + 2 \zeta_2) \Bigr)
                + 2 (1-\yn^2) H_2\\
&\quad\hskip2.5cm
     + \yn (1-\yn) (L^2 + 2 \zeta_2)
     - 2 (1+\yn) ( (1+\yn) H_1 + \yn L ) \biggr]\\
&\quad\hskip1cm
 - 2 \ln T \, (1+\yn) ( (1+\yn) H_1 + \yn L )\\
&\quad\hskip1cm
 + (1+\yn)^2 \Bigl[ L H_2 - H_1 \Bigl( \frac{1}{2} L^2 + \zeta_2 - L - 3 \Bigr)
    + H_1^2 \Bigr]
 + 3 \yn (1+\yn) L \biggr\} + {\cal O}(T^3) \,.
\label{Om2wuv}
\esp
\ee
We determine these expansions by integrating the differential
equations in $F$, $S$, and $T$, as described in the previous
subsection. For parity-even functions, it is necessary to fix the
constants of integration. Here we present one technique for doing
so. Suppose we set $S=T$ in \eqn{BSVparam}.  Then the limit $T\to0$
corresponds to the EMRK limit, $u=v \to 0$, $w\to1$, approached along
the line $(u,u,1)$.  As an example, let us consider applying this
limit to the expansion of $\Omegauvw$, \eqn{Om2uvw}. We only need to
keep the $T^0$ terms, and among them we find that the $H_{\vec{w}}$
terms vanish, $L \to \ln u$, and $\ln T \to\frac{1}{2} \ln u$ (since
$u\sim T^2$).  Therefore, as $u\to0$ we obtain,
\be
\Omega^{(2)}(u,u,1) = \frac{1}{4} \ln^4 u + \zeta_2 \ln^2 u 
+ 4 \zeta_3 \ln u + \frac{5}{2} \zeta_4 + {\cal O}(u) \,.
\label{Om2uu1_smallu}
\ee
The constant of integration, $\tfrac{5}{2} \zeta_4$, clearly survives
in this limit. So, assuming we did not know its value, it could be
fixed if we had an independent way of examining this limit.

This independent method comes from the line $(u,u,1)$, on which all the hexagon
functions have simple representations. This can be seen from the form of the
integration contour parametrized by $v_t$ and $w_t$ in \eqn{eq:y_v_contour}.
Setting $v=u$ and $w=1$, it collapses to
\be
\label{eq:y_v_contour_collapse}
v_t = u_t \,, \qquad\qquad
w_t = 1 \,.
\ee
The integral~(\ref{Om2firstint}) then becomes
\be
\Omega^{(2)}(u,u,1) = - 6\zeta_4 
- \int^u_1\frac{du_t\, \omega^u(u_t,u_t,1)}
               {u_t(u_t-1)} \,,
\label{Om2firstint_uu1}
\ee
where
\be
\omega^u(u,u,1) = [\Omega^{(2),u} + (u\leftrightarrow v)](u,u,1)
= 2 \Bigl[ H_3^u + H_{2,1}^u + \ln u \, H_2^u + \tfrac{1}{2} \ln^3 u \Bigr] \,.
\label{o_u}
\ee
Such integrals can be computed directly using the definition~(\ref{Hdef})
after a partial fraction decomposition of the factor $1/[u_t(u_t-1)]$.
Expressing the result in terms of the Lyndon basis~(\ref{HPLbasis6}) gives,
\be
\Omega^{(2)}(u,u,1) = - 2 H_4^u - 2 H_{3,1}^u + 6 H_{2,1,1}^u
+ 2 ( H_2^u )^2 + 2 \ln u ( H_3^u + H_{2,1}^u )
+ \ln^2 u H_2^u + \frac{1}{4} \ln^4 u
- 6 \, \zeta_4 \,.
\label{Om2_uu1}
\ee
At the point $u=1$, all the $H_{\vec{w}}^u = H_{\vec{w}}(1-u)$ vanish,
as does $\ln u = -H_1^u$, so we see that \eqn{Om2_uu1} becomes
\be
\Omega^{(2)}(1,1,1) = - 6 \, \zeta_4 \,,
\label{Om2_111}
\ee
in agreement with the explicit $- 6\zeta_4$ in \eqn{Om2firstint_uu1}.
In order to take the limit $u\to0$, we use HPL identities to reexpress
the function in terms of HPLs with argument $u$ instead of $(1-u)$:
\be
\bsp
\Omega^{(2)}(u,u,1) &= \frac{1}{4} \ln^4 u + H_1(u) \ln^3 u
+ \Bigl( - 2 H_2(u) + \frac{1}{2} (H_1(u))^2 + \zeta_2 \Bigr) \ln^2 u\\
&\quad
+ \Bigl( 4 ( H_3(u) - H_{2,1}(u) + \zeta_3 )
     - \frac{1}{3} (H_1(u))^3 + 2 \zeta_2 H_1(u) \Bigr) \ln u\\
&\quad
- 6 H_4(u) + 2 H_{3,1}(u) + 2 H_{2,1,1}(u) + 2 (H_2(u))^2 + H_2(u) (H_1(u))^2\\
&\quad
- 2 ( H_3(u) + H_{2,1}(u) - 2 \zeta_3 ) H_1(u)
- \zeta_2 ( 4 H_2(u) + (H_1(u))^2 ) + \frac{5}{2} \zeta_4 \,.
\label{Om2_uu1_fliparg}
\esp
\ee 
In the limit $u\to0$, the $H_{\vec{w}}(u)$ vanish, leaving only the
zeta values and powers of $\ln u$, which are in complete agreement
with \eqn{Om2uu1_smallu}.  In particular, the coefficient of $\zeta_4$
agrees, and this provides a generic method to determine such constants.

In this example, we inspected the $(u,u,1)$ line, whose $u\to0$ limit matches
the $S\to0$ limit of the $T\to0$ expansion.  One can also use the $(1,v,v)$
line in exactly the same way; its $v\to0$ limit matches the $S\to\infty$
limit of the $T\to0$ expansion.

Continuing on in this fashion, we build up the near-collinear expansions through
order $T^2$ for all of the functions in the hexagon basis, and ultimately for
$R_6^{(3)}$ itself. The expansions are rather lengthy, but we present them in a
computer-readable file attached to this document.

%%%%%%%%%%%%%%%%%%%%%%%%%%%%%%%%%%%%%%%%%%%%%%%
\subsection{Fixing most of the parameters}

In \sect{sec:constructthreeloop} we constructed an ansatz~(\ref{R63fndecomp})
for $R_6^{(3)}$ that contained 13 undetermined rational parameters,
after imposing mathematical consistency and extra-purity of $\Rep$.
Two of the parameters affect the symbol: $\alpha_1$ and $\alpha_2$. 
(They could have been fixed using a dual supersymmetry anomaly
equation~\cite{CaronHuot2011kk}.)  The remaining 11 parameters $c_i$
we refer to as ``beyond-the-symbol'' because they accompany functions
(or constants) with Riemann $\zeta$ value prefactors.
Even before we compare to the OPE expansion, the requirement that $R_6^{(3)}$
vanish at order $T^0$ in the collinear limit is already a powerful constraint.
It represents 11 separate conditions when it is 
organized according to powers of $\ln T$, $\ln S^2$ and $H_{\vec{w}}(-S^2)$, as
well as the Riemann $\zeta$ values.  (There is no dependence on $F$
at the leading power-law order.)  The 11 conditions lead to two surviving
free parameters.  They can be chosen as $\alpha_2$ and $c_9$.  

Within $\Rep$, the coefficient $c_9$ multiplies $\zeta_2 \, \Omegauvw$,
as seen from \eqn{r_ambig}.  However, after summing over
permutations, imposing vanishing in the collinear limit, and using
\eqn{R62decomp}, $c_9$ is found to multiply $\zeta_2 \, R_6^{(2)}$.
It is clear that $c_9$ cannot be fixed at this stage (vanishing at
order $T^0$) because the two-loop remainder function vanishes in all
collinear limits.  Furthermore, its leading discontinuity is of the
form $T^m (\ln T)$, which is subleading with respect to the three-loop
leading discontinuity, terms of the form $T^m (\ln T)^2$.  It is rather
remarkable that there is only one other ambiguity, $\alpha_2$, at this
stage.

The fact that $\alpha_1$ can be fixed at the order $T^0$
stage was anticipated in ref.~\cite{Dixon2011pw}. There the symbol multiplying
$\alpha_1$ was extended to a full function, called $f_1$.  It was observed that
the collinear limit of $f_1$, while vanishing at symbol level, did not vanish
at function level, and the limit contained a divergence proportional to 
$\zeta_2 \, \ln T$ times a particular function of $S^2$.  It was argued
that this divergence should cancel against contributions from completing
the $\alpha_i$-independent terms in the symbol into a function.  Now that
we have performed this step, we can fix the value of $\alpha_1$.
Indeed when we examine the $\zeta_2 \, \ln T$ terms in the collinear
limit of the full $R_6^{(3)}$ ansatz, we obtain $\alpha_1 = -3/8$,
in agreement with refs.~\cite{CaronHuot2011kk,Basso2013vsa}.

%%%%%%%%%%%%%%%%%%%%%%%%%%%%

\subsection{Comparison to flux tube OPE results}

In order to fix $\alpha_2$ and $c_9$, as well as obtain many
additional consistency checks, we examine the expansion of $R_6^{(3)}$ to
order $T$ and $T^2$, and compare with the flux tube OPE results of BSV.

BSV formulate scattering amplitudes in planar $\mathcal{N}=4$
super-Yang-Mills theory, or rather the associated polygonal Wilson loops,
in terms of pentagon transitions.  The pentagon transitions map flux tube 
excitations on one edge of a light-like pentagon, to excitations on another,
non-adjacent edge.  They have found that the 
consistency conditions obeyed by the pentagon transitions can be solved
in terms of factorizable $S$ matrices for two-dimensional scattering of the
flux tube excitations.  These $S$ matrices can in turn be determined
nonperturbatively for any value of the coupling, as well as expanded
in perturbation theory in order to compare with perturbative
results~\cite{Basso2013vsa,Basso2013aha}.
The lowest twist excitations dominate the near-collinear or OPE limit
$\tau\to\infty$ or $T\to0$.  The twist $n$ excitations first appear at
${\cal O}(T^n)$.  In particular, the ${\cal O}(T^1)$ term comes only from
a gluonic twist-one excitation, whereas at ${\cal O}(T^2)$ there can be
contributions of pairs of gluons, gluonic bound states, and pairs
of scalar or fermionic excitations.  As mentioned above, BSV have determined
the full order $T^1$ behavior~\cite{Basso2013vsa}, and an unpublished
analysis gives the $T^2F^2$ or $T^2F^{-2}$ terms, plus
the expansion of the $T^2F^0$ terms around $S=0$ through
$S^{10}$~\cite{BSVPrivate}.

BSV consider a particular ratio of Wilson loops: the basic hexagon Wilson
loop, divided by two pentagons, and then multiplied back by a box (square).
The pentagons and box combine to cancel off all of the cusp divergences
of the hexagon, leading to a finite, dual conformally invariant ratio.
We compute the remainder function, which can be expressed as the hexagon
Wilson loop divided by the BDS ansatz~\cite{Bern2005iz} for Wilson loops.
To relate the two formulations, we need to evaluate the logarithm of
the BDS ansatz for the hexagon configuration, subtract the analogous 
evaluation for the two pentagons, and add back the one for the box.
The pentagon and box kinematics are determined from the hexagon
by intersecting a light-like line from a hexagon vertex with an
edge on the opposite side of the hexagon~\cite{Basso2013vsa}.  
For example, if we have 
lightlike momenta $k_i$, $i=1,2,\ldots,6$ for the hexagon, then one
pentagon is found by replacing three of the momenta, say $k_4$, $k_5$, $k_6$,
with two light-like momenta, say $k_4'$ and $k_5'$, having the same sum.
Also, one of the new momenta has to be parallel to one of the three replaced
momenta:
\be
 k_4' + k_5' = k_4 + k_5 + k_6 \,, \qquad\quad k_4' = \xi' k_4 \,.
\ee
The requirement that $k_5'$ is a null vector implies that 
$\xi' = s_{123}/(s_{123}-s_{56})$, where $s_{ij}=(k_i+k_j)^2$,
$s_{ijm}=(k_i+k_j+k_m)^2$.  The five (primed) kinematic variables of the
pentagon are then given in terms of the (unprimed) hexagon variables by
\be
s_{12}^\prime = s_{12} \,, \quad s_{23}^\prime = s_{23} \,, \quad
s_{34}^\prime = \frac{s_{34} s_{123}}{s_{123}-s_{56}} \,, \quad
s_{45}^\prime = s_{123} \,, \quad
s_{51}^\prime = \frac{s_{123} s_{234}-s_{23} s_{56}}{s_{123}-s_{56}} \,.
\label{pentkina}
\ee
The other pentagon replaces $k_1$, $k_2$, $k_3$ with 
$k_1^{\prime\prime}$ and $k_2^{\prime\prime}$ and has $k_1^\prime$ parallel
to $k_1$, which leads to its kinematic variables being given by 
\be
s_{12}^{\prime\prime} = s_{123} \,, \quad 
s_{23}^{\prime\prime} = \frac{s_{123} s_{234}-s_{23} s_{56}}{s_{123}-s_{23}} \,, \quad
s_{34}^{\prime\prime} = s_{45} \,, \quad
s_{45}^{\prime\prime} = s_{56} \,, \quad
s_{51}^{\prime\prime} = \frac{s_{61}s_{123}}{s_{123}-s_{23}} \,.
\label{pentkinb}
\ee
Finally, for the box Wilson loop one makes both replacements
simultaneously; as a result, its kinematic invariants are given by
\be
s_{12}^{\prime\prime\prime} = s_{123} \,, \qquad 
s_{23}^{\prime\prime\prime} = 
\frac{s_{123}(s_{123} s_{234}-s_{23} s_{56})}{(s_{123}-s_{23})(s_{123}-s_{56})} \,.
\label{boxkin}
\ee

The correction term to go between the logarithm of the BSV Wilson loop and 
the six-point remainder function requires the combination of one-loop
normalized amplitudes $V_n$ (from the BDS formula~\cite{Bern2005iz}),
\be
V_6 - V_5^\prime - V_5^{\prime\prime} + V_4^{\prime\prime\prime} \,,
\label{BDScomb}
\ee
which is finite and dual conformal invariant.  There is also 
a prefactor proportional to the cusp anomalous
dimension, whose expansion is known to all orders~\cite{Beisert2006ez},
\be
\gamma_K(a) = 4 a - 4 \zeta_2 a^2 + 22 \zeta_4 a^3
  - 4 \biggl( \frac{219}{8} \zeta_6 + (\zeta_3)^2 \biggr) a^4 + \ldots,
\label{cuspanom}
\ee
where $a = g_{YM}^2 N_c/(32\pi^2) = \lambda/(32\pi^2)$.
Including the proper prefactor, we obtain the following relation between
the two observables,
\be
\label{eq:WL_convert}
\ln\Big[ 1 + \mathcal{W}_{\textrm{hex}}(a/2) \Big]
\ =\ R_6(a)\ +\ \frac{\gamma_K(a)}{8} \, X(u,v,w)\,,
\ee
where
\be
X(u,v,w)\ =\ 
- H^u_2 - H^v_2 - H^w_2 - \ln\biggl(\frac{uv}{w(1-v)}\biggr)\ln (1-v)
- \ln u \ln w + 2 \zeta_2 \,.
\label{Xuvw}
\ee
Here $\mathcal{W}_{\textrm{hex}}$ is BSV's observable (they use the expansion
parameter $g^2 = \lambda/(16\pi^2) = a/2$) and $R_6$ is the remainder function.

In the near-collinear limit, the correction function $X(u,v,w)$ becomes,
\be
\bsp
X(u,v,w)\ &=\ 2 \, T \, \cos\phi\, \Bigl( \frac{H_1}{S} + S\,(H_1+L) \Bigr)\\
&\quad
+ T^2 \biggl[ (1-2\cos^2\phi) \Bigl( \frac{H_1}{S^2} + S^2\,(H_1+L) \Bigr)
             + 2 (H_1+L) \biggr]\ +\ {\cal O}(T^3) \,.
\label{XuvwT}
\esp
\ee
Next we apply this relation in the near-collinear limit, first at order $T^1$.
We find that the $T^1 \ln^2 T$ terms from BSV's formula match perfectly 
the ones we obtain from our expression for $R_6^{(3)}$.  The $T^1 \ln T$
terms also match, given one linear relation between $\alpha_2$ and the
coefficient of $\zeta_2 \, R_6^{(2)}$.  Finally, the $T^1 \ln^0 T$ terms
match if we fix $\alpha_2 = 7/32$, which is the last constant to be fixed.
The value of $\alpha_2$ is in agreement with
refs.~\cite{CaronHuot2011kk,Basso2013vsa}.
The agreement with ref.~\cite{Basso2013vsa} (BSV) is no surprise, because
both are based on comparing the near-collinear limit of $R_6^{(3)}$ with
the same OPE results, BSV at symbol level and here at function level.

Here we give the formula for the leading, order $T$ term in the near-collinear
limit of $R_6^{(3)}$, after fixing all parameters as just described:
\be
\bsp
R_6^{(3)} &= \frac{T}{S} \cos\phi \Bigg\{
\ln^2 T \biggl[ \frac{2}{3} H_1^3 + H_1^2 (L+2)
  + H_1 \Bigl( \frac{1}{4} L^2 + 2 L + \frac{1}{2} \zeta_2 + 3 \Bigr)
  - H_3 + \frac{1}{2} H_2 (L-1) \biggr]\\
&\hskip0.5cm
- \ln T \biggl[ \frac{1}{2} H_1^4 + H_1^3 (L+2)
        + H_1^2 \Bigl( \frac{1}{4} L^2 + 3 L + \frac{3}{2} \zeta_2 + 5 \Bigr)
        + H_1 \Bigl( \frac{1}{2} L^2 + ( H_2 + 2 \zeta_2 + 5 ) L\\
&\hskip1.9cm 
        - \zeta_3 + 3 \zeta_2 + 9 \Bigr)
        + \frac{1}{2} ( H_3 - 2 H_{2,1} ) (L+1)
        + \frac{1}{2} H_2 (L-1) \biggr]\\
&\hskip0.5cm
+ \frac{1}{10} H_1^5 + \frac{1}{4} H_1^4 (L+2)
+ \frac{1}{12}  H_1^3 (L^2 + 12 L + 6\zeta_2 + 20)
+ \frac{1}{4} H_1^2 \Bigl( L^2 + 2 (H_2 + 2\zeta_2 + 5) L\\
&\hskip0.5cm - 2\zeta_3 + 6\zeta_2 + 18 \Bigr)
+ \frac{1}{8}  H_1 \Bigl[ 8  (H_4-H_{3,1}) + 2 H_2^2 + (H_2 + \zeta_2 + 3) L^2
               + \Bigl( 8 (H_2-H_{2,1}) + 4\zeta_3\\
&\hskip0.5cm + 16\zeta_2 + 36 \Bigr)  L
               + 2  \zeta_2  ( H_2 + 9 ) - 39 \zeta_4 - 8 \zeta_3 + 72 \Bigr]
- \frac{1}{4}  H_{2,1}  L^2 
+ \frac{1}{8}  \Bigl( - 6 H_4 + 8 H_{2,1,1} + H_2^2\\
&\hskip0.5cm + 2 H_3 - 12 H_{2,1} + 2 (\zeta_2+2)  H_2 \Bigr) L
+ \frac{1}{8} H_2^2 - \frac{1}{4} H_2 ( 2H_3 + 4H_{2,1} + 2\zeta_3 + \zeta_2 )
- \frac{1}{4} (2\zeta_2 - 3) H_3\\
&\hskip0.5cm
- \frac{1}{2} (\zeta_2+1) H_{2,1}
+ \frac{9}{2} H_5 + H_{4,1} + H_{3,2} + 6 H_{3,1,1} + 2 H_{2,2,1} 
+ \frac{3}{4} H_4 - H_{2,1,1} \Bigg\} + \Big(S\to\frac{1}{S}\Big)\\
&+\ {\cal O}(T^2) \,.
\esp
\ee

The $T^2$ terms are presented in an attached, computer-readable file.
The $T^2$ terms match perfectly with OPE results provided to us
by BSV~\cite{BSVPrivate}, and at this order there are no free parameters
in the comparison.  This provides a very nice consistency check on two
rather different approaches.

Recall that we imposed an extra-pure condition on the terms in \eqn{r_ambig}
that we added to the ansatz for $R_6^{(3)}$.  We can ask what would happen
if we relaxed this assumption.  To do so we consider adding to the solution that
we found a complete set of beyond-the-symbol terms.  Imposing total symmetry,
there are 2 weight-6 constants ($\zeta_6$ and $(\zeta_3)^2$), and 2
weight-5 constants ($\zeta_5$ and $\zeta_2\zeta_3$) multiplying $\ln uvw$.
Multiplying the zeta values $\zeta_4$, $\zeta_3$ and $\zeta_2$ there are
respectively 3, 7 and 18 symmetric functions, for a total of 32 free parameters.
Imposing vanishing of these additional terms at order $T^0$ fixes all
but 5 of the 32 parameters to be zero.  We used constraints from the
multi-Regge limit (see the next section)
to remove 4 of the 5 remaining parameters.  Finally,
the order $T^1$ term in the near-collinear limit fixes the last parameter
to zero.  We conclude that there are no additional ambiguities in $R_6^{(3)}$
associated with relaxing the extra-purity assumption.

\vfill\eject

%%%%%%%%%%%%%%%%%%%%%%%%%%%%%%%%%%%%%%%%%%%%%%%%%%%%%%%%%%%%%%%%%%%%%%%

\section{Multi-Regge limits}
\label{sec:MRK}

The multi-Regge or MRK limit of $n$-gluon scattering
is a $2\to (n-2)$
scattering process in which the $(n-2)$ outgoing gluons are
strongly ordered in rapidity.  It generalizes the Regge limit of
$2\to2$ scattering with large center-of-mass energy at
fixed momentum transfer, $s \gg t$.  Here we are interested in the 
case of $2\to4$ gluon scattering, for which the MRK limit means 
that two of the outgoing gluons are emitted at high
energy, almost parallel to the incoming gluons.  The other two gluons are
also typically emitted at small angles, but they are well-separated in
rapidity from each other and from the leading two gluons, giving them
smaller energies.

The strong ordering in rapidity for the $2\to4$ process leads to
the following strong ordering of momentum invariants:
\be
s_{12} \gg s_{345},s_{123} \gg s_{34},s_{45},s_{56} \gg s_{23},s_{61},s_{234} \,.
\label{mominvstrongorder}
\ee
In this limit, the cross ratio $u = s_{12} s_{45}/(s_{123}s_{345})$ approaches
one.  The other two cross ratios vanish, 
\be
u\to1, \qquad v\to0, \qquad \hat{w}\to0.
\label{MRK_defn}
\ee
In this section, we denote the original cross ratio $w$ by $\hat{w}$,
in order to avoid confusion with another variable which we are about
to introduce.  The cross ratios $v$ and $\hat{w}$ vanish
at the same rate that $u\to1$, so that the ratios $x$ and $y$, defined by
\be
x \equiv \frac{v}{1-u} \,, \qquad y \equiv \frac{\hat{w}}{1-u} \,,
\label{xydef}
\ee
remain fixed.  The variable $y$ in \eqn{xydef} should not be confused with
the variables $y_i$.  In the $y$ variables, the multi-Regge limit
consists of taking $y_u \to 1$, while $y_v$ and $y_w$ are left
arbitrary.  (Their values in this limit are related to $x$ and $y$ by
\eqn{xyEMRK}.)

It is very convenient~\cite{Lipatov2010ad}
to change variables from $x$ and $y$ to the complex-conjugate
pair $(w,\ws)$ defined by,
\be
x = \frac{1}{(1+w)(1+\ws)} \,, \qquad y = \frac{w\ws}{(1+w)(1+\ws)} \,.
\label{wwsdef}
\ee
(Again, this variable $w$ should not be confused with the original cross
ratio called $\hat{w}$ in this section.)  This change of variables
rationalizes the $y$ variables in the MRK limit, so that
\be
y_u \to 1, \qquad
y_v\to\frac{1+\ws}{1+w} \,, \qquad 
y_w\to\frac{(1+w)\ws}{w(1+\ws)} \,.
\label{yMRK}
\ee

As an aside, we remark here that the variables $T,S,F$ in \eqn{BSVparam},
used by BSV to describe the near-collinear limit, are closely related to
the variables $w,w^*$ introduced for the MRK limit.  To establish
this correspondence, we consider (in this paragraph only)
the MRK limit $u\to0$, $v\to0$, $\hat{w}\to1$,
which is related to \eqn{MRK_defn} by a cyclic permutation 
$u_i\to u_{i-1}$, $y_i\to y_{i-1}$.   This limit corresponds to the $T\to0$
limit in \eqn{BSVparam} if we also send $S\to0$ at the same rate,
so that $T/S$ is fixed.  Let's rewrite $y_u$ from \eqn{BSVparam} as
\be
y_u = \frac{1+\frac{T}{SF}}{1+\frac{TF}{S}} \,
\label{yu_rewrite}
\ee
and compare it with the limiting behavior of $y_v$ in \eqn{yMRK}.
(Comparing $y_u$ with $y_v$ is required by the cyclic permutation of the
$u_i$ and $y_i$ variables which we need for the two limits to correspond.)
If we let
\be
w = \frac{T}{S} F \,, \qquad w^* = \frac{T}{S} \frac{1}{F} \,,
\label{w_as_TF_S}
\ee
then $y_v$ in \eqn{yMRK} correctly matches \eqn{yu_rewrite}.
If we start with the variables $T,S,F$ in \eqn{BSVparam},
insert the inverse relations to~\eqn{w_as_TF_S},
\be
T = S \sqrt{ww^*}, \qquad F = \sqrt{\frac{w}{w^*}} \,,
\label{TF_S_as_w}
\ee
and then let $S\to0$ with $w,w^*$ fixed, we can check that
all variables approach the values appropriate for the multi-Regge limit
$u\to0$, $v\to0$, $\hat{w}\to1$.  The cross-ratio $\hat{w}$ approaches unity
as $S$ vanishes, through the relation $\hat{w} = (1+S^2|1+w|^2)^{-1}$.
Finally, we note that the MRK limit interpolates between three 
different limits:
the collinear limit $v\to0$, corresponding to $|w|\to0$;
the endpoint of the line $(u,u,1)$ with $u\to0$, corresponding to $w\to-1$;
and a second collinear limit $u\to0$, corresponding to $|w|\to\infty$.

Now we return to the $u\to1$ version of the MRK limit in \eqn{MRK_defn}.
If this limiting behavior of the cross ratios is approached directly from
the Euclidean region in which all cross ratios are positive, 
we call it the EMRK limit (see also \eqn{EMRK}).
In this limit, the remainder function vanishes, as it does in the Euclidean
collinear limit discussed in the previous section.  However, the physical
region for $2\to4$ scattering is obtained by first analytically
continuing $u\to e^{-2\pi i}u$, then taking $u\to1$, 
$v,\hat{w}\to0$ as above.  The analytic continuation generates 
imaginary terms corresponding to the discontinuity of the function in the
$u$ channel, which survive into the MRK limit; in fact they can be
multiplied by logarithmic singularities as $u\to1$.

The general form of the remainder function at $L$ loops in the MRK limit is
\be
R_6^{(L)}(1-u,w,\ws)\ =\ (2\pi i)\sum_{r=0}^{L-1}\ln^r(1-u)
\left[g_r^{(L)}(w,\ws)+2\pi i h_r^{(L)}(w,\ws)\right] + {\cal O}(1-u)\,,
\label{MRKgeneral}
\ee
where the coefficient functions $g_r^{(L)}(w,\ws)$ are referred to as the
leading-log approximation (LLA) for $r=L-1$, next-to-LLA (NLLA) for
$r=L-2$, and so on.  The coefficient functions $h_r^{(L)}(w,\ws)$ can be
determined simply from the $g_r^{(L)}$, by using a crossing
relation from the $3\to3$ channel~\cite{Bartels2010tx,Dixon2012yy}.

The coefficient functions in this limit are built out of HPLs with
arguments $-w$ and $-\ws$.  Only special combinations of such HPLs
are allowed, with good branch-cut behavior in the $(w,\ws)$ plane,
corresponding to symbols whose first entries are limited
to $x$ and $y$~\cite{Dixon2012yy}.  Such functions may be called
single-valued harmonic polylogarithms (SVHPLs), and were constructed by
Brown~\cite{BrownSVHPLs}.

Using a Fourier-Mellin transformation, Fadin, Lipatov, and Prygarin wrote 
an all-loop expression for the MRK limit in a factorized form depending on 
two quantities, the BFKL eigenvalue $\omega(\nu,n)$ and the impact
factor $\Phi_{\textrm{Reg}}(\nu,n)$ ~\cite{Fadin2011we}:
\be
\bsp
e^{R+i\pi\delta}|_{\textrm{MRK}} = \cos\pi\omega_{ab} 
+ i \, \frac{a}{2} \sum_{n=-\infty}^\infty
(-1)^n\,\left(\frac{w}{\ws}\right)^{\frac{n}{2}}
&\int_{-\infty}^{+\infty}
\frac{d\nu}{\nu^2+{n^2\over 4}}\,|w|^{2i\nu}\,\Phi_{\textrm{Reg}}(\nu,n)\\
&\hskip1cm \times\left(-\frac{1}{1-u}\frac{|1+w|^2}{|w|}\right)^{\omega(\nu,n)}\,.
\esp
\ee
Here
\bea
\omega_{ab} &=& \frac{1}{8}\,\gamma_K(a)\,\log|w|^2\,, 
\label{omegaabdef}\\
\delta &=& \frac{1}{8}\,\gamma_K(a)\,\log\frac{|w|^2}{|1+w|^4}\,,
\label{deltadef}
\eea
where the cusp anomalous dimension $\gamma_K(a)$ is given in \eqn{cuspanom}.

By taking the MRK limit of the symbol of the three-loop remainder
function, it was possible to determine all of the coefficient functions
$g_r^{(l)}$ and $h_r^{(l)}$ through three loops, up to four undetermined
rational numbers, $d_1$, $d_2$, $\gamma^\prime$ and $\gamma^{\prime\prime}$,
representing beyond-the-symbol ambiguities~\cite{Dixon2011pw}.
(Two other parameters, $c$ and $\gamma^{\prime\prime\prime}$, could be fixed
using consistency between the MRK limits in $2\to4$ kinematics
and in $3\to3$ kinematics.)
One of these four constants was fixed by Fadin and Lipatov~\cite{Fadin2011we},
using a direct calculation of the NLLA BFKL eigenvalue: 
$\gamma^\prime=-9/2$.  The 
remaining three undetermined constants, $d_1$, $d_2$ and 
$\gamma^{\prime\prime}$, all appear in the NNLLA coefficient $g_0^{(3)}(w,\ws)$.

In ref.~\cite{Dixon2012yy}, the coefficient functions $g_r^{(3)}(w,w^*)$
and $h_r^{(3)}(w,w^*)$ that appear in the MRK limit~(\ref{MRKgeneral})
of $R_6^{(3)}$ were expressed in terms of the SVHPLs defined in
ref.~\cite{BrownSVHPLs}. More specifically, they were rewritten in terms
of particular linear combinations of SVHPLs,  denoted by $L_{\vec{w}}^\pm$,
that have definite eigenvalues under inversion of $w$ and under its
complex conjugation.  The coefficient function $g_0^{(3)}(w,\ws)$ then
becomes~\cite{Dixon2012yy}:
\be
\bsp
g_0^{(3)}(w,\ws) &\,= 
\rat{27}{8}\,L_5^+ + \rat{3}{4}\,L_{3,1,1}^+ - \rat{1}{2}\,L_3^+\,[L_1^+]^2
- \rat{15}{32}\,L_3^+\,[L_0^-]^2 - \rat{1}{8}\,L_1^+\,L_{2,1}^-\,L_0^-\\
&\, + \rat{3}{32}\,[L_0^-]^2\,[L_1^+]^3 + \rat{19}{384}\,L_1^+\,[L_0^-]^4
+ \rat{3}{8}\,[L_1^+]^2\,\zeta_3 - \rat{5}{32}\,[L_0^-]^2\,\zeta_3
+ \rat{\pi ^2}{96}\,[L_1^+]^3\\
&\, - \rat{\pi ^2}{384}\,L_1^+\,[L_0^-]^2 - \rat{3}{4}\,\zeta_5
- {\pi^2\over 6}\,\gamma''
  \,\Big\{L_3^+-{1\over 6}[L_1^+]^3-{1\over 8}[L_0^-]^2\,L_1^+\Big\}\\
&\, + {1\over 4}\,d_1\,\zeta_3\Big\{[L_1^+]^2 - {1\over 4}[L_0^-]^2\Big\}
- {\pi^2\over 3}d_2\,L_1^+\Big\{[L_1^+]^2 - {1\over 4}[L_0^-]^2\Big\}
+ \rat{1}{30}\,[L_1^+]^5 \,. \\
\esp
\label{g3_0}
\ee

In the remainder of this section we will describe how to extract the MRK
limit of the three-loop remainder function at the full function level.
Comparing this limit with \eqn{g3_0} (as well as the other $g_r^{(3)}$ and
$h_r^{(3)}$ coefficient functions) will serve as a check of our construction
of $R_6^{(3)}$, and it will also provide for us the remaining three-loop MRK
constants, $d_1$, $d_2$ and $\gamma^{\prime\prime}$.

%%%%%%%%%%%%%%%%%%%%%%%%%%%%%%%%%%
\subsection{Method for taking the MRK limit}

Let us begin by discussing a method for taking the multi-Regge limit
of hexagon functions in general, or of $R_6^{(3)}$ in particular,
starting from an expression in terms of multiple polylogarithms.  The
first step is to send $u\to e^{-2\pi i} u$, {\it i.e.} to extract the monodromy around $u=0$. Owing to the
non-linear relationship between the $u_i$ and the $y_i$,~\eqn{u_from_y},
it is not immediately clear what the discontinuity looks like in the
$y$ variables. The correct prescription turns out simply to be to
take $y_u$ around 0. To see this, consider the
$\Delta_{1,n-1}$ component of the coproduct, which can be written as,
\be
\Delta_{1,n-1}(F) \equiv \ln u \otimes\uF
+ \ln v \otimes \vF + \ln w \otimes\wF\,.
\label{eq:copr1nm1_u}
\ee
There are only three terms, corresponding to the three possible
first entries of the symbol. 

Using the coproduct formulas in
appendix~\ref{sec:app_multi_poly}, it is straightforward to extract
the functions $\uF$, $\vF$, and $\wF$ for any given hexagon
function. These functions capture information about the discontinuities
as each of the cross ratios is taken around zero. In particular,
since the monodromy operator acts on the first component of the coproduct, we have
({\it c.f.}~\eqn{eq:monodromy}),
\be
\bsp
\Delta_{1,n-1}\Bigl[\mathcal{M}_{u=0} (F)\Bigr] 
&= \Bigl[\mathcal{M}_{u=0}(\ln u)\Bigr] \otimes \uF \\
&=(\ln u - 2\pi i)\otimes \uF\,.
\label{eq:del1n1}
\esp
\ee
\Eqn{eq:del1n1} is not quite sufficient to deduce $\mathcal{M}_{u=0}(F)$. The
obstruction comes from the fact that all higher powers of $(2\pi i)$ live in the
kernel of $\Delta_{1,n-1}$. On the other hand, these terms can be extracted
from the other components of the coproduct: the $(2\pi i)^k$ terms
come from the piece of $\Delta_{k,n-k}(F)$ with $\ln^k u$ in the first slot.

If we write~\eqn{eq:copr1nm1_u} in terms of the $y_i$ 
variables, we find,
\be
\bsp
\Delta_{1,n-1}(F) &= \Bigl[G(0; y_u) + G\left(1;y_v\right)+ G\left(1;y_w\right)
- G\left(\frac{1}{y_u };y_v\right)- G\left(\frac{1}{y_u};y_w\right)\Bigr]
\otimes\uF\\
&+ \Bigl[G(0; y_v) + G\left(1;y_u\right)+ G\left(1;y_w\right)
- G\left(\frac{1}{y_u };y_v\right)- G\left(\frac{1}{y_v};y_w\right)\Bigr]
\otimes\vF\\
&+ \Bigl[G(0; y_w) + G\left(1;y_u\right)+ G\left(1;y_v\right)
- G\left(\frac{1}{y_u };y_w\right)- G\left(\frac{1}{y_v};y_w\right)\Bigr]
\otimes\wF\, ,
\esp
\label{eq:copr1nm1_y}
\ee
where we have now assumed that we are working in Region I. 
\Eqn{eq:copr1nm1_y} indicates that $\uF$ can be extracted uniquely
from the terms with $G(0;y_u)$ in the first slot. Similarly, the elements
of the full coproduct with $\ln^k u$ in the first slot are given exactly by
the terms with $G(0;y_u)^k$ in the first slot. Therefore the
discontinuity around $u=0$ is the same as the discontinuity around
$y_u=0$. Furthermore, because our basis $\GG_I^L$ exposes all
logarithms $G(0;y_u)$ (by exploiting the shuffle algebra), the only
sources of such discontinuities are powers of $G(0;y_u)$. As a result, we
have a simple shortcut to obtain the monodromy around $u=0$, 
\be
\mathcal{M}_{u=0}(F) = F|_{G(0;y_u)\,\to\,G(0;y_u)-2\pi i}\,.
\label{eq:disc_yu}
\ee

The final step in obtaining the MRK limit is to take $y_u\to 1$. This
limit is trivially realized on functions in the basis $\GG_I^L$
because the only source of singularities is $G(1;y_u)$; all other
functions are finite as $y_u\to1$. Writing the divergence in terms of
$\xi \equiv 1-u$, which approaches 0 in this limit, we take
\be
G(1;y_u)\; \xrightarrow{\scriptscriptstyle y_u\to 1} \;\ln \xi
+ G(1;y_v) + G(1;y_w) - G\left(\frac{1}{y_v};y_w\right)\,,
\label{eq:disc_yu_limit}
\ee
and then set $y_u=1$ in all other terms.

The result of this procedure will be a polynomial in $\ln \xi$ whose
coefficients are multiple polylogarithms in the variables $y_v$ and
$y_w$. On the other hand, we know from general considerations that the
coefficient functions should be SVHPLs. To translate the multiple
polylogarithms into SVHPLs, we use the coproduct bootstrap
of~\sect{sec:coproduct}, seeded by the weight-one identities which
follow from~\eqn{yMRK} and from combining~eqs.~(\ref{xyEMRK}), (\ref{wwsdef})
and (\ref{yMRK}),
\be
\frac{1}{|1+w|^2} = \frac{y_v(1-y_w)^2}{(1-y_v y_w)^2} \,,
\qquad
\frac{|w|^2}{|1+w|^2} = \frac{y_w(1-y_v)^2}{(1-y_v y_w)^2} \,.
\label{extraEMRK}
\ee
We obtain,
\be
\bsp
L_0^- & =\ \ln |w|^2\ =\ -G(0;y_v)+G(0,y_w) + 2 G(1;y_v)-2G(1;y_w)\,,\\
L_1^+ & =\ \ln\frac{|w|}{|1+w|^2} 
\ =\ \frac{1}{2}G(0;y_v) + \frac{1}{2}G(0;y_w) + G(1;y_v)+G(1;y_w) 
- 2 G\left(\frac{1}{y_v};y_w\right)\,,
\label{eq:btsrp_seed1}
\esp
\ee
and,
\be
\ln\left(\frac{1+w^{\phantom{*}}}{1+\ws}\right) 
= -G(0;y_v)\quad~\textrm{and}~\quad\ln\left(\frac{w}{\ws}\right) 
= -G(0;y_v)-G(0;y_w)\,.
\label{eq:btsrp_seed2}
\ee

Alternatively, we can extract the MRK limits of the hexagon functions
iteratively in the weight, by using their definitions in terms of
differential equations.  This procedure is similar to that used
in~\sect{sec:collinear} to find the collinear limits of the hexagon
functions, in that we expand the differential equations around the limiting
region of $u\to1$.

However, first we have to compute the
discontinuities from letting $u\to e^{-2\pi i}u$
in the inhomogeneous (source) terms for the differential equations.
For the lowest weight non-HPL function, $\PhiTilde$, the source terms
are pure HPLs.  For pure HPL functions we use standard HPL identities
to exchange the HPL argument $(1-u)$ for argument $u$, and again use
the Lyndon basis so that the trailing index in the weight vector $\vec{w}$
in each $H_{\vec{w}}(u)$ is 1.  In this new representation, the only
discontinuities come from explicit factors of $\ln u$, which are simply
replaced by $\ln u - 2\pi i$ under the analytic continuation.
After performing the analytic continuation, we take the MRK limit
of the pure HPL functions.

Once these limits are known, we can integrate up the differential
equations for the non-HPL functions in much the same fashion that we did
for the collinear limits, by using a restricted ansatz built from powers
of $\ln \xi$ and SVHPLs.  The Jacobian factors needed 
to transform from differential equations in $(u,v,\hat{w})$ 
to differential equations in the MRK variables $(\xi,w,\ws)$,
are easily found to be:
\be
\bsp
\frac{\partial F}{\partial \xi} 
&= -\frac{\partial F}{\partial u}
+ x\frac{\partial F}{\partial v} + y\frac{\partial F}{\partial w} \,,\\
\frac{\partial F}{\partial w} 
&= \frac{\xi}{w(1+w)}\left[
- w x\frac{\partial F}{\partial v} + y\frac{\partial F}{\partial w}\right]
\,,\\
\frac{\partial F}{\partial \ws} 
&= \frac{\xi}{\ws(1+\ws)}\left[
- \ws x\frac{\partial F}{\partial v} + y\frac{\partial F}{\partial w}\right]
\,.\\
\esp
\label{diff_xiwws}
\ee
We compute the derivatives on the right-hand side of these relations
using the formula for $\partial F/\partial u_i$ in terms of the coproduct
components, \eqn{u_diffeq}.  We also implement the transformation
$u\to e^{-2\pi i}u$ on the coproduct components, as described above
for the HPLs, and iteratively in the weight for the non-HPL hexagon functions.
When we expand as $\xi\to0$, we drop all power-suppressed terms in
$\xi$, keeping only polynomials in $\ln\xi$.  (In $\partial F/\partial\xi$,
we keep the derivatives of such expressions, {\it i.e.}~terms of the form
$1/\xi \times \ln^k\xi$.)

In our definition of the MRK limit, we include any surviving terms
from the EMRK limit. This does not matter
for the remainder function, whose EMRK limit vanishes, but the
individual parity-even hexagon functions can have nonzero, and even
singular, EMRK limits.
%%%%%%%%%%%%%%%%%%%%%%%%%%%%

\subsection{Examples}

We first consider the simplest non-HPL function, $\PhiTilde$. Starting
with the expression for $\PhiTilde$ in Region I,~\eqn{Phimultipoly},
we take the monodromy around $u=0$, utilizing~\eqn{eq:disc_yu},
\be
\bsp
\mathcal{M}_{u=0}(\PhiTilde) &=2\pi i\Bigl[ 
- G\Bigl(0,\frac{1}{y_u y_v};y_w\Bigr) 
- G\Bigl(0,\frac{1}{y_v};y_w\Bigr) 
+ G\Bigl(0,\frac{1}{y_u};y_w\Bigr) 
+ G\Bigl(0,\frac{1}{y_u};y_v\Bigr) \\
&\qquad+   G\left(0,1;y_w\right) +   G\left(0,1;y_v\right) 
- G\left(0,1;y_u\right) +   G\left(0;y_v\right)G\left(0;y_w\right)
+ \zeta_2\Bigr]\, . \\
\esp
\ee
Next, we take the limit $y_u\to 1$. There are no divergent factors, so we
are free to set $y_u=1$ without first applying~\eqn{eq:disc_yu_limit}. 
The result is,
\be
\PhiTilde|_{\textrm{MRK}} = 2 \pi i \Bigl[
- 2  G\Bigl(0,\frac{1}{y_v};y_w\Bigr) + 2  G\left(0,1;y_w\right)
+ 2  G\left(0,1;y_v\right) +   G\left(0;y_v\right)G\left(0;y_w\right)
+ 2 \zeta_2\Bigr]\, .
\label{eq:PhiTildeMRKasG}
\ee
To transform this expression into the SVHPL notation of
ref.~\cite{Dixon2012yy}, we use the coproduct bootstrap to derive
an expression for the single independent SVHPL of weight two, the Bloch-Wigner
dilogarithm, $L_2^-$,
\be
\bsp
\Delta_{1,1}(L_2^-) &=\Delta_{1,1}\left(
{\rm Li}_2(-w) - {\rm Li}_2(-\ws)
+ \frac{1}{2} \ln|w|^2 \, \ln\frac{1+w}{1+\ws}\right)\\
&=\frac{1}{2} L_0^- \otimes 
\left[\ln \left(\frac{1+w}{1+\ws}\right)-\frac{1}{2}\ln\left(\frac{w}{\ws} \right) 
\right] 
+ \frac{1}{2} L_1^+ \otimes \ln\left(\frac{w}{\ws}\right)\\
&= \Delta_{1,1}\left(G\left(0,\frac{1}{y_v};y_w\right)  -  G\left(0,1;y_w\right)
-  G\left(0,1;y_v\right) - \frac{1}{2}  G\left(0;y_v\right)G\left(0;y_w\right)\right)\,.
\label{eq:coprL2m}
\esp
\ee
In the last line we used~\eqns{eq:btsrp_seed1}{eq:btsrp_seed2}.
Lifting~\eqn{eq:coprL2m} from coproducts to functions introduces one 
undetermined rational-number constant, proportional to $\zeta_2$. 
It is easily fixed by specializing to the point $y_v=y_w=1$, yielding,
\be
\label{eq:L2masG}
L_2^- = G\left(0,\frac{1}{y_v};y_w\right)  -  G\left(0,1;y_w\right)
-  G\left(0,1;y_v\right) 
- \frac{1}{2}  G\left(0;y_v\right)G\left(0;y_w\right) -\zeta_2\,,
\ee
which, when compared to~\eqn{eq:PhiTildeMRKasG}, gives,
\be
\PhiTilde|_{\textrm{MRK}}\ =\ -4\pi i \, L_2^-\, .
\ee

Let us derive this result in a different way, using the method based on
differential equations.  Like all parity-odd functions, $\PhiTilde$
vanishes in the Euclidean MRK limit;  however, it survives in the MRK
limit due to discontinuities in the function $\Omega^{(1)}$ given in \eqn{Om1},
which appears on the right-hand side of the $\PhiTilde$ differential
equation~(\ref{duPhi}).  The MRK limits of the three cyclic permutations
of $\Omega^{(1)}$ are given by
\be
\bsp
\Omega^{(1)}(u,v,\hat{w})\Bigl|_{\rm MRK} &= 2\pi i \ln|1+w|^2 \,,\\
\Omega^{(1)}(v,\hat{w},u)\Bigl|_{\rm MRK} &= 2\pi i \ln\xi \,,\\
\Omega^{(1)}(\hat{w},u,v)\Bigl|_{\rm MRK} &= 2\pi i \ln\frac{|1+w|^2}{|w|^2}
\,.\\
\esp
\label{Om1MRK}
\ee
Inserting these values into \eqn{duPhi} for $\partial\PhiTilde/\partial u$
and its cyclic permutations, and then inserting those results into
\eqn{diff_xiwws}, we find that
\be
\bsp
\frac{\partial\PhiTilde}{\partial\xi}\bigg|_{\xi^{-1}} &= 0, \\
\frac{\partial\PhiTilde}{\partial w}\bigg|_{\xi^0} &= 
2\pi i \biggl[ - \frac{\ln|w|^2}{1+w} + \frac{\ln|1+w|^2}{w} \biggr] \,, \\
\frac{\partial\PhiTilde}{\partial \ws}\bigg|_{\xi^0} &= 
2\pi i \biggl[ \frac{\ln|w|^2}{1+\ws} - \frac{\ln|1+w|^2}{\ws} \biggr] \,.
\esp
\label{Phitilde_xiwws}
\ee
The first differential equation implies that there is no $\ln\xi$
term in the MRK limit of $\PhiTilde$.  The second two differential
equations imply that the MRK limit is proportional to the Bloch-Wigner
dilogarithm,
\be
\bsp
\PhiTilde\big|_{\rm MRK} 
&= -4\pi i \Bigl[ {\rm Li}_2(-w) - {\rm Li}_2(-\ws)
+ \frac{1}{2} \ln|w|^2 \, \ln\frac{1+w}{1+\ws} \Bigr] \\
&= - 4\pi i \, L_2^- \,.
\esp
\ee

Now that we have the MRK limit of $\PhiTilde$, we can find the limiting
behavior of all the coproduct components of $\Omega^{(2)}$ appearing in
\eqn{Om2_31}, and perform the analogous expansion of the differential
equations in the MRK limit.  For $\Omega^{(2)}(u,v,\hat{w})$ we obtain,
\be
\bsp
\frac{\partial\Omega^{(2)}(u,v,\hat{w})}{\partial\xi}\bigg|_{\xi^{-1}} &= 
\frac{2\pi i}{\xi} \ln|1+w|^2 \, \biggl[ - \ln\xi
  + \frac{1}{2} \ln|1+w|^2 - \pi i \biggr] \,, \\
\frac{\partial\Omega^{(2)}(u,v,\hat{w})}{\partial w}\bigg|_{\xi^0} &= 
\frac{2\pi i}{1+w} \biggl[ - \frac{1}{2} \ln^2\left(\frac{\xi}{|1+w|^2}\right)
+ \frac{1}{2} \ln|w|^2 \, \ln|1+w|^2 \, - \, L_2^- + \zeta_2\\
&\hskip2cm
- \pi i \ln\left(\frac{\xi}{|1+w|^2}\right) \biggr] \,,
\esp
\label{Omega2uvw_xiwws}
\ee
plus the complex conjugate equation for
$\partial\Omega^{(2)}(u,v,\hat{w})/\partial\ws$.

The solution to these differential equations can be expressed in terms of
SVHPLs.  One can write an ansatz for the result as a linear
combination of SVHPLs, and fix the coefficients using the differential
equations.  One can also take the limit first at the level of
the symbol, matching to the symbols of the SVHPLs; then one only has to
fix the smaller set of beyond-the-symbol terms using the differential
equations. The result is
\be
\bsp 
\Omega^{(2)}(u,v,\hat{w})\big|_{\rm MRK} &= 2\pi i\biggl[
\frac{1}{4} \ln^2\xi \, ( 2\, L_1^+ - L_0^- )
+ \frac{1}{8} \ln\xi \, ( 2\, L_1^+ - L_0^- )^2
+ \frac{5}{48} \, [L_0^-]^3 + \frac{1}{8} \, [L_0^-]^2 \, L_1^+ \\
&\hskip1.5cm
+ \frac{1}{4} \, L_0^- \, [L_1^+]^2 + \frac{1}{6} \, [L_1^+]^3
- L_3^+ - 2 \, L_{2,1}^-
- \frac{\zeta_2}{2} ( 2\, L_1^+ - L_0^- ) - 2 \, \zeta_3 \biggr]\\
&\hskip.5cm
- (4\pi)^2 \biggl[ \frac{1}{4} \ln\xi ( 2\, L_1^+ - L_0^- )
+ \frac{1}{16} ( 2\, L_1^+ - L_0^- )^2 \biggr] \,.
\esp
\ee
In this case the constant term, proportional to $\zeta_3$, can be fixed by
requiring vanishing in the collinear-MRK corner where $|w|^2\to0$.
The last set of terms, multiplying $(4\pi)^2$, come from a double
discontinuity.

The MRK limit of $\Omega^{(2)}(\hat{w},u,v)$ is related by symmetry to that of 
$\Omega^{(2)}(u,v,\hat{w})$:
\be
\Omega^{(2)}(\hat{w},u,v)\big|_{\rm MRK} 
= \Omega^{(2)}(u,v,\hat{w})|_{\rm MRK}(w\to1/w,\ws\to1/\ws) \,.
\ee
The final MRK limit of $\Omega^{(2)}$ is,
\be
\bsp 
\Omega^{(2)}(v,\hat{w},u)\big|_{\rm MRK} &=
\frac{1}{4} L_X^4 - \Bigl( \frac{1}{8} [L_0^-]^2 - \zeta_2 \Bigr) L_X^2
+ 4 \, \zeta_3 \, L_X
+ \frac{1}{64} \, [L_0^-]^4 + \frac{1}{4} \, \zeta_2 \, [L_0^-]^2
+ \frac{5}{2} \, \zeta_4 \\
&\hskip.5cm
+ 2\pi i\biggl[ \frac{1}{3} \, L_X^3
- 2  \Bigl( \frac{1}{8} \, [L_0^-]^2 - \zeta_2 \Bigr) L_X
+ \frac{1}{2} \, [L_0^-]^2 \, L_1^+ - 2 ( L_3^+ - \zeta_3 ) \biggr] \,,
\esp
\ee
where $L_X = \ln\xi + L_1^+$.  Note that this orientation of $\Omega^{(2)}$
has a nonvanishing (indeed, singular) EMRK limit, {\it i.e.}~even before 
analytically continuing into the Minkowski region to pick up the imaginary
part.  On the other hand, there is no surviving double discontinuity for
this ordering of the arguments.

As our final (simple) example, we give the MRK limit of the totally symmetric,
weight five, parity-odd function $G(u,v,\hat{w})$.  As was the case for
$\PhiTilde$, the limit of $G$ is again proportional to the Bloch-Wigner
dilogarithm, but with an extra factor of $\zeta_2$ to account for the
higher transcendental weight of $G$:
\be
G(u,v,\hat{w})\big|_{\rm MRK} = 16\pi i \zeta_2 \, L_2^- \,.
\label{G_MRK}
\ee
As usual for parity-odd functions, the EMRK limit vanishes. In this case
the double discontinuity also vanishes.  In general the MRK limits of the
parity-odd functions must be odd under $w \leftrightarrow \ws$, which
forbids any nontrivial constants of integration.

Continuing onward, we build up the MRK limits for all the remaining hexagon
functions. The results are attached to this document in a computer-readable
format.

%%%%%%%%%%%%%%%%%%%%%%%%%%%%

\subsection{Fixing $d_1$, $d_2$, and $\gamma''$}

Using the MRK limit of all the hexagon functions appearing in \eqn{R63fndecomp},
we obtain the MRK limit of $R_6^{(3)}$.  This is a powerful check of the
function, although as mentioned above, much of it is guaranteed by the
limiting behavior of the symbol.  In fact, there are only three
rational parameters to fix, $d_1$, $d_2$ and $\gamma^{\prime\prime}$,
and they all enter the coefficient of the NNLLA imaginary part,
$g_0^{(3)}(w,\ws)$, given in \eqn{g3_0}.  Inspecting the MRK limit of
$R_6^{(3)}$, we find first of all perfect agreement with the 
functions $h_r^{(L)}(w,w^*)$ entering the real part.  (These can
be determined on general grounds using consistency between the
$2\to4$ and $3\to3$ MRK limits.)  We also agree
perfectly with the imaginary part coefficients $g_2^{(3)}$ at LLA
and $g_1^{(3)}$ at NLLA.

Finally, we find for the NNLLA coefficient $g_0^{(3)}$,
\be
\bsp
g_0^{(3)}(w,\ws) &\,= 
\rat{27}{8}\,L_5^+ + \rat{3}{4}\,L_{3,1,1}^+ - \rat{1}{2}\,L_3^+\,[L_1^+]^2
- \rat{15}{32}\,L_3^+\,[L_0^-]^2 - \rat{1}{8}\,L_1^+\,L_{2,1}^-\,L_0^-\\
&\, + \rat{3}{32}\,[L_0^-]^2\,[L_1^+]^3 + \rat{19}{384}\,L_1^+\,[L_0^-]^4
+ \rat{1}{30}\,[L_1^+]^5
+ \rat{1}{2}\,[L_1^+]^2\,\zeta_3 - \rat{3}{16}\,[L_0^-]^2\,\zeta_3\\
&\, + {5\pi^2\over 24}\,L_3^+
- \rat{\pi^2}{48}\,L_1^+\,[L_0^-]^2 - \rat{\pi^2}{18}\,[L_1^+]^3
- \rat{3}{4}\,\zeta_5 \,.\\
\esp
\label{g3_0_fixed}
\ee
Comparing this result with \eqn{g3_0} fixes the three previously
undetermined rational parameters, $d_1$, $d_2$, and $\gamma^{\prime\prime}$.
We find
\be
d_1 = \frac{1}{2} \,,\qquad d_2 = \frac{3}{32} \,,
\qquad \gamma'' = - \frac{5}{4} \,.
\label{d1d2gammappvalues}
\ee

These three parameters were also the only ambiguities in
the expression found in ref.~\cite{Dixon2012yy}
for the two-loop (NNLLA) impact factor $\Phi^{(2)}_{\textrm{Reg}}(\nu,n)$
defined in ref.~\cite{Fadin2011we}.  Inserting \eqn{d1d2gammappvalues}
into that expression, we obtain,
\be\bsp
\Phi^{(2)}_{\textrm{Reg}}(\nu,n) &\,=
{1\over2}\left[\Phi^{(1)}_{\textrm{Reg}}(\nu,n)\right]^2 - E^{(1)}_{\nu,n} \, E_{\nu,n}
+ \frac{1}{8}\,[D_\nu E_{\nu,n}]^2 + \frac{5}{64}\,N^2 \, ( N^2 + 4\,V^2) \\
& \hskip0.5cm \,  - \frac{\zeta_2}{4} \Bigl( 2\,E_{\nu,n}^2 + N^2 + 6\,V^2 \Bigr)
+ \frac{17}{4}\,\zeta_4 \,.
\label{PhiReg2}
\esp\ee
Here $\Phi^{(1)}_{\textrm{Reg}}$ is the one-loop (NLLA) impact factor, and
$E_{\nu,n}$ and $E^{(1)}_{\nu,n}$ are the LLA and NLLA BFKL
eigenvalues~\cite{Fadin2011we,Dixon2012yy}.  These functions all
are combinations of polygamma ($\psi$) functions and their derivatives,
plus accompanying rational terms in $\nu$ and $n$.  For example,
\be
E_{\nu,n} = \psi\left(1+i\nu+\frac{|n|}{2}\right)
        + \psi\left(1-i\nu+\frac{|n|}{2}\right) - 2\psi(1)
        - \frac{1}{2}\frac{|n|}{\nu^2+\frac{n^2}{4}} \,.
\ee
Additional rational dependence on $\nu$ and $n$ enters \eqn{PhiReg2}
via the combinations
\begin{equation}
V \equiv \frac{i \nu}{\nu^2+\frac{|n|^2}{4}}, \qquad
N \equiv  \frac{n}{\nu^2+\frac{|n|^2}{4}} \,.
\end{equation}

We recall that the NNLLA BFKL eigenvalue $E^{(2)}_{\nu,n}$ also has
been determined~\cite{Dixon2012yy}, up to nine rational parameters,
$a_i$, $i=0,1,2,\ldots,8$.  These parameters enter the NNLLA coefficient
function $g_1^{(4)}(w,w^*)$.  If the above exercise can be repeated
at four loops, then it will be possible to fix all of these parameters
in the same way, and obtain an unambiguous result for the
NNLLA approximation to the MRK limit.

Finally, we ask whether we could have determined all coefficients
from the collinear vanishing of $R_6^{(3)}$ and the MRK limit alone,
{\it i.e.}~without using the near-collinear information from BSV.
The answer is yes, if we assume extra purity and if we also take
the value of $\alpha_2$ from ref.~\cite{CaronHuot2011kk}.
After imposing collinear vanishing, we have two parameters left:
$\alpha_2$ and the coefficient of $\zeta_2 \, R_6^{(2)}$. 
We can fix the latter coefficient in terms of $\alpha_2$ using
the known NLLA coefficient $g_3^{(1)}$ in the MRK limit.  (The LLA
coefficient $g_3^{(2)}$ automatically comes out correct.)
Then we compare to the NNLLA coefficient $g_3^{(0)}$.  We find that
we can fix $d_2$ and $\gamma''$ to the values in \eqn{d1d2gammappvalues},
but that $\alpha_2$ is linked to $d_1$ by the equation,
\be
\alpha_2 = \frac{d_1}{8} + \frac{5}{32} \,.
\label{al2_vs_d1}
\ee
If we do take $\alpha_2$ from ref.~\cite{CaronHuot2011kk}, then
the near-collinear limit of our result for $R_6^{(3)}$ provides an
unambiguous test of BSV's approach at three loops,
through ${\cal O}(T^2)$.

\vfill\eject

%%%%%%%%%%%%%%%%%%%%%%%%%%%%%%%%%%%%%%%%%%%%%%%%%%%%%%%%%%%%%%%%%%%%

\section{Final formula for $R_6^{(3)}$ and its quantitative behavior}
\label{sec:Final}

Now that we have used the (near) collinear limits to fix all
undetermined constants in \eqn{R63fndecomp} for $R_6^{(3)}$, we can write 
an expression for the full function, either in terms of multiple 
polylogarithms or integral representations.  We absorb the $c_i r_i(u,v)$
terms in \eqn{R63fndecomp} into $\Rep$.  In total we have,
\be
R_6^{(3)}(u,v,w) = \Rep(u,v,w) + \Rep(v,w,u) + \Rep(w,u,v)
 + P_6(u,v,w) + \frac{413}{24} \, \zeta_6 + (\zeta_3)^2 \,.
\label{R63Final}
\ee
Expressions for $R_6^{(3)}$ in terms of multiple polylogarithms,
valid in Regions I and II, are too lengthy to present here, but they
are attached to this document in computer-readable format. 
To represent $\Rep$ as an integral, we make use of its extra purity and
similarity to $\Omegauvw$, writing a formula similar to \eqn{Om2firstint}:
\be
\Rep(u,v,w) = - \int^u_1 du_t\,
\frac{[\Rep^{u} + (u\leftrightarrow v)](u_t,v_t,w_t)}{u_t(u_t-1)} \,,
\label{RepRepu}
\ee
with $v_t$ and $w_t$ as defined in~\eqn{eq:y_w_contour}. 
Note that the function $Q_\phi$ in \eqn{Om2firstint} is given, via 
\eqn{QasminusUplusV}, as $-[\Omega^{(2),u}+ (u\leftrightarrow v)]$,
the analogous combination of coproduct components entering \eqn{RepRepu}.
The function $R_{\rm ep}^{u}$ is defined in appendix~\ref{sec:app_Rep}.

We may also define $R_6^{(3)}$ via the $\{5,1\}$ component of its coproduct,
which is easily constructed from the corresponding coproducts of $\Rep$ in
appendix~\ref{sec:app_Rep}, and of the product function $P_6$.
The general form of the $\{5,1\}$ component of the coproduct is,
\be
\bsp
\Delta_{5,1}\left(R_6^{(3)}\right) &= 
R_6^{(3),u}\otimes\ln u+R_6^{(3),v}\otimes\ln v+R_6^{(3),w}\otimes\ln w\\
&+R_6^{(3),1-u}\otimes\ln (1-u)+R_6^{(3),1-v}\otimes\ln (1-v)
+R_6^{(3),1-w}\otimes\ln(1-w)\\
&+R_6^{(3),y_u}\otimes\ln y_u+R_6^{(3),y_v}\otimes\ln y_v
+R_6^{(3),y_w}\otimes\ln y_w\,.
\esp
\label{R63coproddecomp}
\ee
Many of the elements are related to each other, {\it e.g.}~by the total
symmetry of $R_6^{(3)}$:
\be
\bsp
R_6^{(3),1-u} &= - R_6^{(3),u} \,, \qquad
R_6^{(3),1-v} = - R_6^{(3),v} \,, \qquad
R_6^{(3),1-w} = - R_6^{(3),w} \,, \\
R_6^{(3),v}(u,v,w) &= R_6^{(3),u}(v,w,u)\,, \qquad
R_6^{(3),w}(u,v,w) = R_6^{(3),u}(w,u,v)\,,\\
R_6^{(3),y_v}(u,v,w) &= R_6^{(3),y_u}(v,w,u)\,, \qquad
R_6^{(3),y_w}(u,v,w) = R_6^{(3),y_u}(w,u,v)\,.
\esp
\label{R63coprodsym}
\ee
The two independent functions may be written as,
\be
\bsp
R_6^{(3),y_u} &= \frac{1}{32} \biggl\{
- 4 \, \Bigl( H_1(u,v,w) + H_1(u,w,v) \Bigr) - 2 \, H_1(v,u,w)\\
&\hskip1.3cm\null
+ \frac{3}{2} \, \Bigl( J_1(u,v,w) + J_1(v,w,u) + J_1(w,u,v) \Bigr)\\
&\hskip1.3cm\null
- 4 \, \Bigl[ H_2^u + H_2^v + H_2^w
    + \frac{1}{2} \, \Bigl( \ln^2u + \ln^2v + \ln^2w \Bigr)
    - 9 \, \zeta_2 \Bigr] \, \PhiTilde(u,v,w) \biggr\} \,,
\esp
\label{R63_yu}
\ee
and
\be
R_6^{(3),u} = \frac{1}{32} \Bigl[ A(u,v,w) + A(u,w,v) \Bigr]\,,
\label{R63_u}
\ee
where
\be
\bsp
A &= 
% nonrational:
M_1(u,v,w) - M_1(w,u,v)
+ \frac{32}{3} \Bigl( \Qep(v,w,u) - \Qep(v,u,w) \Bigr)\\
&\hskip0.5cm\null
+ ( 4 \ln u - \ln v + \ln w ) \, \Omegauvw
+ ( \ln u + \ln v ) \, \Omegavwu\\
% pure u
&\hskip0.5cm\null
+ 24 H_{5}^u - 14  H_{4,1}^u + \frac{5}{2}  H_{3,2}^u + 42  H_{3,1,1}^u
+ \frac{13}{2}  H_{2,2,1}^u - 36  H_{2,1,1,1}^u
+ H_{2}^u  \Bigl[ - 5  H_{3}^u + \frac{1}{2}  H_{2,1}^u + 7  \zeta_3 \Bigr]\\
&\hskip0.5cm\null
+ \ln u \Bigl[ - 14  ( H_{4}^u - \zeta_4 ) + 19  H_{3,1}^u
     - \frac{57}{2} H_{2,1,1}^u + \frac{1}{4} (H_{2}^u)^2
     + \frac{7}{4} \zeta_2 H_{2}^u \Bigr]
\! + \! \frac{1}{2}  \ln^2u ( H_{3}^u - 12  H_{2,1}^u + 3 \zeta_3 )\\
&\hskip0.5cm\null
+ \frac{1}{4} \ln^3u \, ( H_{2}^u - \zeta_2 )
+ \zeta_2  \Bigl( \frac{33}{4} H_{3}^u + H_{2,1}^u \Bigr)
% pure v
- 2 H_{4,1}^v - \frac{5}{2} H_{3,2}^v + 30 H_{3,1,1}^v + \frac{19}{2} H_{2,2,1}^v
- 12 H_{2,1,1,1}^v\\
&\hskip0.5cm\null
+ H_{2}^v  \Bigl( H_{3}^v - \frac{9}{2}  H_{2,1}^v + \frac{9}{4} \zeta_2 \ln v
          - 7 \zeta_3 \Bigr)
+ \ln v \Bigl[ 2  H_{4}^v + 5  H_{3,1}^v - \frac{15}{2} H_{2,1,1}^v 
         - \frac{5}{4} (H_{2}^v)^2 + 6 \zeta_4 \Bigr] \\
&\hskip0.5cm\null
- \frac{1}{2} \ln^2v \, ( H_{3}^v + 4  H_{2,1}^v + 3  \zeta_3 )
- \frac{1}{4}  \ln^3v \, ( H_{2}^v - \zeta_2 )
- \frac{1}{4} \zeta_2 \, ( H_{3}^v - 28 H_{2,1}^v )\\
% Mixed u,v:
&\hskip0.5cm\null
+ \frac{1}{6} \Bigl( H_{2}^v + \frac{1}{2}  \ln^2v \Bigr)
   \Bigl( - 43 H_{3}^u + 41 H_{2,1}^u - 5  \ln u  H_{2}^u
          - \frac{21}{2}  \ln^3u \Bigr)\\
&\hskip0.5cm\null
+ \frac{1}{6} \Bigl( H_{2}^u + \frac{1}{2}  \ln^2u \Bigr)
  \Bigl( - 5  H_{3}^v - 17 H_{2,1}^v - 7 \ln v H_{2}^v 
         + \frac{3}{2} \ln^3v \Bigr)\\
&\hskip0.5cm\null
+ \ln u \Bigl[ 16 H_{4}^v - 4 H_{3,1}^v - 5 (H_{2}^v)^2
            - 6  \ln v ( 2 H_{3}^v - H_{2,1}^v )
            + 3  \ln^2v ( H_{2}^v - 2  \zeta_2 ) + 12 \zeta_2 H_{2}^v \Bigr] \\
&\hskip0.5cm\null
+ \frac{1}{2} \ln^2u \Bigl[ 4 ( H_{3}^v + H_{2,1}^v ) + \ln v H_{2}^v \Bigr]
+ \ln v  \Bigl[ 2 H_{3,1}^u - \frac{1}{2} (H_{2}^u)^2 + 2 \zeta_2 H_{2}^u \Bigr] 
+ 2 \ln^2v ( H_{2,1}^u + \ln u  H_{2}^u )\\
% Mixed v,w:
&\hskip0.5cm\null
+ \ln w \, \Bigl[ - 6 H_{3,1}^v - \frac{1}{2} (H_{2}^v)^2 
               - 2  \ln v  ( 2  H_{3}^v - H_{2,1}^v ) + \ln^2v  H_{2}^v
               - 2  \zeta_2  ( 3  H_{2}^v + \ln^2v ) \Bigr] \\
&\hskip0.5cm\null
+ \frac{1}{2} \ln^2w \, ( 4  H_{3}^v - \ln v  H_{2}^v )
+ \frac{1}{2} \Bigl( H_{2}^v + \frac{1}{2} \ln^2v \Bigr) 
   ( 8  H_{2,1}^w + 4  \ln w  H_{2}^w - \ln^3w )
% Mixed u,v,w:
- 4  \ln u  H_{2}^v  H_{2}^w\\
&\hskip0.5cm\null
- \ln u  \ln w \Bigl[ 4  H_{3}^v + 2  H_{2,1}^v
      + \frac{3}{2} \ln u  \Bigl( H_{2}^v + \frac{1}{2} \ln^2v \Bigr) \Bigr]\\
&\hskip0.5cm\null
+ \ln v  \ln w  \Bigl[ - 2  H_{2,1}^u - \frac{1}{2}  \ln v  H_{2}^u
             - 2  \ln u \Bigl( H_{2}^u + 2  H_{2}^v
                 + \frac{3}{8} \ln v  \ln w - 6  \zeta_2 \Bigr) \Bigr] \,.
\esp
\label{Aeqn}
\ee
Since the $\{5,1\}$ component of the coproduct specifies all the first derivatives
of $R_6^{(3)}$, \eqns{R63_yu}{R63_u} should be supplemented by the value
of $R_6^{(3)}$ at one point.  For example, the value at $(u,v,w)=(1,1,1)$ will
suffice (see below), or the constraint that it vanishes in all collinear limits.

In the remainder of this section, we use the multiple polylogarithmic
and integral representations to obtain
numerical values for $R_6^{(3)}$ for a variety of interesting contours
and surfaces within the positive octant of the $(u,v,w)$ space.  We also
obtain compact formulae for $R_6^{(3)}$ along specific lines through the space.

%%%%%%%%%%%%%%%%%%%%%%%%%%%%
\subsection{The line $(u,u,1)$}
\label{sec:uu1sec}

On the line $(u,u,1)$, the two- and three-loop remainder functions
can be expressed solely in terms of HPLs of a single argument, $1-u$.
The two-loop function is,
\be
R_6^{(2)}(u,u,1) = H_{4}^u - H_{3,1}^u + 3 \, H_{2,1,1}^u
 + H_{1}^u ( H_{3}^u - H_{2,1}^u ) - \frac{1}{2} \, (H_{2}^u)^2 - (\zeta_2)^2 \,,
\label{R62uu1}
\ee
while the three-loop function is,
\be
\bsp
R_6^{(3)}(u,u,1) &= - 3 \, H_{6}^u + 2 \, H_{5,1}^u - 9 \, H_{4,1,1}^u 
- 2 \, H_{3,2,1}^u + 6 \, H_{3,1,1,1}^u - 15 \, H_{2,1,1,1,1}^u\\
&\quad- \frac{1}{4} \, (H_{3}^u)^2 - \frac{1}{2} \, H_{3}^u \, H_{2,1}^u
+ \frac{3}{4} \, (H_{2,1}^u)^2 - \frac{5}{12} \, (H_{2}^u)^3
+ \frac{1}{2} \, H_{2}^u \, 
\Bigl[ 3 \, \bigl( H_{4}^u + H_{2,1,1}^u \bigr) + H_{3,1}^u \Bigr]\\
&\quad- H_{1}^u \, \bigl( 3 \, H_{5}^u - 2 \, H_{4,1}^u + 9 \, H_{3,1,1}^u
+ 2 \, H_{2,2,1}^u
            - 6 \, H_{2,1,1,1}^u - H_{2}^u \, H_{3}^u \bigr)\\
&\quad- \frac{1}{4} \, (H_{1}^u)^2 \, \Bigl[ 3 \, ( H_{4}^u + H_{2,1,1}^u )
- 5 \, H_{3,1}^u + \frac{1}{2} \, (H_{2}^u)^2 \Bigr]\\
&\quad- \zeta_2 \, \Bigl[ H_{4}^u + H_{3,1}^u + 3 \, H_{2,1,1}^u 
            + H_{1}^u \, ( H_{3}^u + H_{2,1}^u)
            - (H_{1}^u)^2 \, H_{2}^u - \frac{3}{2} \, (H_{2}^u)^2 \Bigr]\\
&\quad- \zeta_4 \, \Bigl[ (H_{1}^u)^2 + 2 \, H_{2}^u \Bigr]
+ \frac{413}{24} \, \zeta_6 + (\zeta_3)^2  \,.
\esp
\label{R63uu1}
\ee
Setting $u=1$ in the above formula leads to
\be
R_6^{(3)}(1,1,1) = \frac{413}{24} \, \zeta_6 + (\zeta_3)^2 \,.
\label{R63_111}
\ee
We remark that the four-loop cusp anomalous dimension in planar
${\cal N}=4$ SYM,
\be
\gamma_K^{(4)} = - \frac{219}{2} \, \zeta_6 - 4 (\zeta_3)^2 \,,
\ee
has a different value for the ratio of the $\zeta_6$ coefficient
to the $(\zeta_3)^2$ coefficient.

The value of the two-loop remainder function at this same point is
\be
R_6^{(2)}(1,1,1) = - (\zeta_2)^2 = - \frac{5}{2} \zeta_4 \,.
\label{R62_111}
\ee
The numerical value of the three-loop to two-loop ratio at the point
$(1,1,1)$ is:
\be
\frac{R_6^{(3)}(1,1,1)}{R_6^{(2)}(1,1,1)}\ =\ -7.004088513718\ldots.
\label{ratio_111}
\ee
We will see that over large swaths of the positive octant, the ratio 
$R_6^{(3)}/R_6^{(2)}$ does not stray too far from $-7$.

%%%%%%%%%%%%%%%%%%%%%%%%%%%%
\begin{figure}
\begin{center}
\includegraphics[width=4.9in]{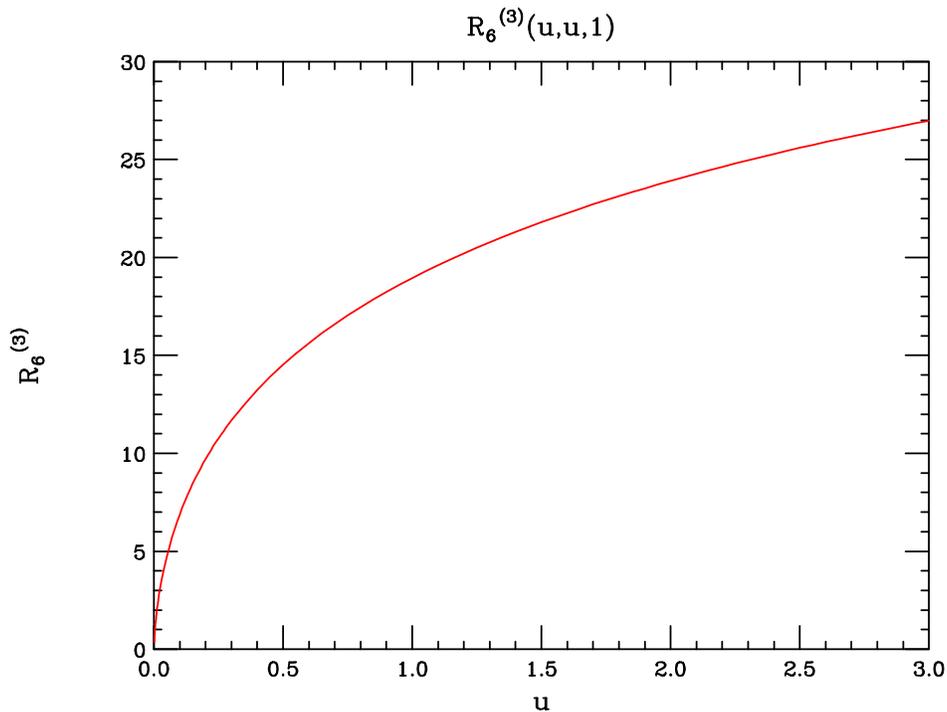}
\end{center}
\caption{$R_6^{(3)}(u,u,1)$ as a function of $u$.}
\label{fig:uu1}
\end{figure}
%%%%%%%%%%%%%%%%%%%%%%%%%%%%

We plot the function $R_6^{(3)}(u,u,1)$ in \fig{fig:uu1}.
We also give the leading term in the expansions of $R_6^{(2)}(u,u,1)$ and
$R_6^{(3)}(u,u,1)$ around $u=0$,
\be
\bsp
R_6^{(2)}(u,u,1) &= u \biggl[ -\frac{1}{2} \ln^2u + 2 \ln u + \zeta_2 - 3
 \biggr] + {\cal O}(u^2)\,,\\
R_6^{(3)}(u,u,1) &= u \biggl[ -\frac{1}{4} \ln^3u
 + \Bigl( \zeta_2 + \frac{9}{4} \Bigr) \ln^2u
 - \Bigl( \frac{5}{2} \zeta_2 + 9\Bigr) \ln u
 - \frac{11}{2} \zeta_4 - \zeta_3 + \frac{3}{2} \zeta_2 + 15 \biggr]\\
&\hskip0.5cm + {\cal O}(u^2)\,.
\label{R623uu1_smallu}
\esp
\ee
Hence the ratio $R_6^{(3)}/R_6^{(2)}$ diverges logarithmically as $u\to0$
along this line:
\be
\frac{R_6^{(3)}(u,u,1)}{R_6^{(2)}(u,u,1)}\ \sim\ \frac{1}{2} \ln u, \qquad
\hbox{as $u\to0$.}
\label{ratiouu1_smallu}
\ee
This limit captures a piece of the near-collinear limit $T\to0$,
the case in which $S\to0$ at the same rate, as discussed in 
\sect{sec:collinear} near \eqn{Om2uu1_smallu}.  The fact that
$R_6^{(3)}$ has one more power of $\ln u$ than does $R_6^{(2)}$
is partly from its extra leading power of $\ln T$
(the leading singularity behaves like $(\ln T)^{L-1}$),
but also from an extra $\ln S^2$ factor in a subleading $\ln T$ term.

\begin{figure}
\begin{center}
\includegraphics[width=4.9in]{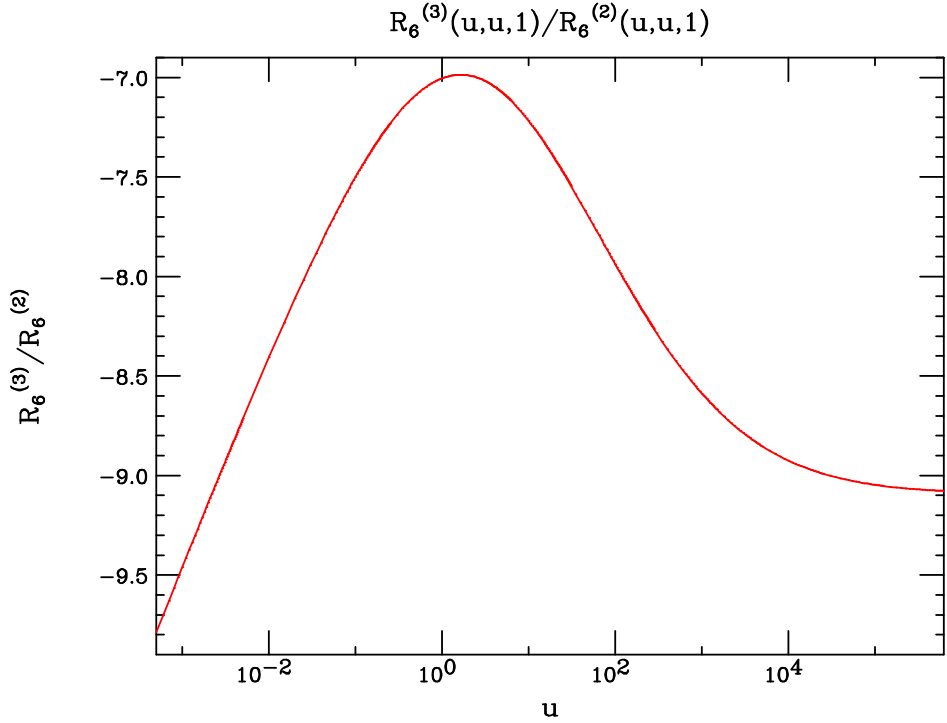}
\end{center}
\caption{$R_6^{(3)}/R_6^{(2)}$ on the line $(u,u,1)$.}
\label{fig:ratio_uu1}
\end{figure}

As $u\to\infty$, the leading behavior at two and three loops is,
\be
\bsp
R_6^{(2)}(u,u,1) &= -\frac{27}{4}\zeta_4
+ \frac{1}{u} \biggl[ \frac{1}{3} \ln^3u + \ln^2u + (\zeta_2 + 2) \ln u
 + \zeta_2 + 2 \biggr] + {\cal O}\left(\frac{1}{u^2}\right)\,,\\
R_6^{(3)}(u,u,1) &= \frac{6097}{96}\zeta_6 + \frac{5}{4} (\zeta_3)^2
+ \frac{1}{u} \biggl[ - \frac{1}{10} \ln^5u - \frac{1}{2} \ln^4 u
 - \frac{1}{3} ( 5 \zeta_2 + 6 ) \ln^3 u \\
&\hskip1cm\null
 + \Bigl( \frac{1}{2} \zeta_3 - 5 \zeta_2 - 6 \Bigr) \ln^2 u
 - \Bigl( \frac{141}{8} \zeta_4 - \zeta_3 + 10 \zeta_2 + 12 \Bigr) \ln u \\
&\hskip1cm\null
 - 2 \zeta_5 + 2 \zeta_2 \zeta_3 - \frac{141}{8} \zeta_4 + \zeta_3
 - 10\zeta_2 - 12
\biggr] + {\cal O}\left(\frac{1}{u^2}\right)\,.
\label{R623uu1_largeu}
\esp
\ee
As $u\to\infty$ along the line $(u,u,1)$, the two- and three-loop remainder
functions, and thus their ratio $R_6^{(3)}/R_6^{(2)}$, approach a constant.
For the ratio it is:
\be
\frac{R_6^{(3)}(u,u,1)}{R_6^{(2)}(u,u,1)}\ \sim\ 
- \biggl[ \frac{50}{3} \frac{(\zeta_3)^2}{\pi^4}
         + \frac{871}{972} \pi^2 \biggr] = - 9.09128803107\ldots,
\qquad \hbox{as $u\to\infty$.}
\label{ratiouu1_largeu}
\ee
We plot the ratio $R_6^{(3)}/R_6^{(2)}$ on the line $(u,u,1)$ in 
\fig{fig:ratio_uu1}.  The logarithmic scale for $u$ highlights
how little the ratio varies over a broad range in $u$.

%%%%%%%%%%%%%%%%%%%%%%%%%%%%
%%%%%%%%%%%%%%%%%%%%%%%%%%%%

The line $(u,u,1)$ is special in that the remainder function is
extra pure on it.  That is, applying the operator $u(1-u)\,d/du$
returns a pure function for $L=2,3$:
\be
\bsp
u(1-u) \frac{dR_6^{(2)}(u,u,1)}{du} &= H_{2,1}^u - H_{3}^u \,,\\
u(1-u) \frac{dR_6^{(3)}(u,u,1)}{du} &= 
3 H_{5}^u - 2 H_{4,1}^u + 9 H_{3,1,1}^u + 2 H_{2,2,1}^u - 6 H_{2,1,1,1}^u
- H_{2}^u H_{3}^u \\
&\hskip0.4cm\null
+ H_{1}^u \biggl[ \frac{3}{2} ( H_{4}^u + H_{2,1,1}^u ) - \frac{5}{2} H_{3,1}^u
               + \frac{1}{4} (H_{2}^u)^2 \biggr]
+ \zeta_2 \Bigl[ H_{3}^u + H_{2,1}^u - 2 H_{1}^u H_{2}^u \Bigr]\\
&\hskip0.4cm\null
+ 2 \zeta_4 H_{1}^u \,.
\label{R623uu1extrapure}
\esp
\ee
The extra-pure property is related to the fact that the asymptotic behavior
as $u\to\infty$ is merely a constant, with no $\ln u$ terms. Indeed,
if one applies $u(1-u)\,d/du$ to any positive power of $\ln u$,
the result diverges at large $u$ like $u$ times a power of $\ln u$,
which is not the limiting behavior of any combination of HPLs in $\mathcal{H}_u$.

%%%%%%%%%%%%%%%%%%%%%%%%%%%%%%%%%%%%%%%%%%%

\subsection{The line $(1,1,w)$}
\label{sec:11w_section}

We next consider the line $(1,1,w)$.  As was the case for the line $(u,u,1)$,
we can express the two- and three-loop remainder functions on the line $(1,1,w)$
solely in terms of HPLs of a single argument.  However, in contrast to
$(u,u,1)$, the expressions on the line $(1,1,w)$ are not extra-pure functions
of $w$.

The two-loop result is,
\be
\bsp
R_6^{(2)}(1,1,w) &= 
\frac{1}{2} \biggl[ H_{4}^w - H_{3,1}^w + 3 \, H_{2,1,1}^w
   - \frac{1}{4} (H_{2}^w)^2 + H_{1}^w ( H_{3}^w - 2 \, H_{2,1}^w )\\
&\hskip1.0cm\null
   + \frac{1}{2} ( H_{2}^w - \zeta_2 ) (H_{1}^w)^2 - 5 \zeta_4 \biggr] \,.
\esp
\label{R62_11w}
\ee
It is not extra pure on this line, because the quantity
\be
w(1-w) \frac{dR_6^{(2)}(1,1,w)}{dw} =
\frac{1}{4}(2-w) ( 2 \, H_{2,1}^w -  H_1^w \, H_2^w ) - \frac{1}{2} H_3^w
+ \frac{\zeta_2}{2} (1-w) H_1^w 
\label{R62_11w_impure}
\ee
contains explicit factors of $w$.

\begin{figure}
\begin{center}
\includegraphics[width=4.9in]{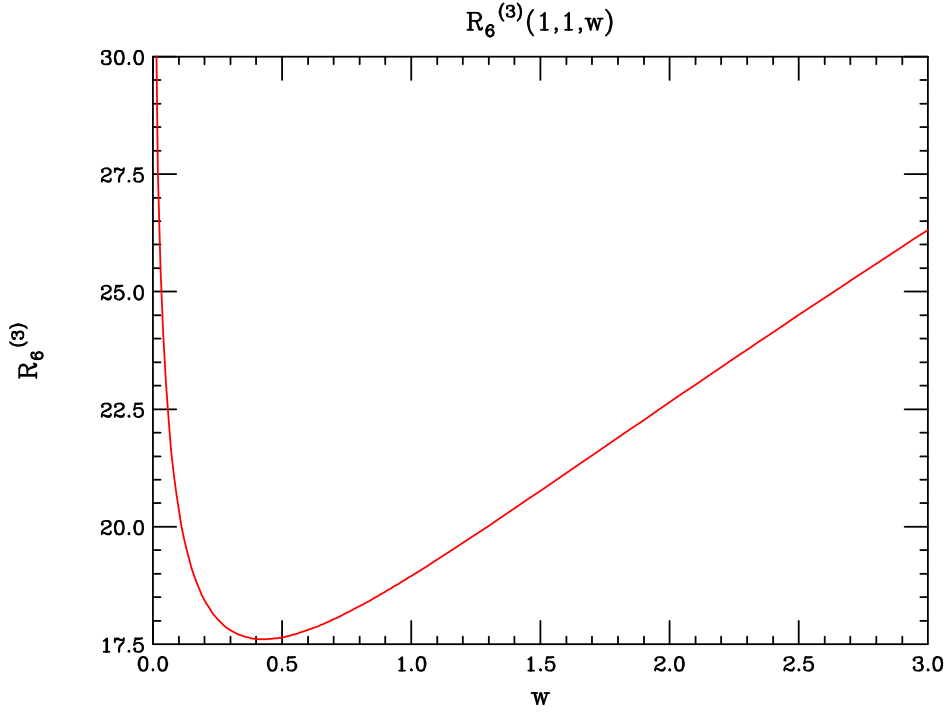}
\end{center}
\caption{$R_6^{(3)}(1,1,w)$ as a function of $w$.}
\label{fig:11w}
\end{figure}

The three-loop result is,
\be
\bsp
R_6^{(3)}(1,1,w) &= - \frac{3}{2} \, H_{6}^w + H_{5,1}^w - \frac{9}{2} \, H_{4,1,1}^w
 - H_{3,2,1}^w + 3 \, H_{3,1,1,1}^w - \frac{15}{2} \, H_{2,1,1,1,1}^w \\
&\hskip0.5cm\null
- \frac{1}{8} \, H_{3}^w \, ( H_{3}^w + 2 \, H_{2,1}^w )
+ \frac{3}{8} \, (H_{2,1}^w)^2
+ \frac{1}{2} \, H_{2}^w \, \Bigl( H_{4}^w + H_{3,1}^w
                             - \frac{1}{6} \, (H_{2}^w)^2 \Bigr)\\
&\hskip0.5cm\null
+ H_{1}^w \biggl[ - \frac{3}{2} \, H_{5}^w - \frac{1}{2} \, H_{3,2}^w
             - 3 \, H_{3,1,1}^w - \frac{1}{2} \, H_{2,2,1}^w
             + \frac{9}{2} \, H_{2,1,1,1}^w + \frac{1}{2} \, H_{2}^w \, H_{3}^w
             - \frac{1}{4} \, H_{2}^w \, H_{2,1}^w \\
&\hskip1.8cm\null
             + \frac{1}{8} \, H_{1}^w \, \Bigl( - 5 \, H_{4}^w + 5 \, H_{3,1}^w
                          - 9 \, H_{2,1,1}^w + (H_{2}^w)^2
                          - H_{1}^w \, ( H_{3}^w - H_{2,1}^w ) \Bigr) \biggr]\\
&\hskip0.5cm\null
- \frac{1}{2} \, \zeta_2 \Bigl[ H_{4}^w + H_{3,1}^w + 3 \, H_{2,1,1}^w - (H_{2}^w)^2 
       + H_{1}^w \, \Bigl( H_{3}^w - 2 \, H_{2,1}^w
                        + \frac{1}{2} \, H_{1}^w \, H_{2}^w \Bigr) \Bigl]\\
&\hskip0.5cm\null
+ \zeta_4 \Bigl[ - H_{2}^w + \frac{17}{8} \, (H_{1}^w)^2 \Bigr]
+ \frac{413}{24} \, \zeta_6 + (\zeta_3)^2 \,.
\esp
\label{R63_11w}
\ee
It is easy to check that it is also not extra pure.
We plot the function $R_6^{(3)}(1,1,w)$ in \fig{fig:11w}.

At small $w$, the two- and three-loop remainder functions diverge
logarithmically,
\be
\bsp
R_6^{(2)}(1,1,w) &= \frac{1}{2} \, \zeta_3 \, \ln w - \frac{15}{16} \, \zeta_4
 + {\cal O}(w)\,,\\
R_6^{(3)}(1,1,w) &= \frac{7}{32} \, \zeta_4 \, \ln^2 w
 - \Bigl( \frac{5}{2} \, \zeta_5 + \frac{3}{4} \, \zeta_2 \, \zeta_3 \Bigr)
   \, \ln w
 + \frac{77}{12} \, \zeta_6 + \frac{1}{2} \, (\zeta_3)^2 
 + {\cal O}(w)\,.
\label{R62311w_smallw}
\esp
\ee
At large $w$, they also diverge logarithmically,
\be
\bsp
R_6^{(2)}(1,1,w) &= - \frac{1}{96} \, \ln^4w
 - \frac{3}{8} \, \zeta_2 \, \ln^2w + \frac{\zeta_3}{2} \ln w 
 - \frac{69}{16} \, \zeta_4
 + {\cal O}\left(\frac{1}{w}\right)\,,\\
R_6^{(3)}(1,1,w) &= \frac{1}{960} \, \ln^6w + \frac{\zeta_2}{12} \, \ln^4w
 - \frac{\zeta_3}{8} \, \ln^3w + 5 \, \zeta_4 \, \ln^2w
 - \Bigl( \frac{13}{4} \, \zeta_5 + 2 \, \zeta_2 \, \zeta_3 \Bigr) \ln w\\
&\hskip0.5cm\null
 + \frac{1197}{32} \zeta_6 + \frac{9}{8} \, (\zeta_3)^2
 + {\cal O}\left(\frac{1}{w}\right)\,.
\label{R62311w_largew}
\esp
\ee
As discussed in the previous subsection, the lack of extra purity
on the line $(1,1,w)$ is related to the logarithmic divergence
in this asymptotic direction.

%%%%%%%%%%%%%%%%%%%%%%%%%%%%%%%%%%%%%%
%%%%%%%%%%%%%%%%%%%%%%%%%%%%%%%%%%%%%%

%%%%%%%%%%%%%%%%%%%%%%%%%%%%%%%%%

\subsection{The line $(u,u,u)$}
\label{sec:uuusec}

The symmetrical diagonal line $(u,u,u)$ has the nice feature that
the remainder function at strong coupling can be written analytically.
Using AdS/CFT to map the problem to a minimal area one, and applying
integrability, Alday, Gaiotto and Maldacena obtained
the strong-coupling result~\cite{Alday2009dv},
\be
R_6^{(\infty)}(u,u,u) = - \frac{\pi}{6} + \frac{\phi^2}{3\pi}
+ \frac{3}{8} \, \Bigl[ \ln^2 u + 2 \, {\rm Li}_2(1-u) \Bigr]
- \frac{\pi^2}{12} \,,
\label{R6strong}
\ee
where $\phi = 3 \, \cos^{-1}(1/\sqrt{4u})$.  The extra constant term
$-\pi^2/12$ is needed in order for $R_6^{(\infty)}(u,v,w)$ to vanish
properly in the collinear limits~\cite{PedroPrivate}.\footnote{We
thank Pedro Vieira for providing us with this constant.}

In perturbation theory, the function $R_6^{(L)}(u,v,w)$ is less simple
to represent on the line $(u,u,u)$ than on the lines $(u,u,1)$ and $(1,1,w)$.
It cannot be written solely in terms of HPLs with argument $(1-u)$.
At two loops, using \eqn{R62decomp}, the only obstruction is the
function $\Omega^{(2)}(u,u,u)$,
\be
\bsp
R_6^{(2)}(u,u,u) &= \frac{3}{4} \, \biggl[ \Omega^{(2)}(u,u,u)
    + 4 \, H_{4}^u - 2 \, H_{3,1}^u - 2 \, (H_{2}^u)^2
    + 2 \, H_{1}^u \, ( 2 \, H_{3}^u - H_{2,1}^u ) \\
&\hskip1.5cm\null
    - \frac{1}{4} \, (H_{1}^u)^4
    - \zeta_2 \Bigl( 2 \, H_{2}^u + (H_{1}^u)^2 \Bigr)
    + \frac{8}{3} \, \zeta_4 \biggr] \,.
\esp
\label{R62_uuu}
\ee

One way to proceed is to convert the first-order partial differential
equations for all the hexagon functions of $(u,v,w)$ into ordinary
differential equations in $u$ for the same functions evaluated
on the line $(u,u,u)$.  The differential equation for the
three-loop remainder function itself is,
\be
\bsp
\frac{d R_6^{(3)}(u,u,u)}{du} &=
 \frac{3}{32} \biggl\{ 
    \frac{1-u}{u\sqrt{\Delta}} 
  \Bigl[ - 10 H_1(u,u,u) + \frac{9}{2} J_1(u,u,u)
         - 4 \PhiTilde(u,u,u) \Bigl( 3 H_{2}^u + \frac{3}{2} (H_1^u)^2
                 - 9 \zeta_2 \Bigr) \Bigr] \\
&\hskip0.5cm\null
 + \frac{8}{u(1-u)} \biggl[ - \frac{3}{2} H_1^u \, \Omega^{(2)}(u,u,u)
   + 6 H_{5}^u - 4 H_{4,1}^u + 18 H_{3,1,1}^u + 4 H_{2,2,1}^u
   - 12 H_{2,1,1,1}^u \\
&\hskip2.8cm\null
 + H_{2}^u \, ( H_{2,1}^u - 3 H_{3}^u )
   - H_1^u \, \Bigl( H_{4}^u + 4 H_{3,1}^u - 9 H_{2,1,1}^u
   - \frac{11}{4} (H_{2}^u)^2 \Bigr) \\
&\hskip2.8cm\null
   + (H_1^u)^2 \, ( H_{2,1}^u - 5 H_{3}^u ) + (H_1^u)^3 \, H_{2}^u
                  + \frac{5}{8} \, (H_1^u)^5 \\
&\hskip2.8cm\null
   + \zeta_2 \, \Bigl( 2 H_{3}^u + 2 H_{2,1}^u - 3 H_1^u H_{2}^u 
            - (H_1^u)^3 \Bigr)
   - 5 \zeta_4 \, H_1^u \biggr] \biggr\} \,,
\esp
\label{d_du_R63_uuu}
\ee
with similar differential equations for $\Omega^{(2)}(u,u,u)$, 
$H_1(u,u,u)$ and $J_1(u,u,u)$.  Interestingly, the parity-even weight-five
functions $M_1$ and $\Qep$ do not enter \eqn{d_du_R63_uuu}.

We can solve the differential equations by using series expansions around
three points: $u=0$, $u=1$, and $u=\infty$.
If we take enough terms in each expansion (of order 30--40 terms suffices),
then the ranges of validity of the expansions will overlap.
At $u=1$, $\Delta$ vanishes, and so do all the parity-odd functions,
so we divide them by $\sqrt{\Delta}$ before series expanding in $(u-1)$.
These expansions, and those of the parity-even functions, are regular,
with no logarithmic coefficients, as expected for a point in the interior
of the positive octant.  (Indeed, we can perform an analogous 
three-dimensional series expansion of all hexagon functions of $(u,v,w)$
about $(1,1,1)$; this is actually a convenient way to fix the
beyond-the-symbol terms in the coproducts, by using consistency of the
mixed partial derivatives.)

At $u=0$, the series expansions also contain powers of $\ln u$ in their
coefficients.  At $u=\infty$, there are two types of terms in the generic
series expansion: a series expansion in $1/u$ with coefficients that
are powers of $\ln u$, and a series expansion in odd powers of $1/\sqrt{u}$
with an overall factor of $\pi^3$, and coefficients that can contain
powers of $\ln u$.  The square-root behavior can be traced back to
the appearance of factors of $\sqrt{\Delta(u,u,u)} = (1-u)\sqrt{1-4u}$
in the differential equations, such as \eqn{d_du_R63_uuu}.

The constants of integration are easy to determine at $u=1$
(where most of the hexagon function are defined to be zero).
They can be determined numerically (and sometimes analytically)
at $u=0$ and $u=\infty$, either by evaluating the multiple
polylogarithmic expressions, or by matching the series expansions
with the one around $u=1$.

At small $u$, the series expansions at two and three loops
have the following form:
\be
\bsp
R_6^{(2)}(u,u,u) &= 
\frac{3}{4} \, \zeta_2 \, \ln^2 u + \frac{17}{16} \, \zeta_4
 + \frac{3}{4} \, u \, \Bigl[ \ln^3u + \ln^2u + ( 5 \, \zeta_2 - 2 ) \, \ln u 
                            + 3 \, \zeta_2 - 6 \Bigr] + {\cal O}(u^2), \\
R_6^{(3)}(u,u,u) &= - \frac{63}{8} \, \zeta_4 \, \ln^2u
 - \frac{1691}{192} \, \zeta_6 + \frac{1}{4} \, (\zeta_3)^2 \\
&\hskip0.5cm\null
 + \frac{3}{16} \, u \, \Bigl[ \ln^5u + \ln^4u
         - 4 \, ( 3 \,\zeta_2 + 1 ) \, \ln^3u
         + 4 \, ( \zeta_3 - 2 \, \zeta_2 - 3 ) \, \ln^2u \\
&\hskip2.0cm\null
            - 2 \, ( 97 \, \zeta_4 - 4 \, \zeta_3 - 4 \, \zeta_2 - 12 ) \, \ln u
         - 60 \, \zeta_4 - 8 \, \zeta_3 + 120 \Bigr] + {\cal O}(u^2),
\esp
\label{R623_uuu_small_u}
\ee
while the strong-coupling result is,
\be
R_6^{(\infty)}(u,u,u) = \Bigl( \frac{3}{8} - \frac{3}{4\pi}\Bigr)\, \ln^2 u
 + \frac{\pi^2}{24} - \frac{\pi}{6}
 + u \, \Bigl[ \Bigl( \frac{3}{4} - \frac{3}{\pi} \Bigr) \, \ln u 
      - \frac{3}{4} \Bigr]
 + {\cal O}(u^2).
\label{R6strong_uuu_small_u}
\ee
Note that the leading term at three loops diverges logarithmically,
but only as $\ln^2u$.  Alday, Gaiotto and Maldacena~\cite{Alday2009dv}
observed that this property holds at two loops and at strong coupling,
and predicted that it should hold to all orders.

At large $u$, the two- and three-loop remainder functions behave as,
\be
\bsp
R_6^{(2)}(u,u,u) &= - \frac{5}{8} \zeta_4 
- \frac{3 \, \pi^3}{4 \, u^{1/2}}
+ \frac{1}{16 u} \Bigl[ 2 \, \ln^3u + 15 \, \ln^2u
       + 6 \, ( 6 \zeta_2 + 11 ) \, \ln u
       + 24 \zeta_3 + 126 \zeta_2 + 138 \Bigr]\\
&\hskip0.5cm\null 
- \frac{\pi^3}{32 \, u^{3/2}} + {\cal O}\left(\frac{1}{u^2}\right),\\
R_6^{(3)}(u,u,u) &= -\frac{29}{48}\zeta_6 + \zeta_3^2
+ \frac{3 \, \pi^5}{4 \, u^{1/2}}\\
&\hskip0.5cm
+ \frac{1}{32\,u} \Bigl[ - \frac{3}{10} \ln^5 u - \frac{15}{4} \ln^4 u
          - ( 22 \zeta_2 + 33 ) \ln^3 u 
          + ( 12 \zeta_3 - 159 \zeta_2 - 207 ) \ln^2 u \\
&\hskip2.0cm
          - ( 747 \zeta_4 - 48 \zeta_3 + 690 \zeta_2 + 846 ) \, \ln u
          - 96 \zeta_5 + 72 \zeta_2 \zeta_3 - \frac{4263}{2} \zeta_4 \\
&\hskip2.0cm
          + 96 \zeta_3 - 1434 \zeta_2 - 1710 \Bigr]
+ \frac{\pi^3}{32 \, u^{3/2}} \, ( - 36 \ln u + 6 \zeta_2 - 70 )
+ {\cal O}\left(\frac{1}{u^2}\right),
\esp
\label{R623_uuu_largeu}
\ee 
while the strong-coupling behavior is,
\be
R_6^{(\infty)}(u,u,u) =  - \frac{5\pi^2}{24} + \frac{7\pi}{12}
- \frac{3}{2 \, u^{1/2}}
+ \frac{3}{4u} \Bigl[ \ln u + 1 + \frac{1}{\pi} \Bigr]
- \frac{1}{16 \, u^{3/2}}
+ {\cal O}\left(\frac{1}{u^2}\right).
\label{R6strong_uuu_large_u}
\ee
%

%%%%%%%%%%%%%%%%%%%%%%%%%%%%%%%%%%%%%%%
\begin{figure}
\begin{center}
\includegraphics[width=4.8in]{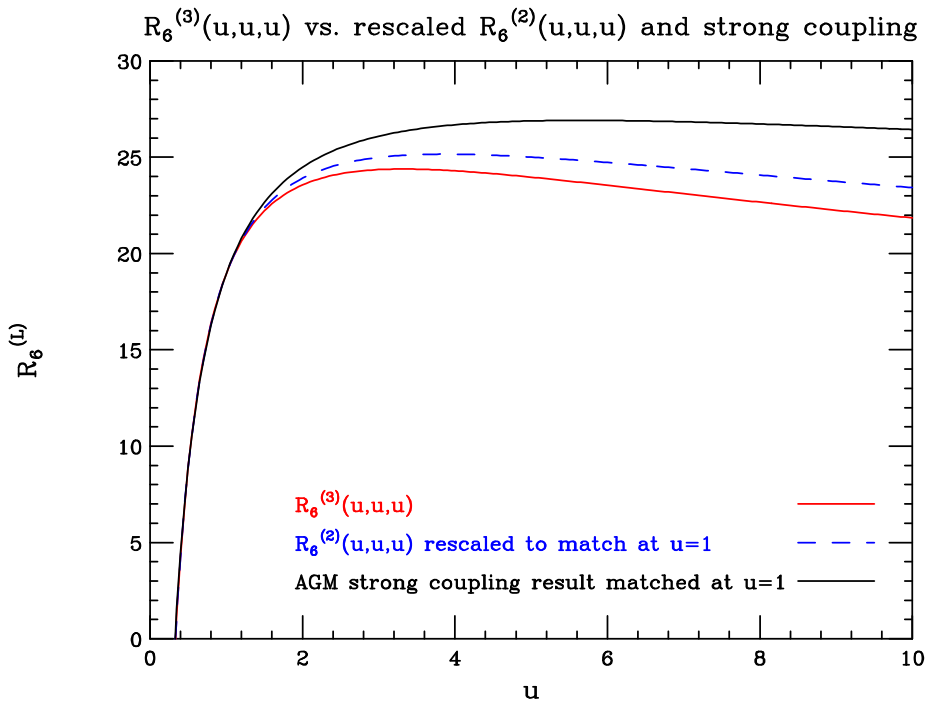}
\end{center}
\caption{Comparison between $R_6^{(2)}$, $R_6^{(3)}$, and the strong-coupling
result on the line $(u,u,u)$.}
\label{fig:uuu}
\end{figure}
%%%%%%%%%%%%%%%%%%%%%%%%%%%%%%%%%%%%%%%

In \fig{fig:uuu} we plot the two- and three-loop and strong-coupling
remainder functions on the line $(u,u,u)$.  In order to highlight
the remarkably similar shapes of the three functions for small and
moderate values
of $u$, we rescale $R_6^{(2)}$ by the constant factor~(\ref{ratio_111}), 
so that it matches $R_6^{(3)}$ at $u=1$.  We perform a similar rescaling
of the strong-coupling result, multiplying it by
\be
\frac{R_6^{(3)}(1,1,1)}{R_6^{(\infty)}(1,1,1)} = -63.4116164\ldots,
\label{strong3ratio}
\ee
where $R_6^{(\infty)}(1,1,1) = \pi/6 - \pi^2/12$.
A necessary condition for the shapes to be so similar is that the
limiting behavior of the ratios as $u\to0$ is almost the same
as the ratios' values at $u=1$.  From \eqn{R623_uuu_small_u},
the three-loop to two-loop ratio as $u\to0$ is, 
\be
\frac{R_6^{(3)}(u,u,u)}{R_6^{(2)}(u,u,u)}
\ \sim\ -\frac{21}{5} \, \zeta_2
\ =\ -6.908723\ldots, \qquad
\hbox{as $u\to0$,}
\label{ratiouuu_smallu}
\ee
which is within 1.5\% of the ratio at $(1,1,1)$, \eqn{ratio_111}.
The three-loop to strong-coupling ratio is,
\be
\frac{R_6^{(3)}(u,u,u)}{R_6^{(\infty)}(u,u,u)}
\ \sim\ -\frac{21}{1-2/\pi} \, \zeta_4
\ =\ -62.548224\ldots, \qquad
\hbox{as $u\to0$,}
\label{strongratiouuu_smallu}
\ee
which is again within 1.5\% of the corresponding
ratio~(\ref{strong3ratio}) at $u=1$.

The similarity of the perturbative and strong-coupling curves
for small and moderate $u$ suggests that if a smooth extrapolation
of the remainder function from weak to strong coupling can be achieved,
on the line $(u,u,u)$ it will have a form that is almost independent of $u$,
for $u<1$.

As mentioned in the introduction, the numerical similarity of two-loop
and strong-coupling remainder functions has been explored previously,
starting with two-dimensional kinematics at eight
points~\cite{Brandhuber2009da,DelDuca2010zp},
and later for the general $2n$-point case~\cite{Hatsuda}.
The rescaling of the remainder functions used to perform those comparisons
is similar to our rescaling in \fig{fig:uuu}.
For the six-point case studied in the present paper, ref.~\cite{Hatsuda2012pb}
has compared the two-loop and strong-coupling rescaled remainder
functions along a curve which runs from $(u,v,w)=(1/4,1/4,1/4)$ to $(1,0,0)$,
as well as analytically in the expansion around $(1/4,1/4,1/4)$
using conformal perturbation theory, and the results are very similar.
The curve runs from the ultraviolet to the infrared region
of the renormalization group flow associated with an integrable
two-dimensional system.
It would be very interesting to perform this comparison with the three-loop
remainder function as well, but we will reserve this exercise
for future work.

We return now to the line $(u,u,u)$.  Whereas all the curves in
\fig{fig:uuu} are very similar for $u<1$, they diverge from each
other at large $u$, although they each approach a constant value
as $u\to\infty$.
The three-to-two-loop ratio at very large $u$,
from \eqn{R623_uuu_largeu}, eventually approaches $-1.227\ldots$, which
is quite different from $-7$.  The three-loop-to-strong-coupling ratio
approaches $-3.713\ldots$, which is very different from $-63.4$.

On the line $(u,u,u)$, all three curves in \fig{fig:uuu}
cross zero very close to $u=1/3$.  The respective
zero crossing points for $L=2,3,\infty$ are:
\be
u_0^{(2)} = 0.33245163\ldots, \qquad u_0^{(3)} = 0.3342763\ldots,
\qquad u_0^{(\infty)} = 0.32737425\ldots.
\label{uuu_zero_crossing}
\ee
Might the zero crossings in perturbation theory somehow converge
to the strong-coupling value at large $L$?  We will return to the issue
of the sign of $R_6^{(L)}$ below.

Another way to examine the progression of perturbation theory,
and its possible extrapolation to strong coupling, is to
use the Wilson loop ratio adopted by BSV, which is related
to the remainder function by \eqn{eq:WL_convert}.
This relation holds for strong coupling as well as weak coupling,
since the cusp anomalous dimension is known exactly~\cite{Beisert2006ez}.
In the near-collinear limit, considering the Wilson loop ratio has the
advantage that the strong-coupling OPE behaves sensibly.  The remainder
function differs from this ratio by the one-loop function $X(u,v,w)$,
whose near-collinear limit does not resemble a strong-coupling OPE at all.
On the other hand, the Wilson loop ratio breaks all of the $S_3$
permutation symmetries of the remainder function.  This is not an
issue for the line $(u,u,u)$, since none of the $S_3$ symmetries
survive on this line.  However, there is also the issue that $X(u,u,u)$
as determined from \eqn{Xuvw} diverges logarithmically as $u\to1$.

\begin{figure}
\begin{center}
\includegraphics[width=4.7in]{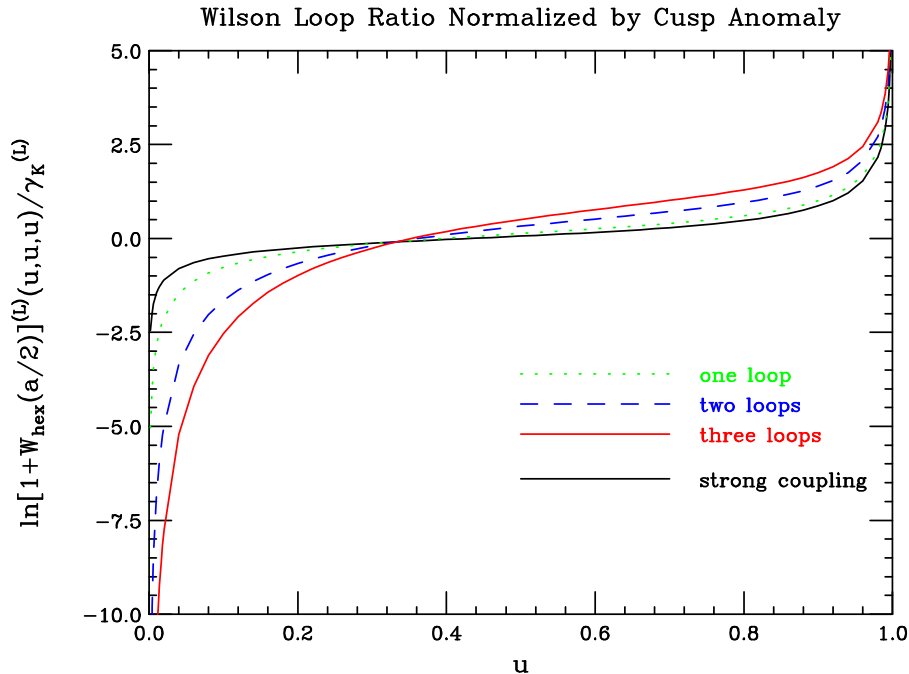}
\end{center}
\caption{Comparison between the Wilson loop ratio at one to three
loops, and the strong coupling value, evaluated on the line $(u,u,u)$.}
\label{fig:WLuuu}
\end{figure}

In \fig{fig:WLuuu} we plot the perturbative coefficients of
$\ln[ 1 + \mathcal{W}_{\textrm{hex}}(a/2) ]$, as well as the strong-coupling
value, restricting ourselves to the range $0 < u < 1$ where $X(u,u,u)$
remains real.  Now there is also a one-loop term, from multiplying $X(u,u,u)$
by the cusp anomalous dimension in \eqn{eq:WL_convert}.
We normalize the results in this case by dividing the coefficient
at a given loop order by the corresponding coefficient of the cusp
anomalous dimension, and similarly at strong coupling.
Equivalently, from \eqn{eq:WL_convert}, we plot
\be
\frac{R_6^{(L)}(u,u,u)}{\gamma_K^{(L)}} + \frac{1}{8} X(u,u,u),
\label{WLdiag}
\ee
for $L=1,2,3,\infty$.

The Wilson loop ratio diverges at both $u=0$ and $u=1$.  The divergence
at $u=1$ comes only from $X$ and is controlled by the cusp anomalous
dimension.  This forces the curves to converge in this region.
The $\ln^2 u$ divergence as $u\to0$ gets contributions
from both $X$ and $R_6$.  The latter contributions are not
proportional to the cusp anomalous dimensions, causing all the curves
to split apart at small $u$.
Because $X(u,u,u)$ crosses zero at $u=0.394\ldots$, which
is a bit different from the almost identical zero crossings
in \eqn{uuu_zero_crossing} and in \fig{fig:uuu},
the addition of $X$ in \fig{fig:WLuuu} splits the zero crossings
apart a little.
However, in the bulk of the range, the perturbative coefficients do
alternate in sign from one to three loops, following the sign alternation
of the cusp anomaly coefficients, and suggesting that a smooth extrapolation
from weak to strong coupling may be possible for this observable as well.

%%%%%%%%%%%%%%%%%%%%%%%%%%%%%%%%%%%%%%%
%%%%%%%%%%%%%%%%%%%%%%%%%%%%%%%%%%%%%%%

%%%%%%%%%%%%%%%%%%%%%%%%%%%%%%%%%%%%%%%%%%

\subsection{Planes of constant $w$}

\begin{figure}
\begin{center}
\includegraphics[width=6.8in]{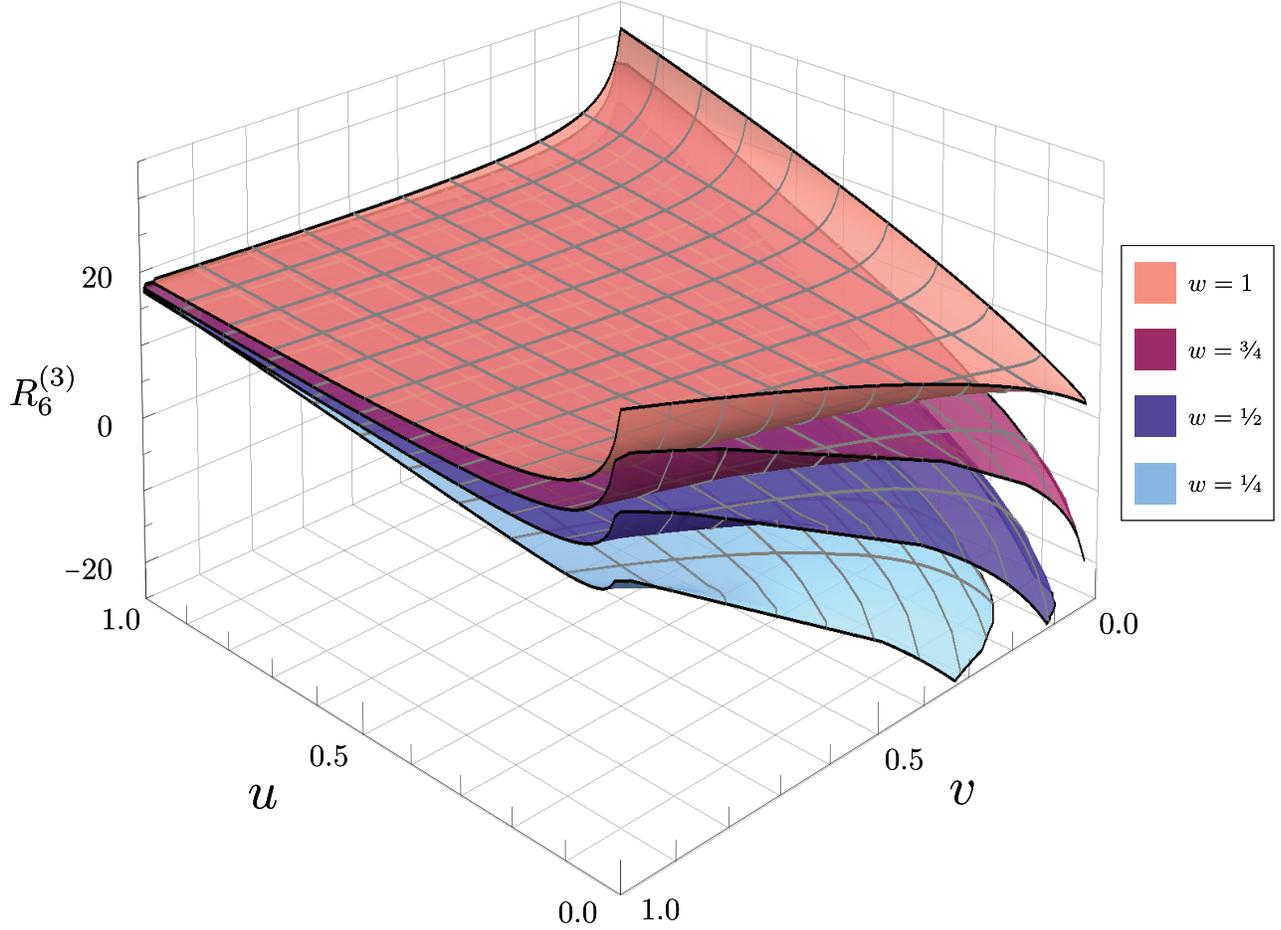}
\end{center}
\caption{The remainder function $R_6^{(3)}(u,v,w)$ on planes of constant $w$, 
plotted in $u$ and $v$. The top surface corresponds to $w=1$, while lower 
surfaces correspond to $w=\frac{3}{4}$, $w=\frac{1}{2}$ and $w=\frac{1}{4}$,
respectively.}
\label{fig:wstack}
\end{figure}

Having examined the remainder function on a few one-dimensional lines,
we now turn to its behavior on various two-dimensional surfaces. We will
now restrict our analysis to the unit cube, $0\leq u,v,w \leq 1$.
To provide a general picture of how the remainder function behaves
throughout this region, we show in~\fig{fig:wstack} the function evaluated
on planes with constant $w$, as a function of $u$ and $v$.  The plane $w=1$
is in pink, $w=\frac{3}{4}$ in purple,
$w=\frac{1}{2}$ in dark blue, and $w=\frac{1}{4}$ in light blue. 
The function goes to zero for the collinear-EMRK corner point
$(u,v,w)=(0,0,1)$ (the right corner of the top sheet).
Except for this point, $R_6^{(3)}(u,v,w)$ diverges as either
$u\to0$ or $v\to0$.  While the plot might suggest that
the function is monotonic in $w$ within the unit cube, our analytic
expression for the $(1,1,w)$ line in section~\ref{sec:11w_section},
and \fig{fig:11w}, shows that at the left corner, where $u=v=1$, the function
does turn over closer to $w=0$.  (In fact, while it cannot be seen clearly
from the plot, the $w=\frac{1}{4}$ surface actually intersects the
$w=\frac{1}{2}$ surface near this corner.)

%%%%%%%%%%%%%%%%%%%%
%%%%%%%%%%%%%%%%%%%

%%%%%%%%%%%%%%%%%%%%%%%%%%%%%%%%%%%%
\subsection{The plane $u+v-w=1$}

Next we consider the plane $u+v-w=1$. Its intersection with the unit cube
is the triangle bounded by the lines $(1,v,v)$ and $(w,1,w)$, which are
equivalent to the line
$(u,u,1)$ discussed in \sect{sec:uu1sec}, and by the collinear limit line
$(u,1-u,0)$, on which the remainder function vanishes.

\begin{figure}
\begin{center}
\includegraphics[width=4.7in]{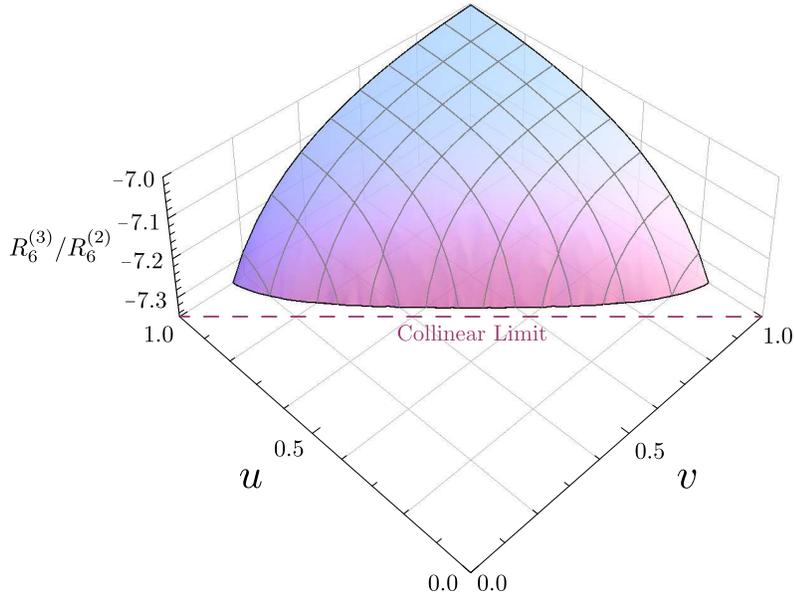}
\end{center}
\caption{The ratio $R_6^{(3)}(u,v,w)\!/\!R_6^{(2)}(u,v,w)$ on the plane
$u+v-w\!=\!1$, as a function of $u$ and $v$.}
\label{fig:ratioplane}
\end{figure}

In \fig{fig:ratioplane} we plot the ratio $R_6^{(3)}/R_6^{(2)}$ on this
triangle.  The back edges can be identified with the $u<1$
portion of \fig{fig:ratio_uu1}, although here they are plotted
on a linear scale rather than a logarithmic scale.
The plot is symmetrical under $u\leftrightarrow v$.
In the bulk of the triangle, the ratio does not stray far from $-7$.
The only place it deviates is in the approach to the collinear limit,
the front edge of the triangle corresponding to $T\to0$
in the notation of \sect{sec:collinear}.  Both $R_6^{(2)}$ and $R_6^{(3)}$
vanish like $T$ times powers of $\ln T$ as $T\to0$.  However,
because the leading singularity behaves like $(\ln T)^{L-1}$ at $L$ loops,
$R_6^{(3)}$ contains an extra power of $\ln T$ in its vanishing, and so the
ratio diverges like $\ln T$.  Otherwise, the shapes of the two functions
agree remarkably well on this triangle.

%%%%%%%%%%%%%%%%%%%%%%%
%%%%%%%%%%%%%%%%%%%%%%%

%%%%%%%%%%%%%%%%%%%%%%%%%%%%%%%%%%%%%%%
\subsection{The plane $u+v+w=1$}

\begin{figure}
\begin{center}
\includegraphics[width=5.5in]{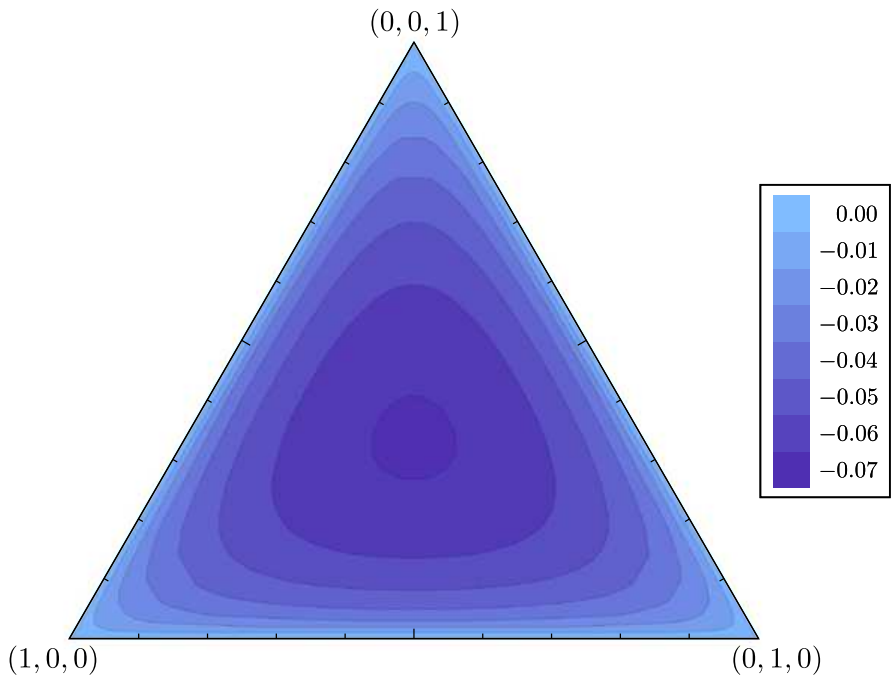}
\end{center}
\caption{Contour plot of $R_6^{(3)}(u,v,w)$ on the plane $u+v+w=1$ and
inside the unit cube.  The corners are labeled with their $(u,v,w)$ values.
Color indicates depth; each color corresponds to roughly a range of $0.01$.
The function must vanish at the edges, each of which corresponds to a 
collinear limit.  Its minimum is slightly under $-0.07$.}
\label{fig:tri}
\end{figure}

The plane $u+v+w=1$ intersects the unit cube along the three collinear
lines.  In \fig{fig:tri} we give a contour plot of $R_6^{(3)}(u,v,w)$
on the equilateral triangle lying between these lines.  The plot has the
full $S_3$ symmetry of the triangle under permutations of $(u,v,w)$.
Because $R_6^{(3)}$ has to vanish on the boundary, one might expect that it
should not get too large in the interior.  Indeed, its furthest deviation
from zero is slightly under $-0.07$, at the center of the triangle.

From the discussion in \sect{sec:uuusec} and \fig{fig:uuu},
we know that along the line $(u,u,u)$ the two- and three-loop remainder
functions almost always have the opposite sign.  The only place
they have the same sign on this line is for a very short interval
$u\in [0.3325, 0.3343]$ (see \eqn{uuu_zero_crossing}). This interval happens
to contain the point $(1/3,1/3,1/3)$, which is the intersection of the
line $(u,u,u)$ with the plane in \fig{fig:tri}, right at the
center of the triangle.  In fact, throughout the entire unit cube, the only
region where $R_6^{(2)}$ and $R_6^{(3)}$
have the same sign is a very thin pouch-like region surrounding this triangle.
In other words, the zero surfaces of $R_6^{(2)}$ and $R_6^{(3)}$
are close to the plane $u+v+w=1$, just slightly on opposite
sides of it in the two cases.
(We do not plot $R_6^{(2)}$ on the triangle here, but it is easy to verify
that it is also uniformly negative in the region of \fig{fig:tri}.
Its furthest deviation from zero is about $-0.0093$, again occurring
at the center of the triangle.)

%%%%%%%%%%%%%%%%%%%%%%%%%%%%%%%%%%%%%%%%
%%%%%%%%%%%%%%%%%%%%%%%%%%%%%%%%%%%%%%%%

\subsection{The plane $u=v$}

\begin{figure}
\begin{center}
\includegraphics[width=6.5in]{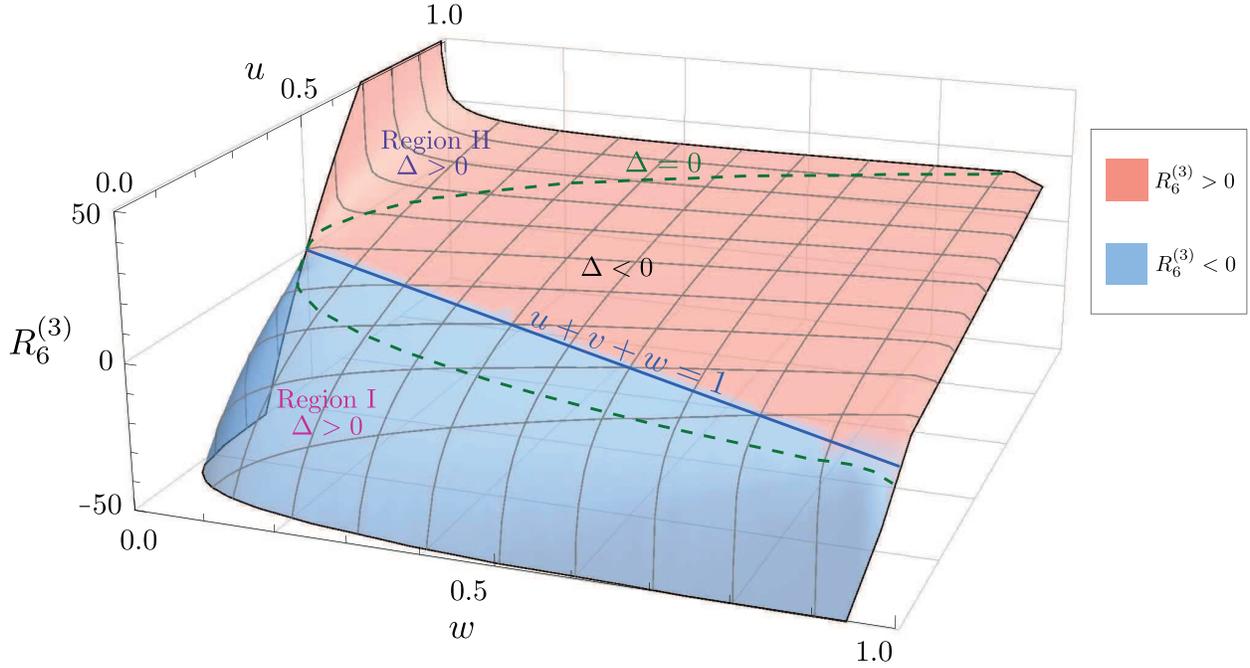}
\end{center}
\caption{Plot of $R_6^{(3)}(u,v,w)$ on the plane $u=v$, as a function
of $u$ and $w$. The region where $R_6^{(3)}$ is positive is shown in pink,
while the negative region is blue. The border between these two regions
almost coincides with the intersection with the $u+v+w=1$ plane,
indicated with a solid line. The dashed parabola shows the intersection
with the $\Delta=0$ surface; inside the parabola $\Delta<0$, while in the
top-left and bottom-left corners $\Delta>0$.}
\label{fig:uvplane}
\end{figure}

In \fig{fig:uvplane} we plot $R_6^{(3)}(u,v,w)$ on the plane $u=v$,
as a function of $u$ and $w$ inside the unit cube.  This plane crosses
the surface $\Delta=0$ on the curve $w=(1-2u)^2$, plotted as the dashed
parabola.  Hence it allows us to observe that $R_6^{(3)}$ is perfectly
continuous across the $\Delta=0$ surface.  We can also see that the
function diverges as $w$ goes to zero, and as $u$ and $v$ go to zero,
everywhere except for the two places that this plane intersects the
collinear limits, namely the points $(u,v,w)=(1/2,1/2,0)$ and
$(u,v,w)=(0,0,1)$.  The line of intersection of the $u=v$ plane and the
$u+v+w=1$ plane passes through both of these points, and \fig{fig:uvplane}
shows that $R_6^{(3)}$ is very close to zero on this line.

Based on considerations related to the positive
Grassmannian~\cite{ArkaniHamed2012nw}, it was recently
conjectured~\cite{AHCHTPrivate} that the three-loop remainder
function should have a uniform sign in the ``positive region'',
or what we call Region I: the portion of the unit cube
where $\Delta>0$ and $u+v+w<1$, which corresponds to positive external
kinematics in terms of momentum twistors.  On the surface $u=v$,
this region lies in front of the parabola shown in \fig{fig:uvplane}.
It was already checked~\cite{AHCHTPrivate} that the two-loop remainder
function has a uniform (positive) sign in Region I.
\Fig{fig:uvplane} illustrates that the uniform sign behavior (with a
negative sign) is indeed true at three loops on the plane $u=v$.
We have checked many other points with $u\neq v$ in Region I,
and $R_6^{(3)}$ was negative for every point we checked, so the conjecture
looks solid.

Furthermore, a uniform sign behavior for $R_6^{(2)}$ and $R_6^{(3)}$ also
holds in the other regions of the unit cube with $\Delta>0$, namely
Regions II, III, and IV, which are all equivalent under
$S_3$ transformations of the cross ratios. In these regions, the overall
signs are reversed:
$R_6^{(2)}$ is uniformly negative and $R_6^{(3)}$ is uniformly positive.
For the plane $u=v$, \fig{fig:uvplane} shows the uniform positive sign of
$R_6^{(3)}$ in Region II,
which lies behind the parabola in the upper-left portion of the figure.

Based on the two- and three-loop data, sign flips in $R_6^{(L)}$
only seem to occur where $\Delta<0$, and in fact very close to $u+v+w=1$.

%%%%%%%%%%%%%%%%%%
%%%%%%%%%%%%%%%%%%

%%%%%%%%%%%%%%%%%%%%%%%%%%%%%%%%%%%%%%%%%%%%
\subsection{The plane $u+v=1$}

\begin{figure}
\begin{center}
\includegraphics[width=4.5in]{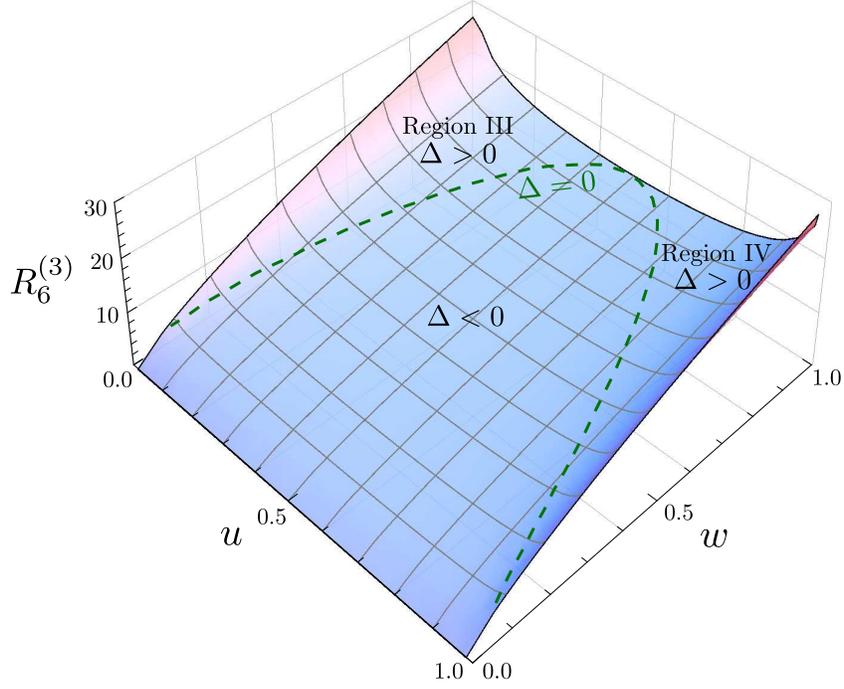}
\end{center}
\caption{$R_6^{(3)}(u,v,w)$ on the plane $u+v=1$, as a function of $u$ and $w$.}
\label{fig:upveq1}
\end{figure}

In \fig{fig:upveq1} we plot $R_6^{(3)}$ on the plane $u+v=1$.  This plane
provides information complementary to that on the plane $u=v$, since
the two planes intersect at right angles.  Like the $u=v$ plane, this
plane shows smooth behavior over the $\Delta=0$ surface, which intersects
the plane $u+v=1$ in the parabola $w=4u(1-u)$.  It also shows that the
function vanishes smoothly in the $w\to0$ collinear limit.

%%%%%%%%%%%%%%%%%%%%%%%%%%%
%%%%%%%%%%%%%%%%%%%%%%%%%%%

\vfill\eject

%%%%%%%%%%%%%%%%%%%%%%%%%%%%%%%%%%%%%%%%%%%%%%%%%%%%%%%%%%%%%%%%%%%%%

\section{Conclusions}
\label{sec:conclusions}

In this paper, we successfully applied a bootstrap, or set of consistency
conditions, in order to determine the three-loop remainder function
$R_6^{(3)}(u,v,w)$ directly from a few assumed analytic properties.  We
bypassed altogether the problem of constructing and integrating
multi-loop integrands.  This work represents the completion of a
program begun in ref.~\cite{Dixon2011pw}, in which the symbol
$\mathcal{S}(R_6^{(3)})$ was determined via a Wilson loop OPE and certain conditions
on the final entries, up to two undetermined rational numbers that were fixed soon
thereafter~\cite{CaronHuot2011kk}.

In order to promote the symbol to a function, we first had to characterize
the space of globally well-defined functions of three variables
with the correct analytic properties, which we call hexagon functions.
Hexagon functions are in one-to-one correspondence with the integrable
symbols whose entries are drawn from the nine letters $\{ u_i, 1-u_i, y_i\}$,
with the first entry restricted to $\{u_i\}$.  We specified the hexagon
functions at function level, iteratively in the transcendental weight,
by using their coproduct structure. In this approach, integrability of the
symbol is promoted to the function-level constraint of consistency of mixed
partial derivatives.  Additional constraints prevent
branch-cuts from appearing except at physical locations ($u_i=0,\infty$).
These requirements fix the beyond-the-symbol terms in the $\{n-1,1\}$
coproduct components of the hexagon functions, and hence they
fix the hexagon functions themselves (up to the arbitrary addition
of lower-weight functions multiplied by zeta values).
We found explicit representations of all the hexagon functions
through weight five, and of $R_6^{(3)}$ itself at weight six, in terms of
multiple polylogarithms whose arguments involve simple combinations of the
$y$ variables.
We also used the coproduct structure to obtain systems of coupled
first-order partial differential equations, which could be integrated
numerically at generic values of the cross ratios, or solved analytically
in various limiting kinematic regions.

Using our understanding of the space of hexagon functions, we constructed
an ansatz for the function $R_6^{(3)}$ containing 11 rational numbers,
free parameters multiplying lower-transcendentality hexagon functions.
The vanishing of $R_6^{(3)}$ in the collinear limits fixed all but one of
these parameters. The last parameter was fixed using the near-collinear limits,
in particular the $T^1 \ln T$ terms which we obtained from the
OPE and integrability-based predictions of Basso, Sever and
Vieira~\cite{Basso2013vsa}.  (The $T^1 \ln^0 T$ terms are also needed
to fix the last symbol-level parameter~\cite{Basso2013vsa}
independently of ref.~\cite{CaronHuot2011kk}.)

With all parameters fixed, we could unambiguously extract further terms
in the near-collinear limit.  We find perfect agreement with Basso,
Sever and Vieira's results through order $T^2$~\cite{BSVPrivate}.
We have also evaluated the remainder function in the multi-Regge limit.
This limit provides additional consistency checks, and allows us to fix
three undetermined parameters in an
expression~\cite{Dixon2012yy} for the NNLLA impact parameter
$\Phi^{(2)}_{\textrm{Reg}}(\nu,n)$ in the BFKL-factorized form of the
remainder function~\cite{Fadin2011we}.

Finally, we found simpler analytic representations for $R_6^{(3)}$
along particular lines in the three-dimensional $(u,v,w)$ space;
we plotted the function along these and other lines, and on some
two-dimensional surfaces within the unit cube $0\leq u_i \leq 1$.
Throughout much of the unit cube, and sometimes much further
out from the origin, we found the approximate numerical relation
$R_6^{(3)} \approx -7 \, R_6^{(2)}$.  The relation has only been observed
to break down badly in regions where the functions vanish:
the collinear limits, and very near the plane $u+v+w=1$.
On the diagonal line $(u,u,u)$, we observed that
the two-loop, three-loop, and strong-coupling~\cite{Alday2009dv}
remainder functions are almost indistinguishable in shape for $0<u<1$.

We have verified numerically a conjecture~\cite{AHCHTPrivate} that
the remainder function should have a uniform sign in the ``positive''
region $\{u,v,w > 0; \Delta > 0; u+v+w < 1\}$.  It also appears to have
an (opposite) uniform sign in the complementary region
$\{u,v,w > 0; \Delta > 0; u+v+w > 1\}$.  The only zero-crossings we
have found for either $R_6^{(2)}$ or $R_6^{(3)}$ in the positive octant
are very close to the plane $u+v+w=1$, in a region where $\Delta$ is
negative.

Our work opens up a number of avenues for further research.
A straightforward application is to the NMHV ratio function.
Knowledge of the complete space of hexagon functions through weight five
allowed us to construct the six-point MHV remainder function at three loops.
The components of the three-loop six-point NMHV ratio function are also
expected~\cite{Dixon2011nj} to be weight-six hexagon functions.
Therefore they should be constructible just as $R_6^{(3)}$ was, provided that
enough physical information can be supplied to fix all the free parameters.

It is also straightforward in principle to push the remainder function
to higher loops.  At four loops, the symbol of the remainder function
was heavily constrained~\cite{Dixon2012yy} by the same information
used at three loops~\cite{Dixon2011pw}, but of order 100 free
parameters were left undetermined.  With the knowledge of the
near-collinear limits provided by Basso, Sever and
Vieira~\cite{Basso2013vsa,BSVPrivate}, those parameters can now all be
fixed.  Indeed, all the function-level ambiguities can be fixed as
well~\cite{DDDPToAppear}.  This progress will allow many of the
numerical observations made in this paper at three loops, to be
explored at four loops in the near future.  Going beyond four loops
may also be feasible, depending primarily on computational issues ---
and provided that no inconsistencies arise related to failure of an
underlying assumption.

It is remarkable that scattering amplitudes in planar $\mathcal{N}=4$
super-Yang-Mills --- polygonal (super) Wilson loops --- are so heavily
constrained by symmetries and other analytic properties, that a full
bootstrap at the integrated level is practical, at least in
perturbation theory. We have demonstrated this practicality explicitly
for the six-point MHV remainder function.  The number of cross ratios increases
linearly with the number of points.  More importantly, the number of
letters in the symbol grows quite rapidly, even at two
loops~\cite{CaronHuot2011ky}, increasing the complexity of the problem.
However, with enough understanding of the relevant transcendental
functions for more external legs~\cite{Golden2013xva,Golden2013lha},
it may still be possible to implement a similar procedure in these
cases as well.  In the longer term, the existence of near-collinear
boundary conditions, for which there is now a fully nonperturbative
bootstrap based on the OPE and integrability~\cite{Basso2013vsa},
should inspire the search for a fully nonperturbative formulation that
also penetrates the interior of the kinematical phase space for
particle scattering.
\\ \\ \\
{\bf Acknowledgments}
\\ \\
We thank Johannes Henn for contributions made at an early stage of this
research.  We are indebted to Benjamin Basso, Amit Sever and Pedro Vieira
for sharing their results on the near-collinear limit with us prior to
publication, and for other useful discussions and comments
on the manuscript.
We are also grateful to Nima Arkani-Hamed, Simon Caron-Huot,
John Joseph Carrasco, Claude Duhr, Sasha Goncharov, Diego Hofman,
Yuji Satoh, Jaroslav Trnka and Cristian Vergu for helpful discussions.
LD, MvH and JP thank the Perimeter Institute for hospitality during
the completion of this article.
This research was supported by the US Department of Energy under
contracts DE--AC02--76SF00515 and DE--FG02--92ER40697, and in part by
Perimeter Institute for Theoretical Physics.  Research at Perimeter Institute
is supported by the Government of Canada through Industry Canada and by
the Province of Ontario through the Ministry of Economic Development and
Innovation.

\vfill\eject

%%%%%%%%%%%%%%%%%%%%%%%%%%%%%%%%%%%%%%%%%%%%%%%%%%%%%%%%%%%%%%%%%%%%%%%

\appendix

\section{Multiple polylogarithms and the coproduct}
\label{sec:app_multi_poly}

\subsection{Multiple polylogarithms}
\label{sec:app_multi_poly1}

Multiple polylogarithms are a general class of multi-variable iterated 
integrals, of which logarithms, polylogarithms, harmonic polylogarithms, 
and various other iterated integrals are special cases. They are defined 
recursively by $G(z)=1$, and,
\be
\label{eq:G_def}
G(a_1,\ldots,a_n; z) = \int_0^z\; \frac{dt}{t-a_1}\,G(a_2,\ldots,a_n;t)\,, 
\quad\quad G(\vec{0}_p; z) = \frac{\ln^p z}{p!}\, ,
\ee
where we have introduced the vector notation 
$\vec{a}_n = (\underbrace{a,\ldots,a}_{n})$.

For special values of the weight vector $(a_1,\ldots,a_n)$, multiple 
polylogarithms reduce to simpler functions. For example, if $a\neq 0$,
\be
G(\vec{0}_{p-1},a;z) = -\textrm{Li}_p(z/a)\,, 
\quad\quad G(\vec{0}_p, \vec{a}_q; z) = (-1)^q S_{p,q}(z/a)\, ,
\ee
where $S_{p,q}$ is the Nielsen polylogarithm. More generally, if
$a_i \in \{-1,0,1\}$, then
\be
G(a_1,\ldots,a_n; z) = (-1)^{w_1}\, H_{a_1,\ldots,a_n}(z)\,,
\ee
where $w_1$ is the number of $a_i$ equal to one.

Multiple polylogarithms are not all algebraically independent. 
One set of relations, known as the \emph{shuffle relations}, derive
from the definition~(\ref{eq:G_def}) in terms of iterated integrals,
\be
\label{eq:G_shuffle}
G(w_1;z)\,G(w_2;z) \;= \sum_{{w}\in{w_1}\ssha {w_2}}G(w;z)\,,
\ee
where ${w_1}\sha{w_2}$ is the set of mergers of the sequences $w_1$
and $w_2$ that preserve their relative ordering. Radford's
theorem~\cite{Radford1979} allows one to solve all of the
identities~(\ref{eq:G_shuffle}) simultaneously in terms of a
restricted subset of multiple polylogarithms $\{G(l_w;z)\}$, where
$l_w$ is a \emph{Lyndon word}. The Lyndon words are those words $w$
such that for every decomposition into two words $w=\{u,v\}$, the left
word is smaller (based on some ordering) than the right, {\it i.e.}~$u<v$.

One may choose whichever ordering is convenient; for our purposes, we
choose an ordering so that zero is smallest. In this case, no zeros
appear on the right of a weight vector, except in the special case of
the logarithm, $G(0;z) = \ln z$. Therefore, we may adopt a Lyndon
basis and assume without loss of generality that $a_n \neq 0$ in
$G(a_1,\ldots, a_n, z)$. Referring to \eqn{eq:G_def}, it is then
possible to rescale all integration variables by a common factor and
obtain the following identity,
\be
G(c\,a_1,\ldots, c\,a_n; c\,z) = G(a_1,\ldots,a_n;z) \,,
\quad\quad\quad a_n \neq 0, \; c\neq 0\, .
\ee

Specializing to the case $c=1/z$, we see that the algebra of multiple
polylogarithms is spanned by $\ln z$ and $G(a_1,\ldots,a_n;1)$ where
$a_n\neq 0$. This observation allows us to establish a one-to-one
correspondence between multiple polylogarithms and particular multiple
nested sums, provided those sums converge. In particular, if for
$|x_i|<1$ we define,
\be
\textrm{Li}_{m_1,\ldots,m_k}(x_1,\ldots,x_k) = 
\sum_{n_1 < n_2 < \cdots < n_k} \frac{x_1^{n_1} x_2^{n_2}
\cdots x_k^{n_k}}{n_1^{m_1} n_2^{m_2} \cdots n_k^{m_k}}\, ,
\ee
then,
\be
\label{eq:LieqG}
\textrm{Li}_{m_1,\ldots,m_k}(x_1,\ldots,x_k) 
= (-1)^k\,G\big(\underbrace{0,\ldots,0}_{m_k-1} , \frac{1}{x_k} , \ldots, 
\underbrace{0,\ldots,0}_{m_1 - 1},\frac{1}{x_1\cdots x_k} ; 1\big)\, .
\ee
\Eqn{eq:LieqG} is easily established by expanding the measure
$dt/(t-a_i)$ in~\eqn{eq:G_def} in a series and
integrating. Furthermore, a convergent series expansion for
$G(a_1,\ldots,a_n;z)$ exists if $|z| \le |a_i|$ for all $i$;
otherwise, the integral representation gives the proper analytic
continuation.

The relation to multiple sums endows the space of multiple
polylogarithms with some additional structure. In particular, the
freedom to change summation variables in the multiple sums allows one
to establish \emph{stuffle} or \emph{quasi-shuffle} relations,
\be
\textrm{Li}_{\vec{m}_1}(\vec{x}) \textrm{Li}_{\vec{m}_2}(\vec{y}) 
 = \sum_{\vec{n}} \textrm{Li}_{\vec{n}}(\vec{z}) \, .
\ee
The precise formula for $\vec{n}$ and $\vec{z}$ in terms of
$\vec{m}_1$, $\vec{m}_2$, $\vec{x}$, and $\vec{y}$ is rather
cumbersome, but can be written explicitly; see, {\it e.g.},
ref.~\cite{Borwein1999js}. For small depth, however, the stuffle
relations are quite simple. For example,
\be
\textrm{Li}_{a}(x) \textrm{Li}_{b}(y) =\textrm{Li}_{a,b}(x,y) 
+ \textrm{Li}_{b,a}(y,x) + \textrm{Li}_{a+b}(x y)\, .
\ee
Beyond the shuffle and stuffle identities, there are additional
relations between multiple polylogarithms with transformed arguments
and weight vectors. For example, one such class of identities follows
from H\"{o}lder convolution~\cite{Borwein1999js},
\be
G(a_1,\ldots,a_n;1) = \sum_{k=0}^n \,(-1)^k\, 
G\left(1-a_k,\ldots,1-a_1;1-\frac{1}{p}\right)
G\left(a_{k+1},\ldots,a_n;\frac{1}{p}\right)\,,
\ee
which is valid for any nonzero $p$ whenever $a_1\neq 1$ and $a_n\neq 0$.

One way to study identities among multiple polylogarithms is via the
symbol, which is defined recursively as,
\begin{align}
\label{eq:G_symb_app}
\mathcal{S}\big(G(a_{n-1},\ldots,a_1;a_n)\big) 
= \sum_{i=1}^{n-1} \biggl[&\mathcal{S}\big(G(a_{n-1},\ldots,\hat{a}_i,\ldots,a_1;a_n)\big)
\otimes\left(a_{i}-a_{i+1}\right) \notag\\
-& \mathcal{S}\big(G(a_{n-1},\ldots,\hat{a}_i,\ldots,a_1;a_n)\big)
\otimes (a_{i}-a_{i-1}) \biggr]
\,,
\end{align}
While the symbol has the nice property that all relations result from
simple algebraic manipulations, it has the drawback that its kernel
contains all transcendental constants. To obtain information about
these constants, one needs some more powerful machinery.

\subsection{The Hopf algebra of multiple polylogarithms}
\label{sec:app_multi_poly2}

When equipped with the shuffle product~(\ref{eq:G_shuffle}), the space
of multiple polylogarithms forms an algebra, graded by weight.
In ref.~\cite{Gonch3}, it was shown how to further equip
the space with a coproduct so that it forms a bialgebra, and, moreover, with an antipode so that it forms a
Hopf algebra. The weight of the multiple polylogarithms also defines a
grading on the Hopf algebra. In the following we will let
$\mathcal{A}$ denote the Hopf algebra and $\mathcal{A}_n$ the
weight-$n$ subspace, so that,
\be
\mathcal{A} = \bigoplus_{n=0}^{\infty} \mathcal{A}_n\,.
\ee
The coproduct is defined most naturally on a slight variant of~\eqn{eq:G_def},
\be
\label{eq:I_def}
I(a_0;a_1,\ldots ,a_n;a_{n+1}) 
= \int_{a_0}^{a_{n+1}}\frac{dt}{t-a_n} I(a_0;a_1,\ldots, a_{n-1};t) \, .
\ee
The two definitions differ only in the
ordering of indices and the choice of basepoint. However, as shown
in ref.~\cite{Duhr2012fh}, it is possible to reexpress any multiple polylogarithm
with a generic basepoint as a sum of terms with basepoint zero. This
manipulation is trivial at weight one, where we have,
\be
I(a_0;a_1;a_2) = I(0;a_1;a_2) - I(0;a_1;a_0) = G(a_1;a_2) - G(a_1;a_0)\,.
\ee
To build up further such relations at higher weights, one must
simply apply the lower-weight identity to the integrand in
\eqn{eq:I_def}. In this way, it is easy to convert between the
two different notations for multiple polylogarithms.

The coproduct on multiple polylogarithms is given by~\cite{Gonch3},
\be
\Delta(I(a_0;a_1,\ldots ,a_n;a_{n+1}) ) 
= \sum_{0<i_1<\cdots<i_k =n}I(a_0;a_{i_1},\ldots ,a_{i_k};a_{n+1}) 
\otimes \Big[\prod_{p=0}^{k} I(a_{i_p};a_{i_p+1},\ldots,a_{i_{p+1}-1};a_{i_{p+1}})\Big]\,.
\label{eq:coprI}
\ee 
Strictly speaking, this definition is only valid when the $a_i$ are
nonzero and distinct; otherwise, one must introduce a regulator to
avoid divergent integrals. We refer the reader to
refs.~\cite{Gonch3,Duhr2012fh} for these technical details.

It is straightforward to check a number of important properties of the
coproduct. First, it respects the grading of $\mathcal{A}$ in the
following sense. If $G_n \in \mathcal{A}_n$, then,
\be
\Delta(G_n) = \sum_{p+q=n} \Delta_{p,q}(G_n)\, ,
\ee
where $\Delta_{p,q}\in \mathcal{A}_p\otimes\mathcal{A}_q$.  Next, if
we extend multiplication to tensor products so that it acts on each
component separately,
\be
(a_1\otimes a_2)\cdot (b_1 \otimes b_2) = (a_1\cdot b_1)\otimes(a_2\cdot b_2)\,,
\ee
one can verify the compatibility of the product and the coproduct,
\be
\Delta(a \cdot b) = \Delta(a)\cdot \Delta(b)\, .
\ee
Finally, the coproduct is coassociative, 
\be
(\textrm{id}\otimes\Delta)\Delta = (\Delta\otimes \textrm{id})\Delta \,,
\ee
meaning that one may iterate the coproduct in any order and always reach a 
unique result.

This last property allows one to unambiguously define components of
the coproduct corresponding to all integer compositions of the
weight. Consider $G_n\in \mathcal{A}_n$ and a particular integer
composition of $n$, $\{w_1,\ldots,w_k\}$, such that $w_i>0$ and
$\sum_{i=1}^k w_i = n$. The component of the coproduct corresponding
to this composition, $\Delta_{w_1,\ldots,w_k}(G_n)$, is defined as the
unique element of the $(k-1)$-fold iterated coproduct in the space
$\mathcal{A}_{w_1}\otimes\cdots\otimes\mathcal{A}_{w_k}$. For our
purposes it is sufficient to consider $k=2$, although other components
have been useful in other contexts.

Consider the weight-$n$ function $f^{(n)}(z_1,\ldots,z_m)$ of $m$
complex variables $z_1,\ldots,z_m$ with symbol,
\be
\mathcal{S}(f^{(n)}) = \sum_{i_1,\ldots,i_n}\, c_{i_1,\ldots,i_n} \phi_{i_1} 
\otimes\cdots\otimes\phi_{i_n}\,.
\ee
The monodromy of $f^{(n)}$ around the point $z_k=z_0$ is encoded by
the first entry of the symbol,
\be
\mathcal{S}\big(\mathcal{M}_{z_k=z_0} f^{(n)}\big) 
=  \sum_{i_1,\ldots,i_n}\,  \mathcal{M}_{z_k=z_0}(\ln \phi_{i_1}\big)
\,c_{i_1,\ldots,i_n}\, \phi_{i_2}\otimes\cdots\otimes\phi_{i_n}\,,
\ee
where $\mathcal{M}_{z_k=z_0}(\ln \phi_{i_1}\big)$ is defined in \eqn{eq:del1n1},
and we have ignored higher powers of $(2\pi i)$ (see \sect{sec:MRK}).
Similarly, derivatives act on the last entry of the symbol,
\be
\mathcal{S}\Big(\frac{\partial}{\partial z_k} f^{(n)}\Big) 
=  \sum_{i_1,\ldots,i_n}\, c_{i_1,\ldots,i_n}
\, \phi_{i_1}\otimes\cdots\otimes\phi_{i_{n-1}}
\,\Big(\frac{\partial}{\partial z_k} \ln \phi_{n}\Big)\, .
\ee
In the same way, the monodromy operator acts only on the first
component of the coproduct and the derivative operator only on the
last component,
\be
\bsp
\Delta\left(\mathcal{M}_{z_k=z_0} f^{(n)}\right) 
&= \left(\mathcal{M}_{z_k=z_0} \otimes \textrm{id}\right)\, \Delta(f^{(n)})\,,\\
\Delta\left(\frac{\partial}{\partial z_k} f^{(n)}\right) 
&= \left(\textrm{id}\otimes\frac{\partial}{\partial z_k}\right)
\,\Delta(f^{(n)})\,.
\esp
\ee

One may trivially extend the definition of the coproduct to include
odd $\zeta$ values,
\be
\Delta(\zeta_{2n+1}) = 1\otimes \zeta_{2n+1} + \zeta_{2n+1}\otimes 1\,
\ee
but including even $\zeta$ values and factors of $\pi$ is more subtle. 
It was argued in ref.~\cite{Brown2011ik,Duhr2012fh} that it is consistent
to define,
\be
\label{eq:even_zeta}
\Delta(\zeta_{2n}) = \zeta_{2n}\otimes 1
\,\quad\quad~\textrm{and}~\quad\quad \Delta(\pi) = \pi\otimes 1\, .
\ee
\Eqn{eq:even_zeta} implies that powers of $\pi$ are absent from all 
factors of the coproduct except for the first one. Finally, we remark that the symbol may be recovered from the 
maximally-iterated coproduct if we drop all factors of $\pi$,
\be
\mathcal{S} \equiv \Delta_{1,\ldots,1} \; \textrm{mod}\; \pi \, .
\ee

\vfill\eject

%%%%%%%%%%%%%%%%%%%%%%%%%%%%%%%%%%%%%%%%%%%%%%%%%%%%%%%%%%%%%

\section{Complete basis of hexagon functions through weight five}
\label{sec:app_basis}

We present the basis of hexagon functions through weight five by providing
their $\{n-1,1\}$ coproduct components. For a hexagon function $F$ of weight 
$n$, we write, 
\be
\Delta_{n-1,1}(F) \equiv \sum_{i=1}^3 F^{u_i} \otimes \ln u_i 
+ F^{1-u_i} \otimes \ln (1-u_i) + F^{y_i} \otimes \ln y_i\, ,
\ee
where the nine functions $\{F^{u_i},F^{1-u_i}, F^{y_i}\}$ are of
weight $n-1$ and completely specify the $\{n-1,1\}$ component of the
coproduct. They also specify all of the first derivatives of $F$,
\be
\label{eq:der_F}
\bsp
\frac{\partial F}{\partial u}\bigg|_{v,w} &= 
\frac{F^u}{u} -\frac{F^{1-u}}{1-u} + \frac{1-u-v-w}{u\sqrt{\Delta}} F^{y_u}
+ \frac{1-u-v+w}{(1-u)\sqrt{\Delta}}F^{y_v}
+ \frac{1-u+v-w}{(1-u)\sqrt{\Delta}} F^{y_w}\,,\\
\sqrt{\Delta} \, y_u \frac{\partial F}{\partial y_u}\bigg|_{y_v,y_w}
&= \, (1-u)(1-v-w) F^u-u(1-v) F^v-u(1-w)F^w - u(1-v-w)F^{1-u} \\
&\quad + uv\,F^{1-v}+uw\,F^{1-w} + \sqrt{\Delta}\, F^{y_u}\, .
\esp
\ee
The other derivatives can be obtained from the cyclic images of \eqn{eq:der_F}. 
These derivatives, in turn, define integral representations for the function.
Generically, we define the function $F$ by (see~\eqn{yv_int_rep}),
\be
F(u,v,w) =  F(1,1,1) - \sqrt{\Delta}\int^u_1\frac{du_t}{v_t[u(1-w)+(w-u)u_t]}
\frac{\partial F}{\partial\ln y_v}(u_t,v_t,w_t) \,,
\ee
where,
\be
\label{eq:y_v_contour_app}
v_t = 1-\frac{(1-v)u_t(1-u_t)}{u(1-w)+(w-u)u_t} \,, \qquad\qquad
w_t = \frac{(1-u)wu_t}{u(1-w)+(w-u)u_t} \,.
\ee
We choose $F(1,1,1)=0$ for all functions except for the special case
$\Omega^{(2)}(1,1,1)=-6\zeta_4$. Other integral representations of the function
also exist, as discussed in~\sect{sec:general_setup}.

We remark that the hexagon functions
$\PhiTilde$, $G$, $N$ and $O$ are totally symmetric under
exchange of all three arguments; $\Omega^{(2)}$ is symmetric under exchange
of its first two arguments; $F_1$ is symmetric under exchange of its
last two arguments; and $H_1$, $J_1$ and $K_1$ are symmetric under exchange
of their first and third arguments.
 
%%%%%%%%%%%%%%%%%%%%%%%%

\subsection{$\PhiTilde$}

The only parity-odd hexagon function of weight three is $\PhiTilde$.
We may write the $\{2,1\}$ component of its coproduct as,
\be
\bsp
\Delta_{2,1}\bigl(\PhiTilde\bigr) &= 
\PhiTilde^{u}\otimes\ln u + \PhiTilde^{v}\otimes\ln v 
+ \PhiTilde^{w}\otimes\ln w\\
& + \PhiTilde^{1-u}\otimes\ln (1-u) + \PhiTilde^{1-v}\otimes\ln (1-v)
+ \PhiTilde^{1-w}\otimes\ln (1-w)\\
& + \PhiTilde^{y_u}\otimes\ln y_u + \PhiTilde^{y_v}\otimes\ln y_v
  + \PhiTilde^{y_w}\otimes\ln y_w\,,
\esp
\ee
where
\be
\PhiTilde^{u} = \PhiTilde^{v} = \PhiTilde^{w}
= \PhiTilde^{1-u} = \PhiTilde^{1-v} = \PhiTilde^{1-w}=0\,.
\ee
Furthermore, $\PhiTilde$ is totally symmetric, which implies,
\be
\PhiTilde^{y_v} = \PhiTilde^{y_u}(v,w,u)\,, 
\quad~\textrm{and}~\quad \PhiTilde^{y_w} = \PhiTilde^{y_u}(w,u,v)\,.
\ee
The one independent function, $\PhiTilde^{y_u}$, may be identified with a 
finite, four-dimensional one-loop hexagon integral, $\Omega^{(1)}$, which is 
parity even and of weight two,
\be
\PhiTilde^{y_u} = -\Omega^{(1)}(v,w,u)
= -H_2^u - H_2^v - H_2^w - \ln{v}\, \ln{w} + 2\,\zeta_2\,.
\ee

%%%%%%%%%%%%%%%%%%%%%

\subsection{$\Omega^{(2)}$}

Up to cyclic permutations, the only non-HPL parity-even hexagon function
of weight three is $\Omega^{(2)}$. We may write the $\{3,1\}$ component of
its coproduct as,
\be
\bsp
\Delta_{3,1}\left(\Omega^{(2)}\right) &= 
\Omega^{(2), u}\otimes\ln u + \Omega^{(2), v}\otimes\ln v
+ \Omega^{(2), w}\otimes\ln w \\
&+ \Omega^{(2), 1-u}\otimes\ln(1-u) + \Omega^{(2), 1-v}\otimes\ln(1-v)
+ \Omega^{(2), 1-w}\otimes\ln(1-w)\\
&+ \Omega^{(2), y_u}\otimes\ln y_u + \Omega^{(2), y_v}\otimes\ln y_v
+ \Omega^{(2), y_w}\otimes\ln y_w\,,
\esp
\ee
where the vanishing components are 
\be
\Omega^{(2), w} = \Omega^{(2), 1-w} = \Omega^{(2), y_w} = 0,
\ee
and the nonvanishing components obey,
\be
\Omega^{(2), v} = -\Omega^{(2), 1-v} 
=  -\Omega^{(2), 1-u}(u\leftrightarrow v) =  \Omega^{(2), u}(u\leftrightarrow v)
\quad~\textrm{and}~\quad  \Omega^{(2), y_v} = \Omega^{(2), y_u}\,.
\ee
The two independent functions are
\be
\Omega^{(2),y_u} = -\frac{1}{2}\PhiTilde \,,
\ee 
and
\be
\bsp
\Omega^{(2),u} &= H_3^{u} + H_{2,1}^v - H_{2,1}^w 
- \frac{1}{2} \ln(u w/v) \big(H_2^u + H_2^w - 2\,\zeta_2\big) 
+\frac{1}{2}\,\ln(uv/w)\,H_2^v\\
&\quad + \frac{1}{2} \ln{u}\,\ln{v}\,\ln(v/w)\, .\\
\esp
\ee

%%%%%%%%%%%%%%%%%%%%%

\subsection{$F_1$}

Up to cyclic permutations, the only parity-odd function of weight four 
is $F_1$. We may write the $\{4,1\}$ component of its coproduct as,
\be
\bsp
\Delta_{3,1}\left(F_1\right) &= 
F_1^{u}\otimes\ln u+F_1^{v}\otimes\ln v+F_1^{w}\otimes\ln w\\
&+F_1^{1-u}\otimes\ln (1-u)+F_1^{1-v}\otimes\ln (1-v)+F_1^{1-w}\otimes\ln (1-w)\\
&+F_1^{y_u}\otimes\ln y_u+F_1^{y_v}\otimes\ln y_v+F_1^{y_w}\otimes\ln y_w\,,
\esp
\ee
where
\be
F_1^{y_w} = F_1^{y_v}(v\leftrightarrow w) 
\quad~\textrm{and}~\quad F_1^u=F_1^v=F_1^w =F_1^{1-v}=F_1^{1-w}=0\, .
\ee
Of the three independent functions, one is parity odd,
$F_1^{1-u} = \PhiTilde$, and two are parity even,
\be
F_1^{y_u} = -2H_3^u + 2\,\zeta_3
\ee
and
\be
F_1^{y_v} = -2 H_3^u - 2 H_{2,1}^w 
+ \ln{w}\, \Big(H_2^u - H_2^v - H_2^w + 2\,\zeta_2\Big) 
+2\,\zeta_3\, .
\ee

In ref.~\cite{Dixon2011nj} the pure function entering the parity-odd part
of the six-point NMHV ratio function was determined to be
\be 
\tilde{V} = \frac{1}{8} ( \tilde{V}_X + \tilde{f} ) \,,
\label{Vtilde}
\ee
where $\tilde{V}_X + \tilde{f}$ satisfied an integral of the
form~(\ref{yv_int_rep}) with
\be
\bsp
\frac{\partial(\tilde{V}_X + \tilde{f})}{\partial \ln y_v} &= \tilde{Z}(u,v,w)\\
&= 2 \Bigl[ H_3^u - H_{2,1}^u - \ln u \, \Bigl( H_2^u + H_2^v - 2\zeta_2
                                 - \frac{1}{2} \ln^2 w \Bigr) \Bigr] - (u\lr w) \,.
\label{Ztilde}
\esp
\ee
This integral can be expressed in terms of $F_1$ and $\PhiTilde$ as,
\be
\tilde{V}_X + \tilde{f} 
= - F_1(u,v,w) +  F_1(w,u,v) + \ln(u/w) \, \tilde\Phi_6(u,v,w) \,.
\label{FromF1}
\ee

%%%%%%%%%%%%%%%%%%%

\subsection{$G$}

The $\{4,1\}$ component of the coproduct of the parity-odd weight five
function $G$ can be written as,
\be
\bsp
\Delta_{4,1}\left(G\right) &= G^{u}\otimes\ln u+G^{v}\otimes\ln v
+G^{w}\otimes\ln w\\
&+G^{1-u}\otimes\ln (1-u)+G^{1-v}\otimes\ln (1-v)+G^{1-w}\otimes\ln (1-w)\\
&+G^{y_u}\otimes\ln y_u+G^{y_v}\otimes\ln y_v+G^{y_w}\otimes\ln y_w\,,
\esp
\ee
where
\be
G^u=G^v=G^w=G^{1-u}=G^{1-v}=G^{1-w}=0\,.
\ee
Furthermore, $G$ is totally symmetric. In particular,
\be
G^{y_v}(u,v,w)= G^{y_u}(v,w,u)\,,\quad~\textrm{and}~\quad 
G^{y_w}(u,v,w)= G^{y_u}(w,u,v)\, .
\ee
Therefore, it suffices to specify the single independent function, $G^{y_u}$,
\be
\bsp
G^{y_u} &= - 2\,\big(H_{3,1}^u + H_{3,1}^v + H_{3,1}^w - \ln{w}\,H_{2,1}^v 
- \ln{v}\,H_{2,1}^w\big)
+ \frac{1}{2}\Big(H_2^u + H_2^v + H_2^w + \ln{v}\, \ln{w}\Big)^2\\
&\quad - \frac{1}{2}\,\ln^2{v}\ln^2{w} - 4\, \zeta_4\, .
\esp
\ee

%%%%%%%%%%%%%%%%%%%%%%

\subsection{$H_1$}

The function $\Huvw$ is parity-odd and has weight five. 
We may write the $\{4,1\}$ component of its coproduct as,
\be
\bsp
\Delta_{4,1}\left(\Huvw\right) &= 
\hat{H}_1^{u}\otimes\ln u+\hat{H}_1^{v}\otimes\ln v+\hat{H}_1^{w}\otimes\ln w\\
&+\hat{H}_1^{1-u}\otimes\ln (1-u)+\hat{H}_1^{1-v}\otimes\ln (1-v)
+\hat{H}_1^{1-w}\otimes\ln (1-w)\\
&+\hat{H}_1^{y_u}\otimes\ln y_u+\hat{H}_1^{y_v}\otimes\ln y_v
+\hat{H}_1^{y_w}\otimes\ln y_w\,,
\esp
\ee
where we put a hat on $\hat{H}_1^{u}$, {\it etc.}, to avoid confusion with 
the HPLs with argument $1-u$.  The independent functions are 
$\hat{H}_1^{u}$, $\hat{H}_1^{y_u}$, and $\hat{H}_1^{y_v}$,
\be
\bsp
\hat{H}_1^u &= 
-\frac{1}{4}\Big(\Fuvw -\ln{u}\,\PhiTilde \Big) - (u \leftrightarrow w)\, ,\\
\hat{H}_1^{y_u} &=\bigg[\frac{1}{2}\big(\Omegavwu+\Omegawuv\big) 
+\frac{1}{2}\big(H_4^u+H_4^v\big)-\frac{1}{2}\big(H_{3,1}^u-H_{3,1}^v\big)\\
&\quad-\frac{3}{2}\big(H_{2,1,1}^u+H_{2,1,1}^v\big)-\Big(\ln{u} 
+ \frac{1}{2} \ln(w/v)\Big)H_3^u-\frac{1}{2}\ln{v}\,H_3^v 
- \frac{1}{2}\ln(w/v)H_{2,1}^u\\
&\quad-\frac{1}{2}\ln{v}\,H_{2,1}^v-\frac{1}{4}\big((H_2^u)^2+(H_2^v)^2\big)
+\frac{1}{4}\Big(\ln^2{u}-\ln^2(w/v)\Big)H_2^u\\
&\quad-\frac{1}{8}\ln^2{u}\ln^2(w/v)-\zeta_2\Big(H_2^u+\frac{1}{2}\ln^2{u}\Big)
+3\zeta_4\bigg] + (u \leftrightarrow w)\, ,\\
\hat{H}_1^{y_v} &= \Omegawuv\, .
\esp
\ee
Of the remaining functions, two vanish, $\hat{H}_1^{v}=\hat{H}_1^{1-v}=0$, 
and the others are simply related,
\be
\hat{H}_1^{1-u}=\hat{H}_1^{w}=-\hat{H}_1^{1-w}=-\hat{H}_1^{u},
\quad\quad\textrm{and}\quad\quad \hat{H}_1^{y_w} = \hat{H}_1^{y_u} \,.
\ee

%%%%%%%%%%%%%%%%%%%%%%%%%

\subsection{$J_1$}

We may write the $\{4,1\}$ component of the coproduct of
the parity-odd weight-five function $\Juvw$ as,
\be
\bsp
\Delta_{4,1}\left(\Juvw\right) &= 
J_1^{u}\otimes\ln u+J_1^{v}\otimes\ln v+J_1^{w}\otimes\ln w\\
&+J_1^{1-u}\otimes\ln (1-u)+J_1^{1-v}\otimes\ln (1-v)+J_1^{1-w}\otimes\ln (1-w)\\
&+J_1^{y_u}\otimes\ln y_u+J_1^{y_v}\otimes\ln y_v+J_1^{y_w}\otimes\ln y_w\,,
\esp
\ee
where the independent functions are $J_1^{u}$, $J_1^{y_u}$, and $J_1^{y_v}$,
\be
\bsp
J_1^u &= \Big[-\Fuvw +\ln{u}\,\PhiTilde\Big] - (u \leftrightarrow w)\, ,\\
J_1^{y_u} &=\Big[-\Omegawuv-6H_4^u+2H_{3,1}^u-2H_{3,1}^v+2H_{2,1,1}^u
+2\,\big(2\ln{u}-\ln(w/v)\big)H_3^u+\frac{1}{2}\big(H_2^v\big)^2 \\
&\quad\quad+ 2\ln(w/v)H_{2,1}^u -\ln{u}\,\big(\ln{u}-2\ln(w/v)\big)H_2^u
-\frac{1}{2} \ln^2(u/w)H_2^v-\frac{1}{3}\ln{v}\ln^3{u}\\
&\quad\quad+\frac{1}{4}\ln^2{u}\ln^2{w}+\zeta_2\Big(8H_2^u+2H_2^v
+\ln^2(u/w)+4\ln{u}\ln{v}\Big)-14\,\zeta_4 + (u\leftrightarrow w)\Big]\\
&\quad\quad -\ln(u/w)\Big(4H_{2,1}^v+2\ln{v}\,H_2^v
-\frac{1}{3}\ln{v}\ln^2(u/w)\Big)\, ,\\
J_1^{y_v} &= \bigg[-4\Big(H_4^u - H_{3,1}^u + H_{3,1}^v + H_{2,1,1}^u 
- \ln{u}\, (H_3^u - H_{2,1}^u)\Big) - 2\ln^2{u}\, H_2^u \\
&\quad + \Big(H_2^v - 2\ln{u}\,\ln(u/w)\Big)\, H_2^v 
- \frac{1}{3}\ln{u}\, \ln{w}\,\Big(\ln^2(u/w) +\frac{1}{2}\ln{u}\,\ln{w}\Big) \\
&\quad +8\,\zeta_2 \, \Big(H_2^u+\frac{1}{2}\ln^2{u}\Big)
-  8\,\zeta_4  \bigg] + (u \leftrightarrow w) \,.
\esp
\ee
Of the remaining functions, two vanish, $J_1^{v}=J_1^{1-v}=0$, and the others
are simply related,
\be
J_1^{1-u}=J_1^{w}=-J_1^{1-w}=-J_1^{u},
\quad\quad\textrm{and}\quad\quad J_1^{y_w} = J_1^{y_u}(u\leftrightarrow w).
\ee

%%%%%%%%%%%%%%%%%%%%%%%%%%%%

\subsection{$K_1$}

The final parity-odd function of weight five is $\Kuvw$. 
We may write the $\{4,1\}$ component of its coproduct as,
\be
\bsp
\Delta_{4,1}\left(\Kuvw\right) &= 
K_1^{u}\otimes\ln u+K_1^{v}\otimes\ln v+K_1^{w}\otimes\ln w\\
&+K_1^{1-u}\otimes\ln (1-u)+K_1^{1-v}\otimes\ln (1-v)+K_1^{1-w}\otimes\ln (1-w)\\
&+K_1^{y_u}\otimes\ln y_u+K_1^{y_v}\otimes\ln y_v+K_1^{y_w}\otimes\ln y_w\,,
\esp
\ee
where the independent functions are $K_1^{u}$, $ K_1^{y_u}$, and $K_1^{y_v}$,
\be
\bsp
K_1^{u} &= -\Fwuv+\ln{w}\, \PhiTilde\,,\\
K_1^{y_u} &= -2\,\big(H_{3,1}^u+H_{3,1}^v+H_{3,1}^w\big)
-2\ln(v/w)H_3^u  + 2 \ln{u}\, H_3^{w} + 2 \ln{v}\, H_{2,1}^w + 2\ln(uw)H_{2,1}^v \\
&\quad +\frac{1}{2}\,\big(H_2^u+H_2^v+H_2^w-2\,\zeta_2\big)^2  
+\big(\ln{u}\ln(v/w)+\ln{v}\ln{w}\big)\big(H_2^u+H_2^v-2\zeta_2\big)\\
&\quad - \big(\ln{u}\ln(v w)-\ln{v}\ln{w}\big)H_2^w -\ln{u}\ln{v}\ln^2{w} 
- 2\,\zeta_3 \ln(uw/v) + \zeta_4\,,\\
K_1^{y_v} &= \bigg[-4H_{3,1}^u -2\ln(u/w)\,H_3^u+2\ln(u w)\,H_{2,1}^u
+\ln^2{u}\,H_2^u\\
&\quad+2\Big(H_2^u+\frac{1}{2}\ln^2 u\Big)
\Big(H_2^v-\frac{1}{2}\ln^2w-2\zeta_2\Big)+3\zeta_4\bigg] 
+ (u\leftrightarrow w)\, .
\esp
\ee
Of the remaining functions, two vanish, $K_1^{v}=K_1^{1-v}=0$, and the 
others are simply related,
\beq
K_1^{1-u} = -K_1^u, \quad
K_1^{w} = -K_1^{1-w} = K_1^{u}(u\leftrightarrow w)
\quad~\textrm{and}~\quad K_1^{y_w} = K_1^{y_u}(u\leftrightarrow w)\, .
\eeq

%%%%%%%%%%%%%%%%%%%%%%%%%

\subsection{$M_1$}

The $\{4,1\}$ component of the coproduct of the parity-even weight-five
function $M_1$ can be written as,
\be
\bsp
\Delta_{4,1}\left(M_1\right) &= 
M_1^{u}\otimes\ln u+M_1^{v}\otimes\ln v+M_1^{w}\otimes\ln w\\
&+M_1^{1-u}\otimes\ln (1-u)+M_1^{1-v}\otimes\ln (1-v)+M_1^{1-w}\otimes\ln (1-w)\\
&+M_1^{y_u}\otimes\ln y_u+M_1^{y_v}\otimes\ln y_v+M_1^{y_w}\otimes\ln y_w\,,
\esp
\ee
where,
\be
M_1^{1-v}=-M_1^{w}\,,\quad~\textrm{and}~\quad 
M_1^{u} = M_1^{v} = M_1^{1-w} = M_1^{y_u} = M_1^{y_v} = 0 \,.
\ee
The three independent functions consist of one parity-odd function,
\be
M_1^{y_w}  = -F_1(u,v,w),
\ee
and two parity-even functions, 
\be
\bsp
M_1^{1-u} &= \bigg[ -\Omegauvw+ 2\ln{v}\,\Big(H_3^u + H_{2,1}^u\Big) 
+ 2 \ln{u}\, H_{2,1}^v - \Big(H_2^u-\frac{1}{2}\ln^2{u}\Big)
\Big(H_2^v+\frac{1}{2}\ln^2{v}\Big)\\
&\quad + \ln{u}\,\ln{v}\,\Big(H_2^u + H_2^v + H_2^w-2\zeta_2\Big)
+2\zeta_3\,\ln{w} - (v \leftrightarrow w) \bigg] + \Omegavwu+ 2 H_4^w \\
&\quad  + 2 H_{3,1}^w - 6 H_{2,1,1}^w-2\ln{w}\, \Big(H_3^w+H_{2,1}^w\Big) 
- \Big(H_2^v+\frac{1}{2}\ln^2{v}\Big)\Big(H_2^w +\frac{1}{2}\ln^2{w}\Big)\\
&\quad -(H_2^w)^2+ \Big(\ln^2(v/w) -4\,\zeta_2\Big)\,H_2^u + 2\,\zeta_3\,\ln{u} 
+ 6\,\zeta_4 \, ,
\esp
\ee
and,
\be
\bsp
M_1^{w} &= -2\,\Big(H_{3,1}^u - H_{3,1}^v - H_{3,1}^w 
+ \ln(uv/w)\,\big(H_3^u-\zeta_3\big) + \ln{w}\,H_{2,1}^v +\ln{v}\,H_{2,1}^w\Big) \\
&\quad -\frac{1}{2}\,\Big(H_2^u - H_2^v - H_2^w - \ln{v}\, \ln{w} 
+ 2\zeta_2\Big)^2 +\frac{1}{2}\ln^2{v}\ln^2{w}+5 \zeta_4 \, .
\esp
\ee

%%%%%%%%%%%%%%%%%%%%%%%%

\subsection{$N$}
The $\{4,1\}$ component of the coproduct of the parity-even weight-five
function $N$ can be written as,
\be
\bsp
\Delta_{4,1}\left(N\right) &= 
N^{u}\otimes\ln u+N^{v}\otimes\ln v+N^{w}\otimes\ln w\\
&+N^{1-u}\otimes\ln (1-u)+N^{1-v}\otimes\ln (1-v)+N^{1-w}\otimes\ln (1-w)\\
&+N^{y_u}\otimes\ln y_u+N^{y_v}\otimes\ln y_v+N^{y_w}\otimes\ln y_w\,,
\esp
\ee
where,
\be
N^{1-u}=-N^{u}\,,\quad N^{1-v}=-N^v\,,\quad N^{1-w}=-N^{w}\,,
\quad~\textrm{and}~\quad N^{y_u} = N^{y_v} = N^{y_w} = 0 \,.
\ee
Furthermore, $N$ is totally symmetric. In particular,
\be
N^v(u,v,w)= N^u(v,w,u)\,,\quad~\textrm{and}~\quad N^w(u,v,w)= N^u(w,u,v)\, .
\ee
Therefore, it suffices to specify the single independent function, $N^u$,
\be
\bsp
N^{u} &= \bigg[\Omegavwu + 2H_4^v + 2 H_{3,1}^v - 6 H_{2,1,1}^v 
- 2 \ln{v}\,\Big( H_{3}^v + H_{2,1}^v\Big) -(H_2^v)^2\\
&\quad - \Big(H_2^v + \frac{1}{2}\ln^2{v}\Big)
\Big(H_2^w+\frac{1}{2}\ln^2{w}\Big)+6\,\zeta_4\bigg] + (v \leftrightarrow w)\, .
\esp
\ee

%%%%%%%%%%%%%%%%%%%%%%%%%%

\subsection{$O$}

The $\{4,1\}$ component of the coproduct of the parity-even weight-five
function $O$ can be written as,
\be
\bsp
\Delta_{4,1}\left(O\right) &= 
O^{u}\otimes\ln u+O^{v}\otimes\ln v+O^{w}\otimes\ln w\\
&+O^{1-u}\otimes\ln (1-u)+O^{1-v}\otimes\ln (1-v)+O^{1-w}\otimes\ln (1-w)\\
&+O^{y_u}\otimes\ln y_u+O^{y_v}\otimes\ln y_v+O^{y_w}\otimes\ln y_w\,,
\esp
\ee
where
\be
O^{u}=O^v=O^{w}=O^{y_u} = O^{y_v} = O^{y_w} = 0 \,.
\ee
Furthermore, $O$ is totally symmetric. In particular,
\be
O^{1-v}(u,v,w)= O^{1-u}(v,w,u)\,,\quad~\textrm{and}~\quad O^{1-w}(u,v,w)
= O^{1-u}(w,u,v)\, .
\ee
Therefore, it suffices to specify the single independent function, $O^{1-u}$,
\be
\bsp
O^{1-u} &= \bigg[
-\Omega^{(2)}(u,v,w) + 2H_{3,1}^v + (3\ln{u}-2\ln{w}) \, H_{2,1}^v 
+ 2\ln{v}\, H_{2,1}^u - \frac{1}{2}(H_2^v)^2\\
&\quad + \ln{u}\, \ln{v}\, (H_2^u + H_2^v) + \ln(u/v)\, \ln{w}\, H_2^v
 + \frac{1}{2}\ln^2{v}\, H_2^u - \frac{1}{2}\ln^2{w}\, H_2^v \\
&\quad + \frac{1}{4}\ln^2{u}\, \ln^2{v} + (v \leftrightarrow w) \bigg]
+ \Omega^{(2)}(v,w,u) - 2H_2^v\, H_2^w - \ln{v}\, \ln{w}\, H_2^u\\
&\quad - \frac{1}{4}\ln^2{v}\, \ln^2{w} 
+ 2\,\zeta_2\,\Big( H_2^v + H_2^w - \ln{u}\,\ln(vw) + \ln{v}\,\ln{w} \Big) 
- 6\,\zeta_4\,.
\esp
\ee

%%%%%%%%%%%%%%%%%%%%%%%%%%%%%%%%
\subsection{$\Qep$}

The $\{4,1\}$ component of the coproduct of the parity-even weight-five
function $\Qep$ can be written as,
\be
\bsp
\Delta_{4,1}\left(\Qep\right) &= 
\Qep^{u}\otimes\ln u+\Qep^{v}\otimes\ln v+\Qep^{w}\otimes\ln w\\
&+\Qep^{1-u}\otimes\ln (1-u)+\Qep^{1-v}\otimes\ln (1-v)+\Qep^{1-w}\otimes\ln(1-w)\\
&+\Qep^{y_u}\otimes\ln y_u+\Qep^{y_v}\otimes\ln y_v+\Qep^{y_w}\otimes\ln y_w\,,
\esp
\ee
where,
\be
\Qep^{1-v}=-\Qep^v\, ,\quad\; \Qep^{1-w} = -\Qep^{w}\,,
\quad \; \Qep^{y_w}=\Qep^{y_v}\, \quad~\textrm{and}~\quad 
\Qep^u=\Qep^{1-u}=\Qep^{y_u}=0\, .
\ee
The three independent functions consist of one parity-odd function, 
$Q^{y_v}_{\textrm{ep}}$, which is fairly simple,
\begin{dmath*}
Q^{y_v}_{\textrm{ep}} = \frac{1}{64} \biggl[ F_1(u,v,w) + F_1(v,w,u) 
- 2\, F_1(w,u,v) + (\ln{u}-3\,\ln{v}) \tilde{\Phi}_6 \biggr]\, ,
\end{dmath*}
and two parity-even functions, $Q^{v}_{\textrm{ep}}$ and $Q^{w}_{\textrm{ep}}$, 
which are complicated by the presence of a large number of HPLs,
\begin{dmath*}
Q^{v}_{\textrm{ep}} = \frac{1}{32}\Omega^{(2)}(u,v,w) 
+ \frac{1}{16}\Omega^{(2)}(v,w,u) + \frac{1}{32}H_4^u + \frac{3}{32}H_4^v 
+ \frac{1}{16}H_4^w - \frac{3}{32}H_{3,1}^u - \frac{3}{32}H_{2,1,1}^u 
- \frac{9}{64}H_{2,1,1}^v - \frac{3}{16}H_{2,1,1}^w + \frac{1}{32}\ln{u}\, H_3^v 
- \frac{1}{16}\ln{u}\, H_3^w - \frac{3}{32}\ln{u}\, H_{2,1}^v 
+ \frac{1}{16}\ln{u}\, H_{2,1}^w + \frac{1}{32}\ln{v}\, H_3^u 
- \frac{3}{32}\ln{v}\, H_3^v - \frac{7}{32}\ln{v}\, H_{2,1}^u 
+ \frac{1}{16}\ln{v}\, H_{2,1}^w - \frac{1}{32}\ln{w}\, H_3^u 
- \frac{1}{32}\ln{w}\, H_3^v + \frac{3}{32}\ln{w}\, H_{2,1}^u 
+ \frac{3}{32}\ln{w}\, H_{2,1}^v - \frac{1}{16}\ln{w}\, H_{2,1}^w 
+ \frac{1}{32}(H_2^u)^2 - \frac{3}{128}(H_2^v)^2 - \frac{1}{64}H_2^v\, H_2^w 
+ \frac{1}{64}(H_2^w)^2 + \frac{1}{64}\ln^2{u}\, H_2^u 
+ \frac{1}{64}\ln^2{u}\, H_2^w - \frac{3}{32}\ln{u}\, \ln{v}\, H_2^u 
- \frac{3}{32}\ln{u}\, \ln{v}\, H_2^v - \frac{1}{32}\ln{u}\, \ln{v}\, H_2^w 
+ \frac{1}{32}\ln{u}\, \ln{w}\, H_2^u + \frac{1}{32}\ln{u}\, \ln{w}\, H_2^w 
- \frac{1}{16}\ln^2{v}\, H_2^u + \frac{3}{128}\ln^2{v}\, H_2^v 
- \frac{1}{128}\ln^2{v}\, H_2^w + \frac{1}{16}\ln{v}\, \ln{w}\, H_2^u 
+ \frac{3}{32}\ln{v}\, \ln{w}\, H_2^v + \frac{1}{32}\ln{v}\, \ln{w}\, H_2^w 
- \frac{1}{128}\ln^2{w}\, H_2^v - \frac{1}{128}\ln^2{u}\, \ln^2{v} 
+ \frac{1}{64}\ln^2{u}\, \ln{v}\, \ln{w} 
- \frac{3}{64}\ln{u}\, \ln^2{v}\, \ln{w} + \frac{5}{256}\ln^2{v}\, \ln^2{w}
- \zeta_2\Big(\frac{1}{8}H_2^u + \frac{11}{128}H_2^v + \frac{1}{16}H_2^w 
+ \frac{1}{32}\ln^2{u} - \frac{3}{16}\ln{u}\, \ln{v} 
+ \frac{1}{16}\ln{u}\, \ln{w} - \frac{3}{128}\ln^2{v} 
+ \frac{3}{16}\ln{v}\, \ln{w}\Big) + \frac{7}{32}\zeta_3\,\ln{v}\, ,
\end{dmath*}
\begin{dmath}
Q^{w}_{\textrm{ep}} = -\frac{1}{32}\Omega^{(2)}(v,w,u) 
+ \frac{1}{32}\Omega^{(2)}(w,u,v) + \frac{1}{32}H_4^u - \frac{1}{32}H_4^v 
+ \frac{3}{32}H_{3,1}^u - \frac{3}{32}H_{3,1}^v - \frac{3}{32}H_{2,1,1}^u 
+ \frac{3}{32}H_{2,1,1}^v + \frac{1}{32}\ln{u}\, H_3^v 
- \frac{1}{16}\ln{u}\, H_3^w - \frac{1}{32}\ln{u}\, H_{2,1}^v 
- \frac{1}{32}\ln{v}\, H_3^u + \frac{1}{16}\ln{v}\, H_3^w 
- \frac{3}{32}\ln{v}\, H_{2,1}^u + \frac{1}{8}\ln{v}\, H_{2,1}^v 
+ \frac{1}{32}\ln{w}\, H_3^u - \frac{1}{64}\ln{w}\, H_3^v 
- \frac{1}{32}\ln{w}\, H_{2,1}^u + \frac{3}{64}\ln{w}\, H_{2,1}^v 
- \frac{1}{64}(H_2^u)^2 + \frac{1}{64}(H_2^v)^2 + \frac{1}{16}H_2^v\, H_2^w 
+ \frac{1}{64}\ln^2{u}\, H_2^u + \frac{1}{64}\ln^2{u}\, H_2^v 
- \frac{1}{16}\ln{u}\, \ln{v}\, H_2^u - \frac{1}{16}\ln{u}\, \ln{v}\, H_2^v 
- \frac{1}{32}\ln{u}\, \ln{w}\, H_2^v + \frac{3}{64}\ln^2{v}\, H_2^u 
+ \frac{3}{64}\ln^2{v}\, H_2^v + \frac{1}{32}\ln^2{v}\, H_2^w 
- \frac{1}{32}\ln{v}\, \ln{w}\, H_2^u + \frac{3}{64}\ln{v}\, \ln{w}\, H_2^v 
- \frac{1}{64}\ln^2{w}\, H_2^u + \frac{1}{64}\ln^2{w}\, H_2^v 
+ \frac{1}{64}\ln^2{u}\, \ln{v}\, \ln{w} - \frac{1}{128}\ln^2{u}\, \ln^2{w} 
- \frac{3}{64}\ln{u}\, \ln^2{v}\, \ln{w} + \frac{3}{128}\ln^3{v}\, \ln{w} 
+ \frac{1}{128}\ln^2{v}\, \ln^2{w} + \zeta_2\,\Big(\frac{1}{16}H_2^u 
- \frac{1}{16}H_2^v - \frac{1}{16}H_2^w - \frac{1}{32}\ln^2{u} 
+ \frac{1}{8}\ln{u}\, \ln{v} - \frac{3}{32}\ln^2{v}\Big) 
- \frac{1}{16} \zeta_3\, \ln{w}\, .
\end{dmath}

%%%%%%%%%%%%%%%%%%%%%%%%%%%%%%%%
\subsection{Relation involving $M_1$ and $\Qep$}

There is one linear relation between the six permutations of $M_1$
and the six permutations of $\Qep$.
This linear relation involves the totally antisymmetric linear
combination of the $S_3$ permutations of both $M_1$ and $\Qep$.
It can be written as,
\be
\label{eq:M1_combo}
\bigg[\Big(M_1(u,v,w) - \frac{64}{3}\,\Qep(u,v,w)
+2 \ln u \,\Omegauvw + E_{\textrm{rat}}(u,v)\Big) 
- \big(u\leftrightarrow v)\bigg] + \textrm{cyclic} = 0\,,
\ee
where $E_{\textrm{rat}}(u,v)$ is constructed purely from ordinary HPLs,
\be
\bsp
E_{\textrm{rat}}(u,v) &= \Big(H_2^v+\frac{1}{2}\ln^2 v\Big)
\Big(\frac{5}{3} ( H_3^u + H_{2,1}^u) 
 + \frac{1}{3} \ln u \, H_2^u - \frac{1}{2} \ln^3 u \Big)
- 4 \, H_2^v \, \Big( 2 \, H_{2,1}^u + \ln u \, H_2^u\Big)\\
&\quad- 2 \ln v \, \Big( H_4^u + 5\,H_{3,1}^u - 3\,H_{2,1,1}^u
   - \frac{1}{2} (H_{2}^u)^2
   + \ln u\, ( H_{3}^u - H_{2,1}^u ) + 4 \zeta_2 \, H_2^u \Big)\,.
\esp
\ee
Because of this relation, the images of $M_1$ and $\Qep$ under the $S_3$
symmetry group together provide only 11, not 12, of the 13 non-HPL basis
functions for $\mathcal{H}_5^+$.  The totally symmetric functions
$N$ and $O$ provide the remaining two basis elements.

\vfill\eject

%%%%%%%%%%%%%%%%%%%%%%%%%%%%%%%%%%%%%%%%%%%%%%%%%%%%%%%%%%%%%%%%%%%%%%%%%%

\section{Coproduct of $\Rep$}
\label{sec:app_Rep}

We may write the $\{5,1\}$ component of the coproduct of the
parity-even weight-six function $\Rep$ as,
\be
\bsp
\Delta_{5,1}\left(\Rep\right) &= 
\Rep^{u}\otimes\ln u+\Rep^{v}\otimes\ln v+\Rep^{w}\otimes\ln w\\
&+\Rep^{1-u}\otimes\ln (1-u)+\Rep^{1-v}\otimes\ln (1-v)+\Rep^{1-w}\otimes\ln(1-w)\\
&+\Rep^{y_u}\otimes\ln y_u+\Rep^{y_v}\otimes\ln y_v+\Rep^{y_w}\otimes\ln y_w\,,
\esp
\ee
where,
\be
\Rep^{v} = -\Rep^{1-v} =  -\Rep^{1-u}(u\leftrightarrow v) 
=  \Rep^{u}(u\leftrightarrow v)\,,\quad\;  \Rep^{y_v} 
= \Rep^{y_u}\, , \quad~\textrm{and}~\quad \Rep^{w}=\Rep^{1-w}=\Rep^{y_w} = 0\, .
\ee
The two independent functions may be written as,
\be
\bsp
\Rep^{y_u} & = \frac{1}{32} \biggl\{
- H_1(u,v,w) - 3 \, H_1(v,w,u) - H_1(w,u,v)
+ \frac{3}{4} \, ( J_1(u,v,w) + J_1(v,w,u) + J_1(w,u,v) )\\
&\hskip0.5cm\null
+ \Bigl[ - 4 \, ( H_2^u + H_2^v ) - \ln^2u - \ln^2v + \ln^2w
    + 2 \, \Bigl( \ln u \, \ln v -  ( \ln u + \ln v ) \, \ln w \Bigr)
    + 22 \, \zeta_2 \Bigr] \PhiTilde \biggr\}\,,
\esp
\label{Rep_yu}
\ee
and,
\be
\bsp
\Rep^{u} &= 
- \frac{1}{3} \Bigl( 2 \, ( \Qep(u,v,w) - \Qep(u,w,v) + \Qep(v,w,u) )
        + \Qep(v,u,w) - 3 \Qep(w,v,u) \Bigr)\\
&\hskip0.5cm\null
+ \frac{1}{32} \Bigl[ M_1(u,v,w) - M_1(v,u,w)
+ \Bigl( 5 \, ( \ln u - \ln v ) + 4 \, \ln w \Bigr)  \Omegauvw\\
&\hskip1.5cm\null
- ( 3 \, \ln u + \ln v - 2 \, \ln w ) \, \Omegavwu
- ( \ln u + 3 \, \ln v - 4 \, \ln w ) \, \Omegawuv \Bigr]\\
&\hskip0.5cm\null
+  R_{\textrm{ep, rat}}^{u}\, ,
\esp
\label{Rep_u}
\ee
where,
\be\bsp
R_{\textrm{ep, rat}}^{u} &= \frac{1}{32} \biggl\{
24 \, H_5^u - 14 \, (H_{4,1}^u - H_{4,1}^v) - 16 \, H_{4,1}^w
 + \frac{5}{2} \, H_{3,2}^u + \frac{11}{2} \, H_{3,2}^v - 8 \, H_{3,2}^w 
+ 42 \, H_{3,1,1}^u\\
&\hskip-0.8cm\null
+ 24 \, H_{3,1,1}^v + 6 \, H_{3,1,1}^w
 + \frac{13}{2} \, H_{2,2,1}^u + \frac{15}{2} \, H_{2,2,1}^v + 2 \, H_{2,2,1}^w 
- 36 \, H_{2,1,1,1}^u - 36 \, H_{2,1,1,1}^v + 24 \, H_{2,1,1,1}^w\\
&\hskip-0.8cm\null
 +  \Bigl( \frac{15}{2} \, H_{2,1}^v - 5 \, H_3^u + \frac{1}{2} \, H_{2,1}^u 
- \frac{1}{3} \, H_3^w - \frac{1}{2} \, H_3^v - \frac{31}{3} \, H_{2,1}^w \Bigr)  
 \,  H_2^u
 +  \Bigl(  - \frac{5}{3} \, H_3^w + \frac{7}{2} \, H_{2,1}^v 
- \frac{5}{3} \, H_{2,1}^w \\
&\hskip-0.8cm\null
- 3 \, H_3^v + 4 \, H_{2,1}^u - 14 \, H_3^u \Bigr)  \,  H_2^v
 +  \Bigl(  - \frac{7}{6} \, H_3^u - \frac{7}{3} \, H_{2,1}^v + 4 \, H_3^w 
+ \frac{5}{3} \, H_3^v + \frac{17}{6} \, H_{2,1}^u \Bigr)  \,  H_2^w\\
&\hskip-0.8cm\null
 +  \Bigl(  - 14 \, H_4^u + 16 \, H_4^v + 19 \, H_{3,1}^u - 2 \, (H_{3,1}^v + H_{3,1}^w)
     - \frac{57}{2} \, H_{2,1,1}^u - 24 \, (H_{2,1,1}^v - H_{2,1,1}^w)
     + \frac{1}{4} \, (H_2^u)^2 \\
&\hskip-0.8cm\null
  - \frac{5}{2} \, (H_2^v)^2 + \frac{3}{2} \, (H_2^w)^2
     + 6 \, H_2^u \, H_2^v - \frac{17}{6} \, H_2^u \, H_2^w
 - 4 \, H_2^v \, H_2^w \Bigr) \,  \ln u
 +  \Bigl(  - 10 \, H_4^u - 8 \, (H_4^w + H_4^v) 
\esp
\ee
\be
\bsp
&\null
- 4 \, H_{3,1}^u + 3 \, H_{3,1}^v + 2 \, H_{3,1}^w
     + 6 \, H_{2,1,1}^u - \frac{3}{2} \, H_{2,1,1}^v
     + \frac{11}{2} \, (H_2^u)^2 + \frac{19}{4} \, (H_2^v)^2 
     + \frac{1}{2} \, (H_2^w)^2
     + \frac{17}{2} \, H_2^u \, H_2^v \\
&\null
 + 4 \, H_2^u \, H_2^w + \frac{7}{3} \, H_2^v \, H_2^w \Bigr)  \,  \ln v
 +  \Bigl( 10 \, (H_4^u + H_4^w) + 8 \, H_4^v + 6 \, H_{3,1}^u 
- 8 \, H_{3,1}^v + 2 \, H_{3,1}^w\\
&\null
    - 6 \, (H_{2,1,1}^u + H_{2,1,1}^w)  -  6 \, (H_2^u)^2 - 5 \, (H_2^v)^2 
    - 2 \, (H_2^w)^2 - 8 \, H_2^u \, H_2^v - \frac{17}{3} \, H_2^u \, H_2^w 
    - \frac{1}{3} \, H_2^v \, H_2^w \Bigr)  \,  \ln w\\
&\null
 +  \Bigl( \frac{1}{2} \, H_3^u + \frac{3}{4} \, H_3^v + \frac{5}{6} \, H_3^w 
 - 6 \, H_{2,1}^u - \frac{21}{4} \, H_{2,1}^v + \frac{35}{6} \, H_{2,1}^w
 \Bigr)  \,  \ln^2u
 + \Bigl(  - 7 \, H_3^u + \frac{13}{2} \, H_3^v + \frac{1}{6} \, H_3^w\\
&\null
 + 4 \, H_{2,1}^u + 2 \, H_{2,1}^v - \frac{11}{6} \, H_{2,1}^w \Bigr)  \,  \ln^2v
 + \Bigl(  - \frac{7}{12} \, H_3^u + \frac{11}{6} \, H_3^v - 7 \, H_3^w 
+ \frac{17}{12} \, H_{2,1}^u - \frac{1}{6} \, H_{2,1}^v \Bigr)  \,  \ln^2w\\
&\null
 +  \Bigl( 6 H_3^u - 14 H_3^v - 2 H_3^w + 4 H_{2,1}^v - 2 H_{2,1}^w 
  \Bigr)  \,  \ln u  \,  \ln v
 +  \Bigl(  - 6 H_3^u + 2 H_3^w - 2 H_3^v + 2 H_{2,1}^w 
  \Bigr)  \,  \ln u  \,  \ln w\\
&\null
 +  \Bigl(  - 10 \, H_3^v + 6 \, H_3^w - 2 \, H_{2,1}^u + 4 \, H_{2,1}^v 
  - 2 H_{2,1}^w \Bigr)  \,  \ln v  \,  \ln w
 +  \Bigl( \frac{1}{4} \, H_2^u - \frac{5}{2} \, H_2^v + \frac{3}{4} \, H_2^w \Bigr) 
   \,  \ln^3u\\
&\null
 +  \Bigl( \frac{7}{4} H_2^u - \frac{1}{4} H_2^v + H_2^w \Bigr)  \, \ln^3\!v
 +  \Bigl( \frac{1}{2} H_2^u + \frac{1}{2} H_2^v + 2 \, H_2^w  
  \Bigr)  \, \ln^3\!w
 +  \Bigl(  - \frac{1}{2} H_2^w - H_2^u + \frac{3}{4} H_2^v  
    \Bigr)  \,  \ln^2\!u  \,  \ln v\\
&\null
 +  \Bigl( \frac{9}{2} \, H_2^u + 6 \, H_2^v - \frac{3}{2} \, H_2^w 
     \Bigr)  \,  \ln u  \,  \ln^2 v
 +  \Bigl(  - \frac{5}{6} \, H_2^w + H_2^u - H_2^v  \Bigr) \, \ln^2 u \, \ln w\\
&\null
 +  \Bigl(  - \frac{11}{12} \, H_2^u - \frac{1}{2} \, H_2^v - H_2^w 
    \Bigr)  \,  \ln u  \,  \ln^2w
 +  \Bigl(  - \frac{1}{6} \, H_2^w + 2 \, H_2^v  \Bigr)  \,  \ln^2 v  \,  \ln w\\
&\null
 +  \Bigl(  - \frac{5}{2} \, H_2^u - 3 \, H_2^w - \frac{4}{3} \, H_2^v 
    \Bigr)  \,  \ln v  \,  \ln^2w
 -  2  \,  ( H_2^u + H_2^v - H_2^w )  \,  \ln u  \,  \ln v  \,  \ln w\\
&\null
 + \frac{3}{8} \, \ln^3u \, \ln^2w - \frac{3}{4} \, \ln^2 u \, \ln^3w 
 + \ln^2 u \, \ln^2 v \, \ln w
 - \frac{7}{4} \, \ln^2 u \, \ln v \, \ln^2w 
 - \frac{7}{4} \, \ln u \, \ln^2 v \, \ln^2w \\
&\null
 + 2 \, \ln u \, \ln v \, \ln^3w
 - \frac{5}{4} \, \ln^3u \, \ln^2 v + \frac{7}{8} \, \ln^2 u \, \ln^3v 
+ \frac{1}{2} \, \ln^3v \, \ln^2w - \frac{3}{4} \, \ln^2 v \, \ln^3w\\
&\null
 +  \zeta_2  \,  \Bigl[ \frac{33}{4} \, H_3^u - \frac{9}{4} \, H_3^v
   + 2 \, H_3^w + H_{2,1}^u - 17 \, H_{2,1}^v + 24 \, H_{2,1}^w
   +  \Bigl( 14 \, H_2^w + \frac{7}{4} \, H_2^u - 10 \, H_2^v \Bigr) \, \ln u\\
&\hskip1cm\null
   +  \Bigl(  - 6 \, H_2^w - 18 \, H_2^u - \frac{47}{4} \, H_2^v \Bigr)  \,  \ln v
   +  \Bigl( 8 \, H_2^v + 6 \, H_2^w + 20 \, H_2^u \Bigr)  \,  \ln w
   - \frac{1}{4} \, \ln^3u + \frac{1}{4} \, \ln^3v \\
&\hskip1cm\null
   - 4 \, \ln^3w + 2 \, \ln^2 u \, \ln v - 12 \, \ln u \, \ln^2 v
   - 2 \, \ln w \, \ln^2 u + 2 \, \ln u \, \ln^2w - 4 \, \ln^2 v \, \ln w\\
&\hskip1cm\null
   + 6 \, \ln v \, \ln^2w
   + 12 \, \ln u \, \ln v \, \ln w \Bigr]
 +  \zeta_3  \,  \Bigl[ 7 \, H_2^u - 5 \, H_2^v - 2 \, H_2^w 
   + \frac{3}{2} \, \ln^2 u - \frac{1}{2} \, \ln^2 v - \ln^2w \Bigr]\\
&\null
 +  \zeta_4  \,  \Bigl[ 14 \, \ln u + 50 \, \ln v - 44 \, \ln w \Bigr] \biggr\} \,.
\esp
\label{Rep_u_rat}
\ee

\vfill\eject

%%%%%%%%%%%%%%%%%%%%%%%%%%%%%%%%%%%%%%%%%%%%%%%%%%%%%%%%%%%%%%%%%%%%%%%%

\end{document}